\documentclass[iop]{emulateapj}

\usepackage{morefloats}

\shorttitle{The Dependence of C\,{\sc iv} BAL Properties on Accompanying Si\,{\sc iv} 
and Al\,{\sc iii} BALs}
\shortauthors{Filiz Ak et~al.}

\usepackage{natbib}
\begin{document}

\title{The Dependence of C\,{\sc iv} Broad Absorption Line Properties 
on Accompanying Si\,{\sc iv} and Al\,{\sc iii} Absorption:  
Relating Quasar-Wind Ionization Levels, Kinematics,  and Column Densities}
 
\author{N. Filiz Ak\altaffilmark{1,2,3},
W.~N. Brandt\altaffilmark{1,2}, 
P. B. Hall\altaffilmark{4},  
D.~P. Schneider\altaffilmark{1,2}, 
J.~R. Trump\altaffilmark{1,2,5}, 
S. ~F. Anderson \altaffilmark{6}, 
F. Hamann\altaffilmark{7}, 
Adam~D. Myers\altaffilmark{8},
I. P\^aris \altaffilmark{9},
P. Petitjean\altaffilmark{10}, 
Nicholas~P. Ross \altaffilmark{11,12},   
Yue Shen\altaffilmark{5,13},
Don York\altaffilmark{14,15}}

\altaffiltext{1}{Department of Astronomy \&   Astrophysics, Pennsylvania 
State University,  University Park,  PA, 16802, USA}
\altaffiltext{2}{Institute for Gravitation and the Cosmos, Pennsylvania 
State University, University Park, PA 16802, USA}
\altaffiltext{3}{Faculty of Sciences, Department of Astronomy and Space 
Sciences, Erciyes University, 38039 Kayseri, Turkey}
\altaffiltext{4}{Department of Physics and Astronomy, York University, 
4700 Keele St., Toronto, Ontario,  M3J 1P3, Canada}
\altaffiltext{5}{Hubble Fellow}
\altaffiltext{6}{Astronomy Department, University of Washington, Seattle, WA 98195, USA}
\altaffiltext{7}{Department of Astronomy, University of Florida, Gainesville, FL 32611-2055, USA}
\altaffiltext{8}{Department of Physics and Astronomy, University of Wyoming, Laramie, WY 82071, USA}
\altaffiltext{9}{Departamento de Astronom\'ia, Universidad de Chile, Casilla 36-D, Santiago, Chile}
\altaffiltext{10}{Universite Paris 6, Institut d'Astrophysique de Paris, 75014, 
Paris, France}
\altaffiltext{11}{Lawrence Berkeley National Laboratory, 1 Cyclotron Road, Berkeley, CA 94720, USA}
\altaffiltext{12}{Department of Physics, Drexel University, 3141 Chestnut Street, Philadelphia, PA 19104, USA}
\altaffiltext{13}{Carnegie Observatories, 813 Santa Barbara Street, Pasadena, CA 91101, USA}
\altaffiltext{14}{The University of Chicago, Department of Astronomy and Astrophysics, Chicago, IL 60637, USA}
\altaffiltext{15}{The University of Chicago, Enrico Fermi Institute, Chicago, IL 60637, USA}

\email{nfilizak@astro.psu.edu}

\begin{abstract}

We consider how the profile and multi-year variability properties of a
large sample of C\,{\sc iv} Broad Absorption Line (BAL) troughs change when BALs
from Si\,{\sc iv} and/or Al\,{\sc iii} are present at corresponding velocities, indicating
that the line-of-sight intercepts at least some lower ionization gas.
We derive a number of observational results for C\,{\sc iv} BALs
separated according to the presence or absence of accompanying lower
ionization transitions, including measurements of composite profile shapes,
equivalent width (EW), characteristic velocities, composite variation profiles, and EW variability.
We also measure the correlations between EW and fractional-EW variability
for C\,{\sc iv}, Si\,{\sc iv}, and Al\,{\sc iii}.  Our measurements reveal the basic correlated
changes between ionization level, kinematics, and column density expected
in accretion-disk wind models; e.g., lines-of-sight including lower
ionization material generally show deeper and broader C\,{\sc iv} troughs that have 
smaller minimum velocities and 
that are less variable.  Many C\,{\sc iv} BALs with no accompanying Si\,{\sc iv}
or Al\,{\sc iii} BALs may have only mild or no saturation.

\end{abstract}

\section{Introduction}\label{hmlintro}

Broad  Absorption Lines (BALs) in quasar spectra are seen as a result of high-velocity outflows. 
The most commonly used empirical definition of BALs requires an absorption feature to have at 
least a 2000~km\,s$^{-1}$ width at 10\% under the continuum level \citep[e.g.,][]{wey91}. 
BAL quasars  exhibit such broad absorption troughs in a wide variety of species in their rest-frame 
ultraviolet spectra, such as 
P\,{\sc v}  $\lambda$$\lambda$1118, 1128~\AA,
Ly$\alpha$ $\lambda$1216~\AA,  
N\,{\sc v}  $\lambda$$\lambda$1239, 1243~\AA, 
Si\,{\sc iv}  $\lambda$$\lambda$1394, 1403~\AA,
C\,{\sc iv}  $\lambda$$\lambda$1548, 1551~\AA,
Al\,{\sc ii}  $\lambda$1671~\AA,
Al\,{\sc iii}  $\lambda$$\lambda$1855, 1863~\AA, and
Mg\,{\sc ii}  $\lambda$$\lambda$2797, 2804~\AA.  


BAL quasars are classified into three groups based on the observed transitions in their spectra. 
The majority of BAL quasars exhibit  absorption from only  high-ionization transitions 
such as N\,{\sc v}, Si\,{\sc iv}, and C\,{\sc iv} \citep[HiBALs, e.g.,][]{wey91}. Approximately 10\% of BAL 
quasars  in optically selected samples also exhibit absorption from low-ionization transitions such 
as  Al\,{\sc ii}, Al\,{\sc iii}, and  Mg\,{\sc ii}  \citep[LoBALs, e.g.,][]{voit93,gibson09}. 
Only $\approx$~1\% of BAL quasars show absorption from excited states of Fe\,{\sc ii} and/or Fe\,{\sc iii} in 
addition to the high and low-ionization transitions listed above \citep[FeLoBALs, e.g.,][]{becker00,hall02}. 
The existence of these groups 
indicates that quasar outflows can have a wide range of ionization states.  It has been argued that 
the presence of the low-ionization lines is not  the only difference between these groups
\citep[e.g.,][]{boroson92,turnshek94,zhang10}. The generally weak [O\,{\sc iii}]  emission and strong reddening 
of LoBALs suggest that LoBALs tend to be  surrounded by dust and gas that has a larger global covering 
factor compared to HiBALs. 

The details of the structure and geometry of  quasar outflows remain unclear. A commonly adopted and 
well-developed model suggests that many BALs are formed in an equatorial wind that is launched from the 
accretion disk at $\approx 10^{16}$--$10^{17}$~cm  from the central supermassive black hole (SMBH)  
for black-hole masses of ~\hbox{$10^{8}$--$10^{9}$~M$_{\odot}$} 
and is driven by radiation pressure \citep[e.g.,][]{murray95,proga00,higginbottom}. 
This  disk-wind model  successfully explains several important observational facts about 
BAL quasars, such as the presence of  absorption from both high and low-ionization transitions despite the 
luminous ionizing  radiation from the central source, and  the large range of outflow velocities 
that reach from the systemic velocity up to $0.1c$. Numerical hydrodynamical simulations of the disk-wind model 
provide detailed predictions of the structure and dynamics of quasar outflows \citep[e.g.,][]{proga00}.

Previous  studies have shown that  some  BAL troughs  are much more optically thick than they appear 
and that the depths of such troughs only mildly depend on column density.  Several pieces of evidence 
for  this line-saturation interpretation have been  presented, such as  P\,{\sc v} BALs, depth differences in 
unblended  doublet lines, and ``flat-bottom''  
BAL profiles \citep[e.g.,][]{arav97,hamann98,arav99a,arav99,arav01,leighly09,borguet12}. 
For instance, \citet{hamann98} studied spectra of the BAL quasar  
PG~1254+047 which possesses  relatively strong P\,{\sc v} absorption at velocities corresponding to  non-black strong  
C\,{\sc iv} and S\,{\sc iv} BAL troughs. The existence of P\,{\sc v} absorption was taken as evidence for saturated C\,{\sc iv} 
absorption since phosphorus is expected to be $\sim 1000$ times less abundant than carbon (based on solar abundances). 
As another example, \citet{arav99} argued that the depth differences between the unblended Si\,{\sc iv} doublet lines 
of the quasar FIRST~J1603+3002 arise as a result of 
velocity-dependent partial coverage. Calculating the optical depths, they found that the C\,{\sc iv} and Si\,{\sc iv} absorption 
lines  are saturated.  
Such studies  have suggested that the non-black nature of these saturated lines arises due to partial coverage of the 
emission source along the line-of-sight;  BAL troughs do not reach zero intensity due to  photons from the emission source 
that are not absorbed and/or  are  scattered into the observer's  line-of-sight. 
Supporting this argument, spectropolarimetric observations of BAL quasars have shown that the fractional contribution 
from scattered emission often increases at the wavelengths where  BAL troughs are found, indicating an excess of 
scattered light relative to  direct light \citep[e.g.,][]{ogle99}. 
These lines of observational evidence indicate that  some BAL quasars have  highly saturated BAL troughs and that the 
depths of such troughs  are largely set by the line-of-sight covering factor rather than column density. However, 
detailed studies of line saturation are only available for a limited number of objects and have often focused on deep 
C\,{\sc iv} BALs. 
It is possible that some BALs might be weak simply because the column density along the line-of-sight is 
small.

Recent sample-based investigations of BAL variability  have brought new insights about the structure, dynamics, 
and evolution of quasar outflows showing that  variability is common  for most  BAL troughs  on multi-year timescales 
\citep[e.g.,][]{lundgren07,gibson08,gibson10,cap11,cap12,ak12, ak13,vivek12,wildy13}. 
These studies have revealed that the fractional variation of BAL 
troughs in lower ionization transitions (such as Si\,{\sc iv}) 
is generally stronger than that in C\,{\sc iv} \citep[e.g.,][]{cap12,vivek12,ak13}. 
A recent study by 
\citet{ak13} presented a detailed investigation of the variability of C\,{\sc iv} and Si\,{\sc iv} BALs on 
multi-year timescales in a large quasar sample assessing  variation characteristics and the  lifetimes 
of BAL troughs. This study found  coordinated trough variability for BAL quasars showing multiple  C\,{\sc iv}  
troughs; they suggested that global changes in ionization level are  the most straightforward mechanism for 
explaining such coordinated variability of multiple  C\,{\sc iv}  troughs at different velocities.
This mechanism would require at least some BAL troughs not to be highly saturated, as highly saturated 
troughs should not be responsive to the expected changes in ionization level.

The available analytic calculations and numerical simulations 
of quasar disk winds predict the ionization level, kinematics, and 
column density of the outflowing gas along possible lines-of-sight 
to the relevant emission region \citep[e.g.,][]{murray95,proga00,higginbottom}. 
These three quantities are expected generally to show correlated changes as the 
line-of-sight is varied. Thus, we expect correlated 
object-to-object changes of resulting observable phenomena 
such as the BAL transitions present, BAL-profile shapes, and 
BAL variability.  In this paper, we aim to investigate systematically and quantify 
such correlated object-to-object changes for a large sample of BAL quasars with 
uniform high-quality measurements from the Sloan Digital Sky Survey \citep[SDSS,][]{york00}. 
Utilization of a large sample is important to overcome object-to-object scatter 
associated with, e.g., time-variable wind inhomogeneities.

To probe correlated object-to-object changes of ionization 
level, kinematics, and column density, we require a basic
means of identifying lines-of-sight with different average 
ionization levels. We accomplish this using the 
strong BAL transitions of C\,{\sc iv}, Si\,{\sc iv}, and Al\,{\sc iii}. 
These three transitions span a significant range of ionization potential 
 \citep[with creation ionization potentials of 47.9, 33.5, and 18.8 eV, 
respectively, e.g.,][]{hall02},
and their BAL regions can all be measured 
simultaneously in SDSS spectra of quasars with redshift \hbox{$1.9<z<3.9$}. 
Another practical advantage of using these three transitions is 
that their local continuum emission can be modeled more reliably 
than that for, e.g., Mg\,{\sc ii}. Lines-of-sight with C\,{\sc iv}  BALs but 
not Si\,{\sc iv} or Al\,{\sc iii} BALs intercept only relatively highly 
ionized gas. Lines-of-sight with C\,{\sc iv}  and Si\,{\sc iv} (but not 
Al\,{\sc iii}) BALs intercept at least some less ionized gas.
Finally, lines-of-sight with BALs from all three ions intercept 
at least some even less ionized gas. In the numerical simulations
of \citet{proga00}, these three lines-of-sight lie at 
progressively larger inclinations relative to the rotational
axis of the accretion disk. 

In this study,  utilizing multi-epoch observations from the SDSS (Section~\ref{colobs}), 
we classify C\,{\sc iv} BAL troughs into three groups considering the corresponding BAL regions of Si\,{\sc iv} and 
Al\,{\sc iii} (Section~\ref{ident}). We present the observational  results of our investigation in Section~\ref{hmlresults}. 
In Section 5, we present a summary of our results and  a discussion of their implications for disk-wind models.

Throughout this study,  timescales and EWs are given in the rest frame of the quasar unless stated otherwise.
Negative signs for velocities indicate that a BAL trough is blueshifted with respect to the systemic velocity.
We adapt  a cosmology with \linebreak
\hbox{$H_0=70$~km~s$^{-1}$~Mpc$^{-1}$}, 
\hbox{$\Omega_M=0.3$}, and 
\hbox{$\Omega_{\Lambda}=0.7$} \citep[e.g.,][]{spergel03}.

\section{Observations,  Sample Selection, and Data Preparation}\label{colobs}

We utilize spectroscopic observations from the Sloan Digital Sky Survey-I/II \citep[hereafter SDSS,][]{york00} 
and the Baryon Oscillation Spectroscopic Survey of SDSS-III  \citep[hereafter BOSS,][]{eisen11,dawson13}. 
SDSS  is a large multi-filter  imaging and spectroscopic survey using a dedicated 2.5-m optical telescope 
\citep{gunn98,gunn06} at Apache Point Observatory in New Mexico. 
During its first phase of operations, \hbox{2000--2005}, the SDSS imaged more than 
8000~deg$^2$ of the sky in five optical bandpasses, and it obtained spectra of galaxies and quasars.  
The SDSS spectral coverage was continuous from 3800~\AA\, to 9200~\AA\, at a resolution of 1800--2200 
\citep[e.g.,][]{york00}.
In 2005 the survey entered a new phase, the SDSS-II, expanding its spectroscopic samples to over 800{,}000 
galaxies and 100{,}000 quasars. The BOSS, part of the third phase of SDSS operations, is  acquiring 
spectra for approximately  1.5 million luminous galaxies and 160,000 quasars 
\citep[e.g.,][]{anderson12,ross12}.
The BOSS survey started operating in mid-2008 and   is planned to 
continue observations until the end of June 2014.
The BOSS quasar survey provides an outstanding opportunity  for investigating intrinsic UV absorption in 
quasars, owing to its focus upon selection at  $z >$ 2.1 which shifts the important  C\,{\sc iv} and Si\,{\sc iv} 
transitions well into its spectral coverage \citep{smee13}. 
The BOSS spectral coverage is continuous from 3600~\AA\, to 10{,}000~\AA\, at a resolution of 1300--3000 
\citep[e.g.,][]{dawson13}.

An ancillary BOSS project aims  to investigate the dynamics of quasar winds over multi-year timescales
utilizing  second-epoch spectra for 2005 BAL quasars originally identified in the SDSS-I/II spectroscopy by
\citet{gibson09}. These 2005 quasars were initially selected  following \hbox{$i < 19.3$}, 
\hbox{$0.48 < z < 4.65$}, \hbox{SN$_{1700} > 6$}, and \hbox{BI$_0 > 100$ km\,s$^{-1}$} criteria. Here  
SN$_{1700}$ is the average  signal-to-noise ratio in a  4~\AA\, resolution element within \hbox{1650--1750~\AA,} 
and BI$_0$ is the modified ``balnicity''  index defined in Section~2 of \citet{gibson09}.
The details of the project and the target selection are described in \citet{ak12,ak13}.
 
We select a sample of quasars for this study from these 2005 targets that were observed by SDSS  
between MJD~51{,}602 (2000 February 28) and 54{,}557 (2008 January 04) and 
by  BOSS between MJD~55{,}176 (2009 December 11) and 56{,}455 (2013 June 12). 
Observation start dates correspond to completion of hardware commissioning for both 
SDSS and BOSS; post-commissioning observations have the most reliable spectral calibration. 
We applied basic spectral preparation procedures  to these observed  spectra following Section~3.1 of 
\citet{ak12} and select our ``main sample'' for this study considering the following criteria:

 \begin{enumerate}
\item
We select quasars with $1.9 < z < 3.9$ to ensure spectral coverage  of both  the Si\,{\sc iv} 1394, 1403 \AA\, 
and Al\,{\sc iii} 1855, 1863 \AA\, absorption regions (out to $20{,}000~\rm{km\,s^{-1}}$) where blueshifted absorption 
features are often found. 

\item
We select quasars that have a C\,{\sc iv} balnicity index between $-3000$ and $-20{,}000$ km\,s$^{-1}$, 
BI$_{3}^{20}$, greater than 0 for both the SDSS and BOSS spectra; implementing this requirement for 
both SDSS and BOSS spectra avoids biases that could arise from non-uniform SDSS vs.\,BOSS 
BAL-identification thresholds.\footnote{We have verified that only a small percentage of quasars 
($\approx10\%$) have BI$_0 > 100~\rm{km\,s^{-1}}$  but BI$_{3}^{20}=0$.}
Thus the quasars with disappearing BAL troughs in \citet{ak12} are not 
included in this study. We define BI$_{3}^{20}$ using $a = 3$ and $b = 20$ for the generalized 
BI definition, BI$_{a}^{b}$:


\begin{equation}
{\rm BI}_{a}^{b} \equiv \int_{-1000 \times a}^{-1000 \times b} \left( {1-\frac{f(v)} {0.9}} \right) C dv.
\label{ehml1}
\end{equation}

Similar to the original BI definition \citep{wey91}, in this equation $f(v)$ is the normalized flux 
density  as a function of velocity, $v$.  $C$ is a constant which is equal to 1.0 only 
when a trough is wider than $2000~\rm{km\,s^{-1}}$; it is  otherwise 0.0. 
Following the original BI definition, the minimum red-edge velocity limit is chosen to 
minimize confusion between C\,{\sc iv} BALs and the C\,{\sc iv} emission line. 
In this paper we use BI$_{3}^{20}$, rather than BI$_{3}^{30}$ used in \citet{ak13}.

\item
We select only radio-quiet quasars by requiring the radio-loudness
parameter, $R$, to be less than 10; we utilize $R$ parameters from
\citet{shen11}. Considering that radio-loud quasars are a minority part
of the quasar population and may have different BAL properties than
radio-quiet quasars \citep[e.g.,][]{becker95,becker00,brotherton98},
implementing this criterion avoids possible confusion associated with
the presence of an additional radio-loud population.

\end{enumerate}

Differing from the original BI definition of \citet{wey91}, our adapted  BI$_{3}^{20}$ definition for this study 
(see Equation~\ref{ehml1}) limits the maximum blue-edge velocity of a BAL-trough region at  20{,}000~km\,s$^{-1}$,  
where it is 25{,}000~km\,s$^{-1}$  in the original definition.  Given  that this study is  focused on the characteristics  of and 
differences between  C\,{\sc iv} BAL troughs that are accompanied by Si\,{\sc iv} and/or Al\,{\sc iii} BALs in 
corresponding velocity ranges, we adjust the maximum blue-edge velocity limit of the C\,{\sc iv} BAL-trough 
region considering the Si\,{\sc iv} (1394, 1403~\AA) and Al\,{\sc iii} (1855, 1863~\AA) BAL-trough regions. 
Both the Si\,{\sc iv} and Al\,{\sc iii} BAL-trough regions are occasionally  contaminated by the emission lines 
of O\,{\sc i}  1302~\AA\, (at $\approx -21{,}800$~km\,s$^{-1}$  from Si\,{\sc iv} emission),
Si\,{\sc ii}  1304~\AA\, (at $ \approx-21{,}300$~km\,s$^{-1}$  from Si\,{\sc iv} emission),
C\,{\sc ii}  1334~\AA\, (at $ \approx-14{,}500$~km\,s$^{-1}$  from Si\,{\sc iv} emission),
Ni\,{\sc ii}  1741 and 1751~\AA\, (at $ \approx -20{,}200$ and $-$18{,}400~km\,s$^{-1}$  from Al\,{\sc iii} emission), and
Fe\,{\sc ii}  1787~\AA\, (at $\approx -12{,}500$~km\,s$^{-1}$  from Al\,{\sc iii} emission). 
Although these emission lines are usually weak,  a visual inspection showed that in some cases these features  
may bring an end to shallow BAL troughs. Moreover, low-velocity absorption from these other line species 
may lead to confusion in the detection of Si\,{\sc iv} or Al\,{\sc iii} absorption. 
Thus,  we consider a BAL-trough region between $-3000$ and $-20{,}000$~km\,s$^{-1}$  as a 
suitable balance between uncontaminated spectral regions and useful sample size. Both 
C\,{\sc ii} and Fe\,{\sc ii} have low ionization potentials (24.4~eV and 16.2~eV, respectively) and are 
rarely found in quasar spectra \citep[e.g.,][]{hall02}.

From the initial set of 2005 BAL quasars,  the above criteria  select 714 quasars 
that are observed by both the  SDSS and BOSS (the main sample will be reduced to 
671 quasars below via further considerations). 
The median SN$_{1700}$ is 10.7 for  SDSS and 17.4 for  BOSS observations. 
In order to compare multi-epoch observations of our main sample, for each spectrum, we follow 
the basic spectral preparation procedures given in Section~3.1 of \citet{ak12}. These procedures include Galactic 
extinction correction using the $A_V$ values from \citet{schlegel98},   transforming from the observed frame to the 
rest frame using the redshift values from \citet{hw10}, and removing pixels that are flagged by the 
SDSS and BOSS data-reduction pipelines \citep{bolton12}. 
As in \citet{gibson08,gibson10} and \citet{ak12,ak13}, we define relatively line-free (RLF) windows to reconstruct the underlying 
continuum. We fit the RLF windows of each spectrum with an intrinsically reddened power-law using a Small Magellanic 
Cloud type reddening model \citep[e.g.,][]{pei92}. 
This fit is performed by running an iterative sigma-clipping algorithm where in each iteration we perform
a non-linear least squares fit.
We calculate the uncertainties on the continuum model using $\Delta \chi^2$ confidence-region estimation as 
described in \citet{ak12}. We do not model the emission lines. We follow the procedures of \citet{ak12,ak13} for error 
propagation of continuum uncertainties into the subsequent measurements.

\section{Identification of BAL Troughs and Measurements}\label{ident}

\subsection{Identification of BAL Troughs}\label{identity2}

We identify  C\,{\sc iv} BAL troughs in the spectra of the 714  quasars that satisfy our quasar selection criteria 
(see Section~\ref{colobs} using the definition in Equation~\ref{ehml1} and following the BAL-trough identification 
algorithm for multi-epoch observations defined in Section~3.2 of  \citealt{ak13}).  By construction, 
each quasar in our main sample has at least one SDSS and at least one BOSS observation. However, 
$\approx 25\%$ of the main-sample quasars have multiple SDSS and/or BOSS observations. 
These repeat observations sample similar timescales (considering just the multi-year timescales 
of primary interest here) and could cause some of our objects  to be given inordinate weight by allowing the repeat 
examination of BAL troughs. In order to avoid such multi-counting biases in our analyses, we use only 
two-epoch spectra for each quasar in our main sample. We select the one SDSS-I/II spectrum and the one 
BOSS spectrum that have the  highest SN$_{1700}$. Using these two-epoch 
observations, we classify C\,{\sc iv} BAL troughs into three groups (denoted with subscripts) as explained 
below:

\begin{enumerate}
\item
{C\,{\sc iv}$_{\rm 00}$} indicates C\,{\sc iv} BAL troughs with no detection of  BAL or mini-BAL troughs 
at corresponding velocities in the  Si\,{\sc iv} and Al\,{\sc iii} absorption regions in both epochs.  

\item
{C\,{\sc iv}$_{\rm S0}$} indicates C\,{\sc iv} BAL troughs accompanied by a Si\,{\sc iv} BAL trough in 
either epoch but with no detection of  a BAL or mini-BAL trough at  corresponding velocities in the 
Al\,{\sc iii} BAL region.\footnote{We define a mini-BAL  using Equation~\ref{ehml1} but adapting 
the constant,  $C$,   equal to one for absorption lines with $500 < \Delta v < 2000$~km\,s$^{-1}$ 
and zero otherwise \citep[e.g.,][and references therein]{ak13}.}  

\item
{C\,{\sc iv}$_{\rm SA}$} indicates C\,{\sc iv} BAL troughs accompanied by a  Si\,{\sc iv} BAL and also an 
Al\,{\sc iii} BAL  detected at corresponding velocities in either epoch. 

\end{enumerate}

\noindent We did not find any examples of C\,{\sc iv} BAL troughs accompanied by an Al\,{\sc iii} BAL 
but no Si\,{\sc iv} BAL (i.e., C\,{\sc iv}$_{\rm 0A}$); lower ionization absorption (in this case, Al\,{\sc iii}) 
troughs are always accompanied by higher ionization (Si\,{\sc iv} and C\,{\sc iv}) troughs at the same velocities.
As discussed in Section~\ref{hmlintro}, these three C\,{\sc iv} groups serve as a basic means for identifying lines-of-sight 
with different average ionization levels.

We identify 43 quasars showing multiple C\,{\sc iv} BAL troughs that are 
classed in different C\,{\sc iv} groups (38 with {C\,{\sc iv}$_{\rm 00}$} 
and {C\,{\sc iv}$_{\rm S0}$} troughs, and 5 with {C\,{\sc iv}$_{\rm S0}$} and  
{C\,{\sc iv}$_{\rm SA}$} troughs). While the existence of such objects is 
entirely expected within accretion-disk wind models, practically they 
introduce complexity in connecting a given BAL trough to a given 
line-of-sight; e.g., {C\,{\sc iv}$_{\rm 00}$} troughs in quasars that also 
show {C\,{\sc iv}$_{\rm S0}$} troughs likely sample a different part of the
outflow from {C\,{\sc iv}$_{\rm 00}$} troughs in quasars with no other 
BALs (see Section~\ref{hmlsummary} for further discussion). To reduce 
complexity and avoid potentially mixing troughs  in the same C\,{\sc iv} 
group that sample different parts of the outflow, we exclude from our main 
sample these 43 quasars. We have performed all analyses below also 
including such quasars, and the results do not change materially. 
After excluding these 43 quasars, there are 671 quasars in  our main sample 
(see Table~\ref{quasars}). 
We present our  BI$_{3}^{20}$ measurements and  uncertainties on 
BI$_{3}^{20}$, $\sigma_{{\rm BI}_{3}^{20}}$, both for SDSS and BOSS observations 
of BAL quasars in our main sample in Table~\ref{quasars}; $\sigma_{{\rm BI}_{3}^{20}}$ 
is propagated from uncertainties on the estimated continuum and errors on the trough 
measurements.   
We detect a total of 852 C\,{\sc iv} BAL troughs in the spectra of these 671 
main-sample quasars as our main-sample C\,{\sc iv} troughs.

We have identified 113 C\,{\sc iv}$_{\rm 00}$, 246 C\,{\sc iv}$_{\rm S0}$, 
and 95 C\,{\sc iv}$_{\rm SA}$ BAL troughs in our main sample.   
We present our measurements for {C\,{\sc iv}$_{\rm 00}$} troughs in 
Table~\ref{c00},  {C\,{\sc iv}$_{\rm S0}$} and corresponding Si\,{\sc iv} 
BAL troughs in Table~\ref{cs0}, and  {C\,{\sc iv}$_{\rm SA}$} and 
corresponding Si\,{\sc iv} and Al\,{\sc iii} BAL troughs in Table~\ref{csa}. 
In addition to these groups we have detected  271 C\,{\sc iv} BAL 
troughs accompanied by a  Si\,{\sc iv} mini-BAL ({C\,{\sc iv}$_{\rm s0}$}, 
where the lower case subscript ``s'' denotes the mini-BAL nature), 
32 C\,{\sc iv} BAL troughs accompanied by a  Si\,{\sc iv} mini-BAL and 
also an Al\,{\sc iii} mini-BAL ({C\,{\sc iv}$_{\rm sa}$}), and 
94 C\,{\sc iv} BAL troughs accompanied by a  Si\,{\sc iv} BAL  and an 
Al\,{\sc iii} mini-BAL ({C\,{\sc iv}$_{\rm Sa}$}).
We present our measurements 
for {C\,{\sc iv}$_{\rm s0}$} troughs in Table~\ref{mcs0},
{C\,{\sc iv}$_{\rm sa}$} troughs in Table~\ref{mcsa},
and  {C\,{\sc iv}$_{\rm Sa}$} troughs in Table~\ref{mcssa}.
We do not include these troughs in any of the groups defined above to consider 
only strong intrinsic absorption of the given transition; this approach should give 
the strongest distinction between groups.

We have cross-matched our 671 main-sample quasars with the SDSS DR7 
quasar properties catalog of \citet{shen11} to obtain absolute $i$-band magnitudes, 
$M_i$ (see Table~\ref{quasars}). We found that the majority of our main-sample 
quasars lie within ~$2.5$ mag in $M_i$, corresponding  to a factor of $\approx$10 
range in $i$-band luminosity. 
Given this relatively narrow range, we do not have any immediate reason to be concerned
about luminosity effects. 
Moreover, using two-sample Anderson-Darling (AD) tests  \citep[see][]{press}, we found 
that the $M_i$ distributions for  the {C\,{\sc iv}$_{\rm 00}$}, 
{C\,{\sc iv}$_{\rm S0}$}, and {C\,{\sc iv}$_{\rm SA}$} groups are consistent, indicating that 
 luminosity effects should not be causing confusion in group intercomparisons.

\begin{figure*}[t!]
\epsscale{0.85}
\plotone{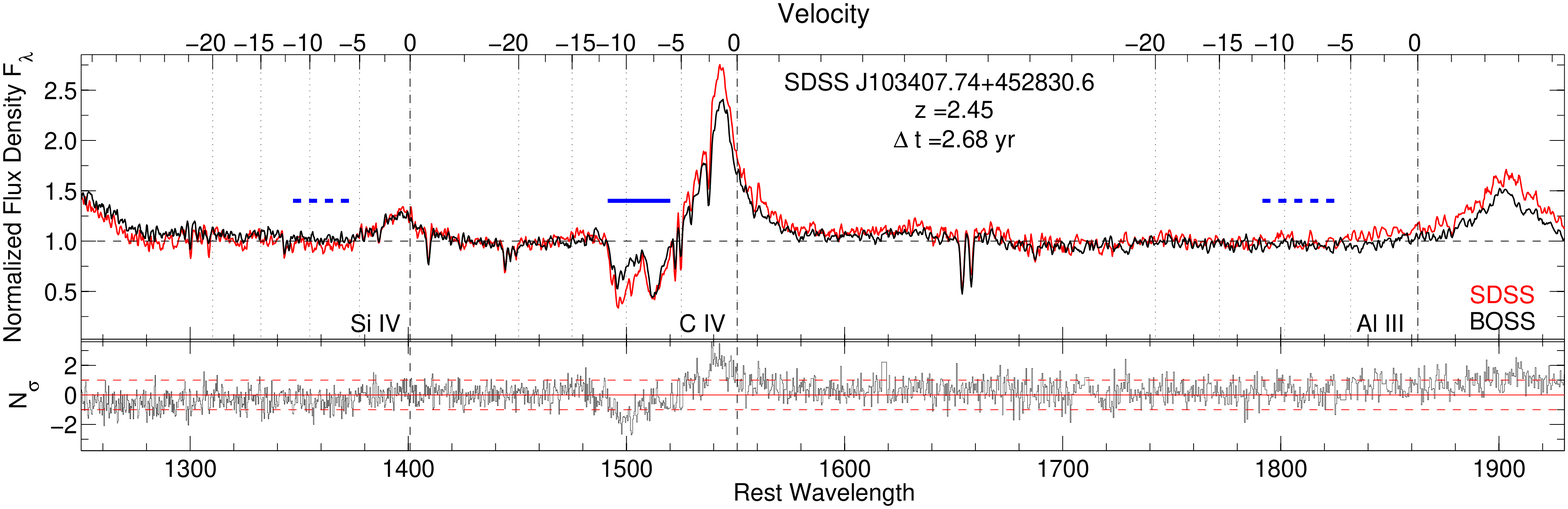} 
\plotone{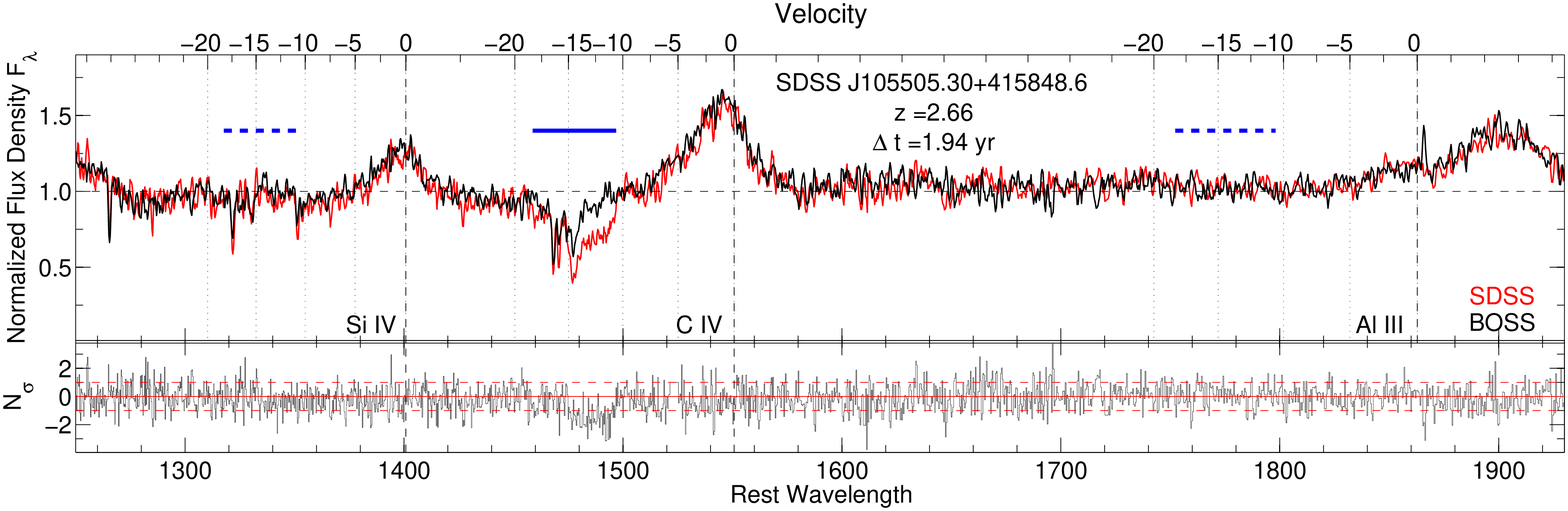}
\plotone{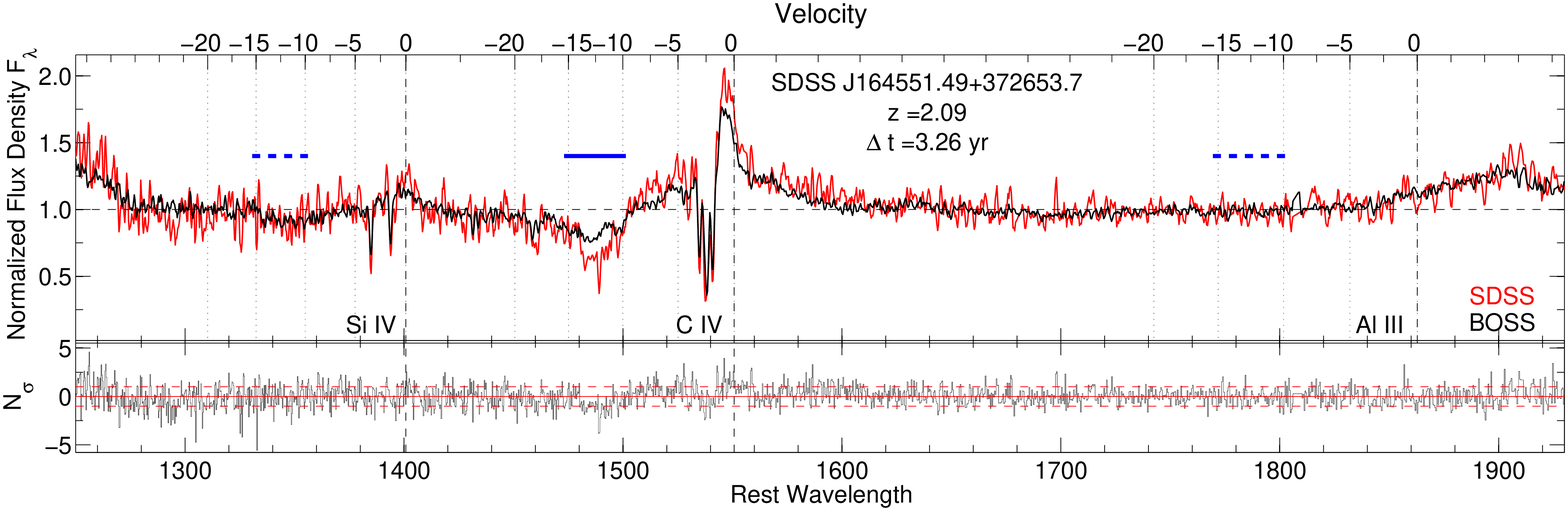} 
\caption{Two-epoch spectra of quasars with  C\,{\sc iv}$_{\rm 00}$ BAL troughs (i.e., C\,{\sc iv} 
BAL troughs with no detection of  BAL or mini-BAL troughs at corresponding velocities in the  
Si\,{\sc iv} and Al\,{\sc iii} absorption regions) from SDSS (red) and BOSS (black). 
The $x$-axes show both the rest-frame wavelength (bottom, in \AA\,) and  the blueshift velocity 
from the Si\,{\sc iv}, C\,{\sc iv}, and Al\,{\sc iii} emission lines (top, in $10^3$~km\,s$^{-1}$). 
The $y$-axes show flux densities normalized by the fitted  continuum model ($F_{\lambda}$). 
The horizontal solid-blue bars designate C\,{\sc iv} BAL troughs, and the dashed blue bars 
indicate corresponding velocities in the Si\,{\sc iv} and  Al\,{\sc iii} BAL regions. The lower section 
of each panel shows deviations between SDSS and BOSS observations for each $\approx4$~\AA\, 
pixel in units of $\sigma$, $N_{\sigma}$, and the dashed-red lines  show $\pm1\sigma$ levels.}
\label{fig1}
\end{figure*}

\begin{figure*}[t!]
\epsscale{0.85}
\plotone{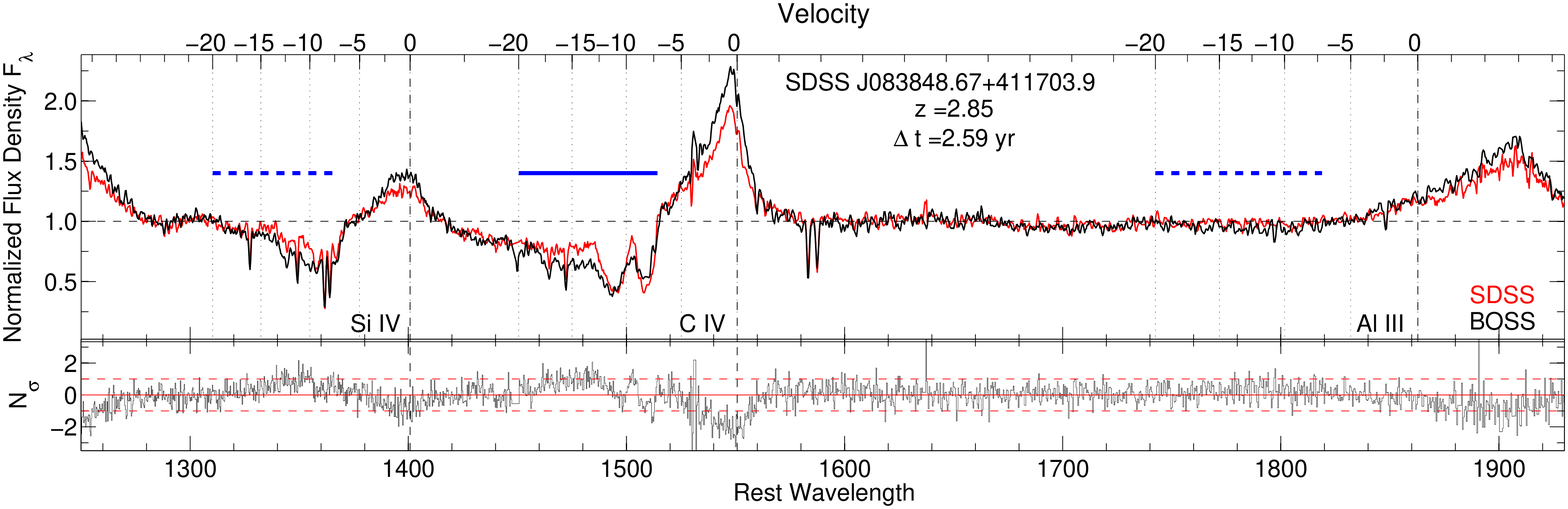} 
\plotone{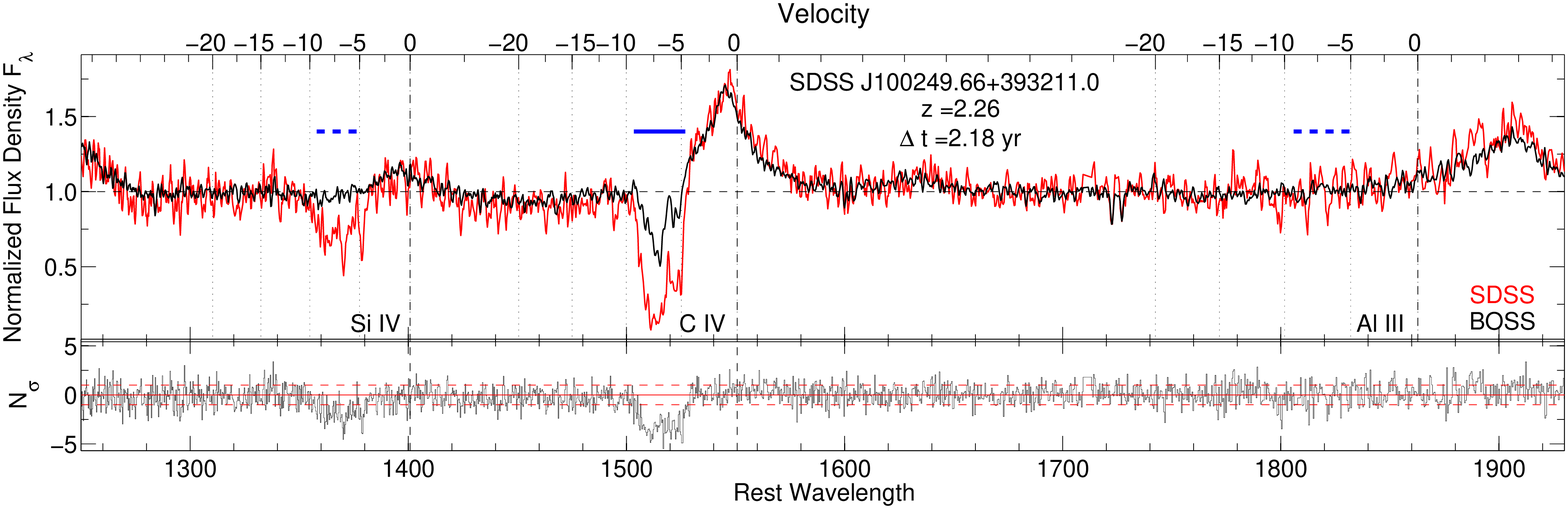} 
\plotone{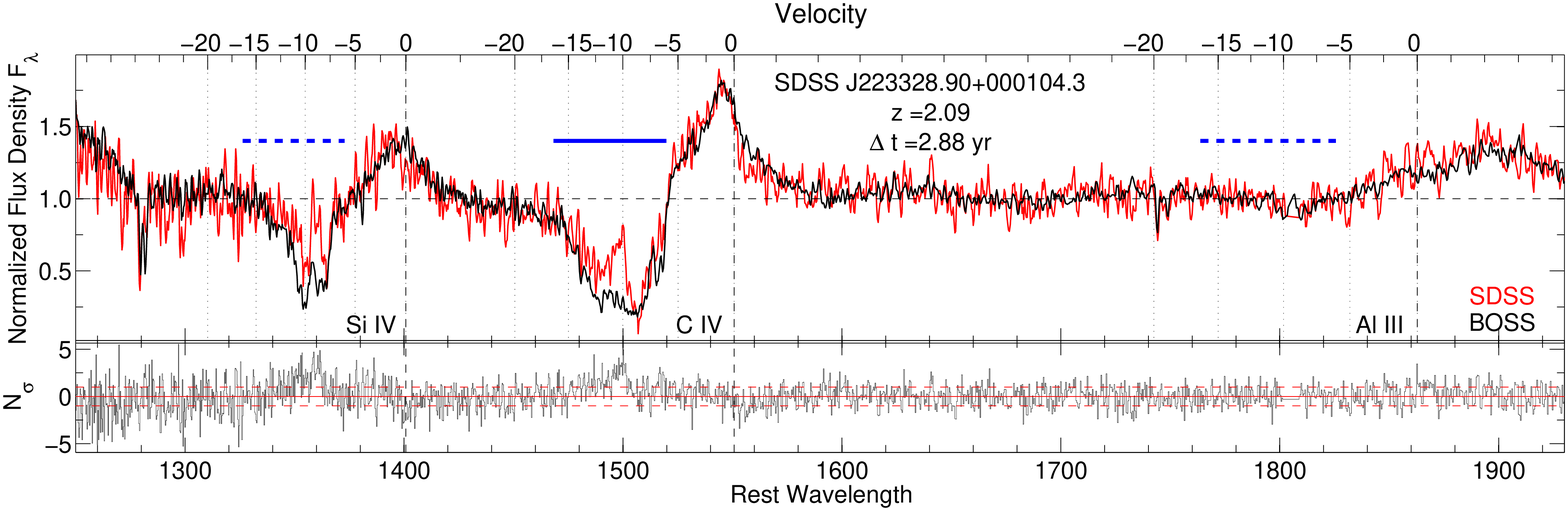} 
\caption{Same as Figure~\ref{fig1} but for C\,{\sc iv}$_{\rm S0}$ BAL troughs (i.e.,  C\,{\sc iv} 
BAL troughs accompanied by a Si\,{\sc iv} BAL trough but with no detection of  a BAL or 
mini-BAL trough at  corresponding velocities in the Al\,{\sc iii} BAL region). }
\label{fig2}
\end{figure*}

\begin{figure*}[t!]
\epsscale{0.85}
\plotone{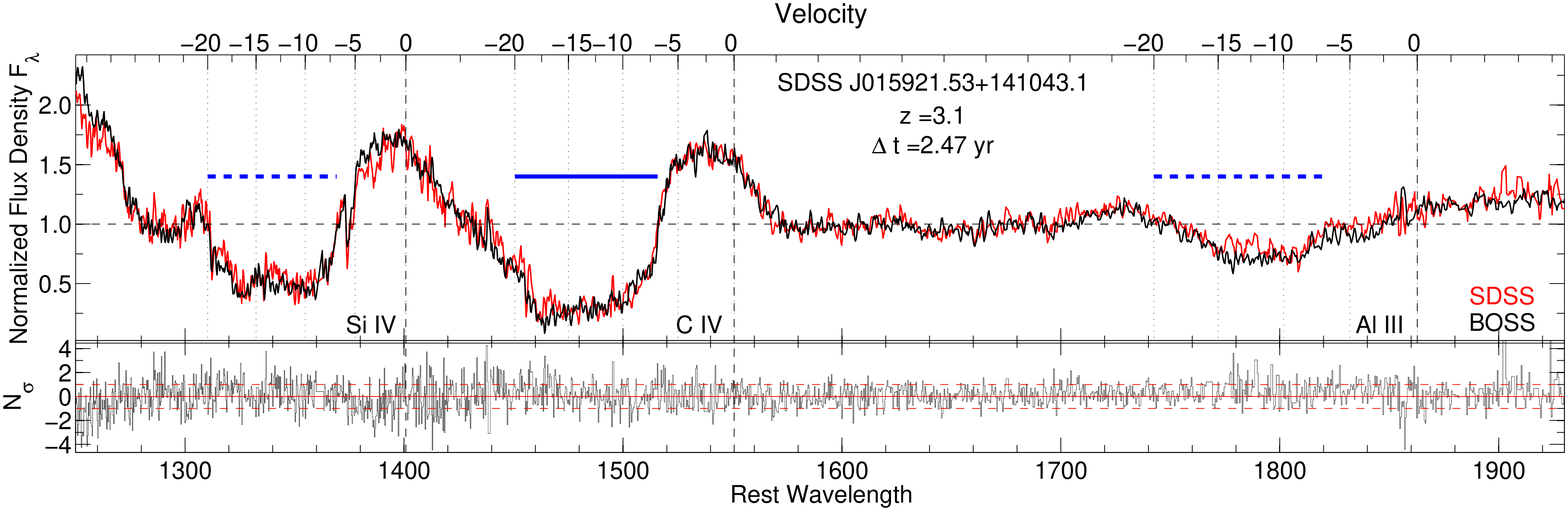} 
\plotone{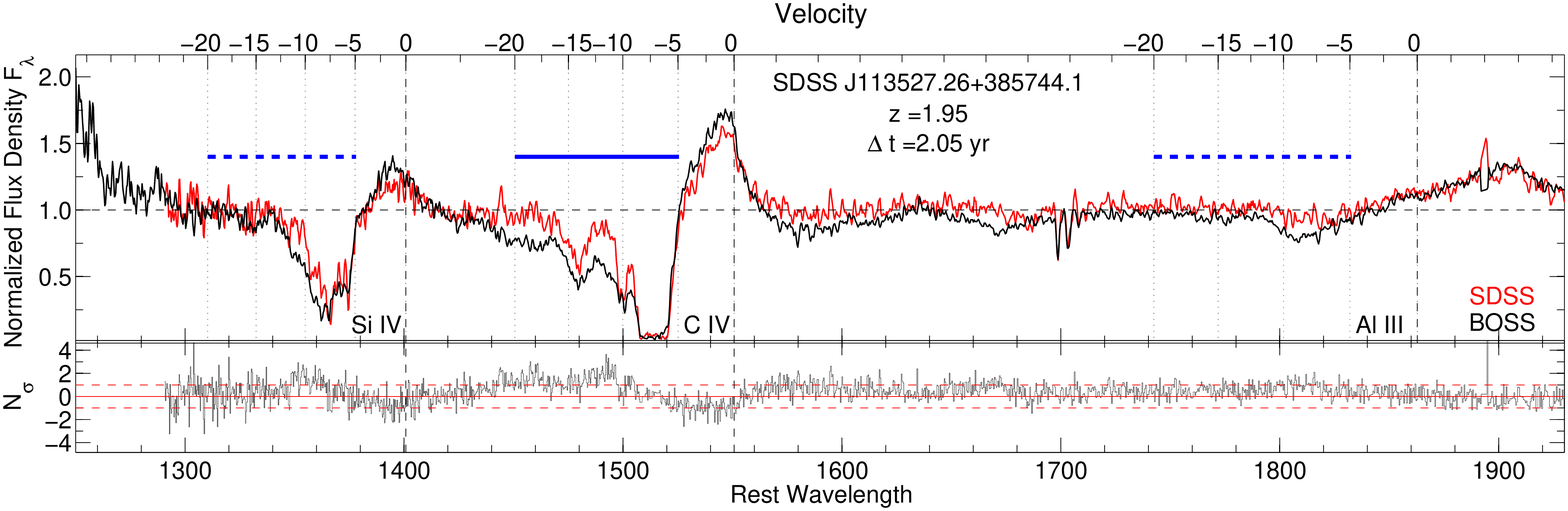}
\plotone{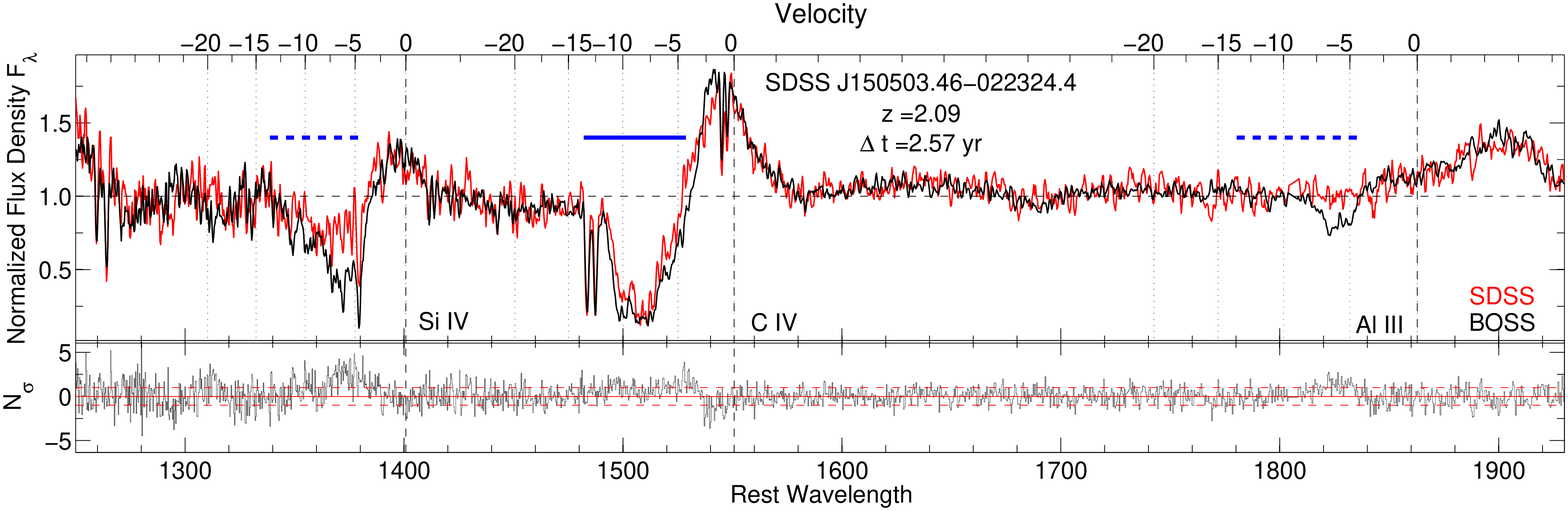} 
\caption{Same as Figure~\ref{fig1} but for C\,{\sc iv}$_{\rm SA}$ BAL troughs (i.e., C\,{\sc iv} 
BAL troughs accompanied by a  Si\,{\sc iv} BAL trough and also an Al\,{\sc iii} BAL trough). }
\label{fig3}
\end{figure*}

Figures~\ref{fig1}, \ref{fig2}, and \ref{fig3} show two-epoch observations for representative 
examples of C\,{\sc iv}$_{\rm 00}$, C\,{\sc iv}$_{\rm S0}$, and C\,{\sc iv}$_{\rm SA}$ BAL 
troughs, respectively. We smoothed each spectrum for display purposes using a Savitzky-Golay 
algorithm \citep[see][]{press} that  performs a local linear regression for three consecutive points 
\citep[see][]{ak12}. In these figures, we indicate C\,{\sc iv} BAL troughs and their corresponding 
velocity ranges in the Si\,{\sc iv} and Al\,{\sc iii} absorption regions. 
The SDSS identification, redshift, and timescale between the two epochs are given in each panel.

{
\begin{table*}
\caption{Main-Sample BAL Quasars} \label{quasars}
\begin{center}
{\scriptsize 
\begin{tabular}{ccrrrrccrrrrrrrrrrrrrrrrrrrrrrrrrrrrrrrrr}
\tableline\tableline \\[-0.3em]
 Quasar ID & {Quasar Name} & \multicolumn{1}{c}{RA} & \multicolumn{1}{c}{Dec} &\multicolumn{1}{c} {$z$} 
&\multicolumn{1}{c}{$\sigma_z$} & \multicolumn{1}{c}{$M_{\rm i}$($z=2$)} & {Plate[1]} &\nodata\\
  &  SDSS & J2000 & J2000 & &  &(mag) \\
 \\[-0.3em]
\tableline\\ 
Q1 & J000119.64+154828.8 & 0.33184 & 15.80800 & 1.921029 & 0.000494   &   $-26.73$ &750   \\  
Q2 & J001025.90+005447.6 & 2.60796 & 0.91324 & 2.859854 & 0.000323     &   $-27.81$  &389   \\  
Q3 & J001502.26+001212.4 & 3.75943 & 0.20346 & 2.852539 & 0.000552     &   $-27.68$  &389    \\  
Q4 & J003135.57+003421.2 & 7.89823 & 0.57257 & 2.236426 & 0.000255     &   $-27.65$  &689    \\  
Q5 & J003312.25+155442.4 & 8.30105 & 15.91178 & 1.955444 & 0.000617   &   $-26.74$ &417     \\
Q6 & J003517.95+004333.7 & 8.82481 & 0.72604 & 2.916893 & 0.000545     &   $-27.61$ &1086     \\
Q7 & J003551.98+005726.4 & 8.96660 & 0.95734 & 1.905903 & 0.000576     &   $-26.75$ &392      \\
Q8 & J003832.26+152515.5 & 9.63446 & 15.42100 & 2.448965 & 0.000500   &   $-27.46$ &418      \\
Q9 & J004732.73+002111.3 & 11.88639 & 0.35315 & 2.873223 & 0.000340    &   $-27.99$ &691       \\
Q10 &J005215.64+003120.5 & 13.06520 & 0.52236 & 2.792256 & 0.000321   &  $-27.48$ &084        \\  
\\[-0.6em]
\hline \\[0.6em]
\end{tabular}

\begin{tabular}{rrrrrrrrrrrrrrrrrrrrrrrrrrrrrrrrrrrrrrrrr}
\tableline\tableline \\[-0.3em]

 {MJD[1]} &  {Fiber[1]} & {SN$_{1700}$[1]}&{Plate[2]} & {MJD[2]} & {Fiber[2]}& {SN$_{1700}$[2]} 
 & BI$_{3}^{20}$[1] & $\sigma_{{\rm BI}_{3}^{20}}$[1]
 & BI$_{3}^{20}$[2] & $\sigma_{{\rm BI}_{3}^{20}}$[2]
&\multicolumn{1}{c}{$\Delta t$}  \\
 \multicolumn{1}{c}{(days)}   & &&&\multicolumn{1}{c}{(days)}&&&( km\,s$^{-1}$)&( km\,s$^{-1}$)&( km\,s$^{-1}$)&( km\,s$^{-1}$)&\multicolumn{1}{c}{(years)} \\
\\[-0.3em]
\tableline\\

 52235 & 566 & 8.201 & 6172 & 56269 & 318 & 18.296     &6378 & 142 & 7291 & 92& 3.781   \\
 51795 & 332 & 6.379 & 4218 & 55479 & 592 & 20.610    &3974 & 212 & 4893 & 77& 2.613    \\
 51795 & 465 & 10.623 & 4218 & 55479 & 818 & 21.430   &2273 & 107 & 2042 & 28& 2.618     \\
 52262 & 502 & 15.060 & 3587 & 55182 & 570 & 24.501  &5305 & 39 & 5855 & 23& 2.470    \\
 51821 & 576 & 8.304 & 6192 & 56269 & 184 & 15.174     &11666 & 168 & 11571 & 91& 4.121    \\
52525 & 481 & 12.271 & 3588 & 55184 & 512 & 14.787 &1811 & 70 & 1406 & 44& 1.859    \\
 51793 & 449 & 8.931 & 3588 & 55184 & 552 & 24.975     &2724 & 66 & 2194 & 27& 3.195    \\
 51817 & 483 & 6.585 & 6197 & 56191 & 760 & 12.239     &7509 & 174 & 9076 & 114& 3.472    \\
 52199 & 559 & 19.051 & 4223 & 55451 & 724 & 26.027  &1165 & 39 & 1045 & 23& 2.299    \\
 52591 & 516 & 9.739 & 4223 & 55451 & 968 & 11.653    &4709 & 118 & 7004 & 146& 2.065    \\
 \\[-0.6em]
\hline \\[0.6em]
\end{tabular}
\\ \textbf{Notes. }Throughout this table [1] indicates the first-epoch spectra and [2] indicates the second-epoch spectra.
\\ (This table is available in its entirety in a machine-readable form in the online journal. A portion is shown here for guidance regarding its form and content.)}

\end{center}
\end{table*}

{
\begin{table*}
\caption{C\,{\sc iv}$_{\rm 00}$ Troughs} \label{c00}
\begin{center}
{\scriptsize 
\begin{tabular}{ccrrrrrrrrrrrrrrrrrrrrrrrrrrrrrrrrrrrrrrr}
\tableline\tableline \\[-0.3em]

Quasar & C\,{\sc iv}$_{00}$ &
\multicolumn{1}{c}{$v_{\rm max}$} &
\multicolumn{1}{c}{$v_{\rm min}$} & 
\multicolumn{1}{c}{$\Delta v$} & 
\multicolumn{1}{c}{$v_{\rm cent}$[1]} & 
\multicolumn{1}{c}{$v_{\rm cent}$[2] }&
$d_{\rm BAL}$[1] & $\sigma_{d_{\rm BAL}}$[1] & 
 \nodata\\
ID & Trough ID &  (km\,s$^{-1}$) & ( km\,s$^{-1}$) &  (km\,s$^{-1}$)&  (km\,s$^{-1}$) &  (km\,s$^{-1}$)& \\

\\[-0.3em]
\tableline\\ 

Q29 & C00-1 & $-10957.2$ & $-$8528.4 & 2428.8 & $-$9712.9 & $-$9676.2 & 0.200 & 0.022         \\
Q29 & C00-2 & $-7593.0$ & $-$4686.9 & 2906.1 & $-$6046.1 & $-$6072.0 & 0.339 & 0.024           \\
Q34 & C00-3 & $-20000.0$ & $-$17884.3 & 2115.7 & $-$18892.9 & $-$18901.7 & 0.124 & 0.009  \\
Q34 & C00-4 & $-17862.3$ & $-$12780.2 & 5082.1 & $-$15274.8 & $-$15295.9 & 0.130 & 0.009  \\
Q40 & C00-5 & $-15763.9$ & $-$9381.9 & 6382.0 & $-$12610.5 & $-$12497.6 & 0.307 & 0.017    \\
Q41 & C00-6 & $-20000.0$ & $-$16100.5 & 3899.5 & $-$17974.4 & $-$17956.4 & 0.348 & 0.020  \\
Q47 & C00-7 & $-19397.8$ & $-$11712.7 & 7685.1 & $-$15810.3 & $-$15687.3 & 0.368 & 0.015  \\
Q47 & C00-8 & $-11537.6$ & $-$8089.2 & 3448.4 & $-$9811.0 & $-$9823.4 & 0.329 & 0.023        \\
Q51 & C00-9 & $-9963.7$ & $-$7962.3 & 2001.5 & $-$8946.8 & $-$8954.2 & 0.144 & 0.021         \\
Q57 & C00-10 & $-20000.0$ & $-$4959.0 & 15041.0 & $-$12291.8 & $-$12402.4 & 0.189 & 0.006  \\

\\[-0.6em]
\hline \\[0.6em]
\end{tabular}

\begin{tabular}{rrrrrrrrrrrrrrrrrrrrrrrrrrrrrrrrrrrrrrrrr}
\tableline\tableline \\[-0.3em]
$d_{\rm BAL}$[2] & $\sigma_{d_{\rm BAL}}$[2] &
EW[1] & $\sigma_{\rm EW}$[1]   & 
EW[2] & $\sigma_{\rm EW}$[2] & 
$\Delta \rm{EW}$ & $\sigma_{\Delta EW}$ & 
$\frac {\Delta\rm{EW}} {\langle \rm{EW} \rangle}$ & 
$\sigma_{\frac {\Delta\rm{EW}}  {\langle \rm{EW} \rangle}}$  \\
& & (\AA) & (\AA)& (\AA)& (\AA)& (\AA)& (\AA)\\
\\[-0.3em]
\tableline\\

 0.147 & 0.015  &2.37 & 0.19 & 1.65 & 0.18 & $-$0.72 & 0.26 & $-$0.36 & 0.18 \\
 0.311 & 0.020  & 5.10 & 0.20 & 4.71 & 0.17 & $-$0.39 & 0.26 & $-$0.08 & 0.08 \\
 0.133 & 0.005  &1.27 & 0.16 & 1.42 & 0.05 & 0.15 & 0.17 & 0.11 & 0.16 \\
 0.145 & 0.005  &3.18 & 0.25 & 3.67 & 0.08 & 0.49 & 0.26 & 0.14 & 0.10 \\
  0.294 & 0.013 & 9.49 & 0.50 & 9.37 & 0.21 & $-$0.12 & 0.54 & $-$0.01 & 0.07 \\
 0.279 & 0.015  &6.76 & 0.22 & 5.45 & 0.18 & $-$1.30 & 0.28 & $-$0.21 & 0.06 \\
 0.118 & 0.010  &13.77 & 0.61 & 3.70 & 0.24 & $-$10.08 & 0.66 & $-$1.15 & 0.07 \\
 0.189 & 0.016  &5.68 & 0.35 & 3.21 & 0.13 & $-$2.47 & 0.37 & $-$0.56 & 0.09 \\
 0.394 & 0.036  & 1.23 & 0.26 & 4.10 & 0.17 & 2.87 & 0.31 & 1.08 & 0.18 \\
 0.207 & 0.004  &13.86 & 0.83 & 15.56 & 0.62 & 1.70 & 1.03 & 0.12 & 0.10 \\

 \\[-0.6em]
\hline \\[0.6em]
\end{tabular}
\\ \textbf{Notes. }Throughout this table [1] indicates the first-epoch spectra and [2] indicates the second-epoch spectra.
\\ (This table is available in its entirety in a machine-readable form in the online journal. A portion is shown here for guidance regarding its form and content.)}
\end{center}
\end{table*}

{
\begin{table*}
\caption{C\,{\sc iv}$_{\rm S0}$ Troughs} \label{cs0}
\begin{center}
{\scriptsize 
\begin{tabular}{lcrrrrrrrrrrrrrrrrrrrrrrrrrrrrrrrrrrrrrrr}
\tableline\tableline \\[-0.3em]

Quasar & C\,{\sc iv}$_{\rm S0}$ &
\multicolumn{1}{c}{$v_{\rm max}$} &
\multicolumn{1}{c}{$v_{\rm min}$} & 
\multicolumn{1}{c}{$\Delta v$} & 
\multicolumn{1}{c}{$v_{\rm cent}$[1]} & 
\multicolumn{1}{c}{$v_{\rm cent}$[2] }&
$d_{\rm BAL}$[1] & $\sigma_{d_{\rm BAL}}$[1] & 
 \nodata\\
ID & Trough ID &  (km\,s$^{-1}$) & ( km\,s$^{-1}$) &  (km\,s$^{-1}$)&  (km\,s$^{-1}$) &  (km\,s$^{-1}$)& \\ 
 \\[-0.3em]
\tableline\\

Q3 & CS0-1 & $-$10477.0 & $-$5841.4 & 4635.7 & $-$8427.4 & $-$8285.0 & 0.491 & 0.032            \\
 Q4 & CS0-2 & $-$12444.7 & $-$4858.5 & 7586.2 & $-$9558.5 & $-$9262.8 & 0.697 & 0.025            \\
 Q10 & CS0-3 & $-$15491.2 & $-$3000.0 & 12491.2 & $-$10869.6 & $-$10629.1 & 0.531 & 0.024  \\
 Q11 & CS0-4 & $-$11596.2 & $-$5638.0 & 5958.2 & $-$8516.9 & $-$8806.4 & 0.217 & 0.010         \\
 Q12 & CS0-5 & $-$10921.6 & $-$3000.0 & 7921.6 & $-$8088.0 & $-$8031.0 & 0.648 & 0.023         \\

\\[-0.6em]
\hline \\[0.6em]
\end{tabular}

\begin{tabular}{rrrrrrrrrrrrrrrrrrrrrrrrrrrrrrrrrrrrrrrrr}
\tableline\tableline \\[-0.3em]

$d_{\rm BAL}$[2] & $\sigma_{d_{\rm BAL}}$[2] &
EW[1] & $\sigma_{\rm EW}$[1]   & 
EW[2] & $\sigma_{\rm EW}$[2] & 
$\Delta \rm{EW}$ & $\sigma_{\Delta EW}$ & 
$\frac {\Delta\rm{EW}} {\langle \rm{EW} \rangle}$ & 
$\sigma_{\frac {\Delta\rm{EW}}  {\langle \rm{EW} \rangle}}$  \\
& & (\AA) & (\AA)& (\AA)& (\AA)& (\AA)& (\AA)\\\\[-0.3em]
\tableline\\

0.442 & 0.029 & 11.51 & 0.56 & 10.29 & 0.15 & $-$1.22 & 0.58 & $-$0.11 & 0.06 \\
0.744 & 0.023 & 26.86 & 0.20 & 28.43 & 0.10 & 1.56 & 0.22 & 0.06 & 0.01 \\
0.543 & 0.021 & 31.37 & 0.92 & 35.28 & 0.73 & 3.91 & 1.18 & 0.12 & 0.05 \\  
0.253 & 0.016 & 6.63 & 0.15 & 7.73 & 0.19 & 1.10 & 0.24 & 0.15 & 0.05 \\      
0.593 & 0.024 & 27.95 & 0.32 & 25.64 & 0.22 & $-$2.31 & 0.39 & $-$0.09 & 0.02 \\

 \\[-0.6em]
\hline \\[0.6em]
\end{tabular}

\begin{tabular}{ccrrrrrrrrrrrrrrrrrrrrrrrrrrrrrrrrrrrrrrr}
\tableline\tableline \\[-0.3em]

Si\,{\sc iv}$_{\rm S0}$ &
\multicolumn{1}{c}{$v_{\rm max}$} &
\multicolumn{1}{c}{$v_{\rm min}$} & 
\multicolumn{1}{c}{$\Delta v$} & 
\multicolumn{1}{c}{$v_{\rm cent}$[1]} & 
\multicolumn{1}{c}{$v_{\rm cent}$[2] }&
$d_{\rm BAL}$[1] & $\sigma_{d_{\rm BAL}}$[1] & 
 \nodata\\
 Trough ID &  (km\,s$^{-1}$) & ( km\,s$^{-1}$) &  (km\,s$^{-1}$)&  (km\,s$^{-1}$) &  (km\,s$^{-1}$)& \\
 \\[-0.3em]
\tableline\\

 SS0-1 & $-$12065.5 & $-$5872.0 & 6193.5 & $-$8972.6 & $-$8908.6 & 0.243 & 0.015 	\\
 SS0-2 & $-$12525.3 & $-$4926.5 & 7598.8 & $-$8905.7 & $-$8832.4 & 0.314 & 0.016 	\\
 SS0-3 & $-$12457.4 & $-$3927.3 & 8530.1 & $-$8587.1 & $-$8601.9 & 0.400 & 0.020 	\\
  SS0-4 & $-$11062.2 & $-$5638.0 & 5424.2 & $-$8258.4 & $-$8293.8 & 0.206 & 0.009 	\\
  SS0-5 & $-$5804.7 & $-$3000.0 & 2804.7 & $-$4427.8 & $-$4462.2 & 0.503 & 0.029 	\\

\\[-0.6em]
\hline \\[0.6em]
\end{tabular}
\begin{tabular}{rrrrrrrrrrrrrrrrrrrrrrrrrrrrrrrrrrrrrrrrr}
\tableline\tableline \\[-0.3em]

$d_{\rm BAL}$[2] & $\sigma_{d_{\rm BAL}}$[2] &
EW[1] & $\sigma_{\rm EW}$[1]   & 
EW[2] & $\sigma_{\rm EW}$[2] & 
$\Delta \rm{EW}$ & $\sigma_{\Delta EW}$ & 
$\frac {\Delta\rm{EW}} {\langle \rm{EW} \rangle}$ & 
$\sigma_{\frac {\Delta\rm{EW}}  {\langle \rm{EW} \rangle}}$  \\
& & (\AA) & (\AA)& (\AA)& (\AA)& (\AA)& (\AA)\\
\\[-0.3em]
\tableline\\

 0.137 & 0.006 & 6.41 & 0.68 & 3.91 & 0.19 & $-$2.50 & 0.71 & $-$0.48 & 0.14\\
 0.361 & 0.018 & 10.84 & 0.24 & 12.61 & 0.13 & 1.76 & 0.27 & 0.15 & 0.03\\
 0.515 & 0.018 & 15.14 & 0.64 & 20.22 & 0.48 & 5.08 & 0.80 & 0.29 & 0.06\\
 0.186 & 0.010 & 5.22 & 0.14 & 4.70 & 0.17 & $-$0.52 & 0.22 & $-$0.11 & 0.06\\
 0.467 & 0.031 & 6.78 & 0.20 & 6.25 & 0.11 & $-$0.52 & 0.23 & $-$0.08 & 0.05\\

 \\[-0.6em]
\hline \\[0.6em]
\end{tabular}
\\ \textbf{Notes. }Throughout this table [1] indicates the first-epoch spectra and [2] indicates the second-epoch spectra.
\\ (This table is available in its entirety in a machine-readable form in the online journal. A portion is shown here for 
guidance regarding its form and content.)}

\end{center}
\end{table*}

{
\begin{table*}
\caption{C\,{\sc iv}$_{\rm SA}$ Troughs} \label{csa}
\begin{center}
{\scriptsize 
\begin{tabular}{ccrrrrccrrrrrrrrrrrrrrrrrrrrrrrrrrrrrrrrr}
\tableline\tableline \\[-0.3em]

Quasar & C\,{\sc iv}$_{\rm SA}$ &
\multicolumn{1}{c}{$v_{\rm max}$} &
\multicolumn{1}{c}{$v_{\rm min}$} & 
\multicolumn{1}{c}{$\Delta v$} & 
\multicolumn{1}{c}{$v_{\rm cent}$[1]} & 
\multicolumn{1}{c}{$v_{\rm cent}$[2] }&
$d_{\rm BAL}$[1] & $\sigma_{d_{\rm BAL}}$[1] & 
 \nodata\\
ID & Trough ID &  (km\,s$^{-1}$) & ( km\,s$^{-1}$) &  (km\,s$^{-1}$)&  (km\,s$^{-1}$) &  (km\,s$^{-1}$)& \\ 
 \\[-0.3em]
\tableline\\ 

 Q1 & CSA-1 & $-$17205.7 & $-$3487.4 & 13718.4 & $-$10850.0 & $-$10770.0 & 0.521 & 0.018 \\
 Q5 & CSA-2 & $-$20000.0 & $-$4805.6 & 15194.4 & $-$13846.9 & $-$13939.7 & 0.755 & 0.011\\
 Q8 & CSA-3 & $-$20000.0 & $-$3000.0 & 17000.0 & $-$11866.5 & $-$11868.8 & 0.450 & 0.015\\
 Q20 & CSA-4 & $-$20000.0 & $-$6837.8 & 13162.2 & $-$13057.2 & $-$12687.1 & 0.607 & 0.014\\
 Q28 & CSA-5 & $-$20000.0 & $-$7283.0 & 12717.0 & $-$14376.7 & $-$14356.4 & 0.291 & 0.014\\
  
\\[-0.6em]
\hline \\[0.6em]
\end{tabular}

\begin{tabular}{rrrrrrrrrrrrrrrrrrrrrrrrrrrrrrrrrrrrrrrrr}
\tableline\tableline \\[-0.3em]
$d_{\rm BAL}$[2] & $\sigma_{d_{\rm BAL}}$[2] &
EW[1] & $\sigma_{\rm EW}$[1]   & 
EW[2] & $\sigma_{\rm EW}$[2] & 
$\Delta \rm{EW}$ & $\sigma_{\Delta EW}$ & 
$\frac {\Delta\rm{EW}} {\langle \rm{EW} \rangle}$ & 
$\sigma_{\frac {\Delta\rm{EW}}  {\langle \rm{EW} \rangle}}$  \\
& & (\AA) & (\AA)& (\AA)& (\AA)& (\AA)& (\AA)\\
\\[-0.3em]
\tableline\\

0.527 & 0.018 & 33.47 & 0.80 & 36.51 & 0.46 & 3.04 & 0.93 & 0.09 & 0.04\\
0.752 & 0.011 & 58.06 & 0.84 & 57.61 & 0.45 & $-$0.45 & 0.95 & $-$0.01 & 0.02\\
0.532 & 0.014 & 38.15 & 1.08 & 45.28 & 0.57 & 7.13 & 1.22 & 0.17 & 0.04\\
0.634 & 0.013 & 39.57 & 0.78 & 41.52 & 0.73 & 1.95 & 1.07 & 0.05 & 0.04\\
0.306 & 0.015 & 19.46 & 0.42 & 20.19 & 0.22 & 0.73 & 0.47 & 0.04 & 0.03\\

 \\[-0.6em]
\hline \\[0.6em]
\end{tabular}

\begin{tabular}{ccrrrrccrrrrrrrrrrrrrrrrrrrrrrrrrrrrrrrrr}
\tableline\tableline \\[-0.3em]

 Si\,{\sc iv}$_{\rm SA}$ &
\multicolumn{1}{c}{$v_{\rm max}$} &
\multicolumn{1}{c}{$v_{\rm min}$} & 
\multicolumn{1}{c}{$\Delta v$} & 
\multicolumn{1}{c}{$v_{\rm cent}$[1]} & 
\multicolumn{1}{c}{$v_{\rm cent}$[2] }&
$d_{\rm BAL}$[1] & $\sigma_{d_{\rm BAL}}$[1] & 
 \nodata\\
Trough ID &  (km\,s$^{-1}$) & ( km\,s$^{-1}$) &  (km\,s$^{-1}$)&  (km\,s$^{-1}$) &  (km\,s$^{-1}$)& 
\\[1em] 
\tableline\\ 

SSA-1 & $-$13895.2 & $-$4807.7 & 9087.5 & $-$8736.4 & $-$8724.1 & 0.434 & 0.021\\
SSA-2 & $-$20000.0 & $-$5385.4 & 14614.6 & $-$14178.8 & $-$14518.4 & 0.617 & 0.017\\
SSA-3 & $-$15485.3 & $-$3361.3 & 12124.0 & $-$9281.0 & $-$8879.0 & 0.432 & 0.016\\
SSA-4 & $-$19792.4 & $-$6932.7 & 12859.7 & $-$13757.4 & $-$13620.0 & 0.426 & 0.012\\
SSA-5 & $-$13809.3 & $-$7290.5 & 6518.8 & $-$10512.4 & $-$10490.5 & 0.357 & 0.018\\
 
\\[-0.6em]
\hline \\[0.6em]
\end{tabular}

\begin{tabular}{rrrrrrrrrrrrrrrrrrrrrrrrrrrrrrrrrrrrrrrrr}
\tableline\tableline \\[-0.3em]
$d_{\rm BAL}$[2] & $\sigma_{d_{\rm BAL}}$[2] &
EW[1] & $\sigma_{\rm EW}$[1]   & 
EW[2] & $\sigma_{\rm EW}$[2] & 
$\Delta \rm{EW}$ & $\sigma_{\Delta EW}$ & 
$\frac {\Delta\rm{EW}} {\langle \rm{EW} \rangle}$ & 
$\sigma_{\frac {\Delta\rm{EW}}  {\langle \rm{EW} \rangle}}$  \\
& & (\AA) & (\AA)& (\AA)& (\AA)& (\AA)& (\AA)
\\[1em]
\tableline\\

0.423 & 0.018 & 15.58 & 0.77 & 17.40 & 0.37 & 1.82 & 0.85 & 0.11 & 0.07\\
0.521 & 0.020 & 41.32 & 0.92 & 31.13 & 0.71 & $-$10.19 & 1.16 & $-$0.28 & 0.04\\
0.524 & 0.017 & 21.58 & 0.91 & 28.62 & 0.43 & 7.05 & 1.01 & 0.28 & 0.06\\
0.467 & 0.010 & 24.58 & 1.03 & 27.40 & 0.82 & 2.83 & 1.32 & 0.11 & 0.07\\
0.415 & 0.018 & 10.61 & 0.24 & 12.43 & 0.11 & 1.83 & 0.26 & 0.16 & 0.03\\

 \\[-0.6em]
\hline \\[0.6em]
\end{tabular}

\begin{tabular}{ccrrrrccrrrrrrrrrrrrrrrrrrrrrrrrrrrrrrrrr}
\tableline\tableline 

  Al\,{\sc iii}$_{\rm SA}$ &
\multicolumn{1}{c}{$v_{\rm max}$} &
\multicolumn{1}{c}{$v_{\rm min}$} & 
\multicolumn{1}{c}{$\Delta v$} & 
\multicolumn{1}{c}{$v_{\rm cent}$[1]} & 
\multicolumn{1}{c}{$v_{\rm cent}$[2] }&
$d_{\rm BAL}$[1] & $\sigma_{d_{\rm BAL}}$[1] &  \nodata\\
Trough ID &  (km\,s$^{-1}$) & ( km\,s$^{-1}$) &  (km\,s$^{-1}$)&  (km\,s$^{-1}$) &  (km\,s$^{-1}$)& \\[1em] 
\tableline\\ 

ASA-1 & $-$13232.8 & $-$8928.5 & 4304.3 & $-$11126.1 & $-$11012.0 & 0.205 & 0.016 \\
ASA-2 & $-$10920.5 & $-$7369.8 & 3550.7 & $-$9144.0 & $-$9100.4 & 0.124 & 0.010\\
ASA-3 & $-$12641.1 & $-$9903.0 & 2738.1 & $-$11232.1 & $-$11231.1 & 0.148 & 0.012\\
ASA-4 & $-$16454.8 & $-$7915.9 & 8538.9 & $-$12067.3 & $-$12139.7 & 0.215 & 0.011\\
ASA-5 & $-$12102.2 & $-$9794.0 & 2308.2 & $-$10935.6 & $-$10939.9 & 0.136 & 0.009\\

\\[-0.6em]
\hline \\[0.6em]
\end{tabular}

\begin{tabular}{rrrrrrrrrrrrrrrrrrrrrrrrrrrrrrrrrrrrrrrrr}
\tableline\tableline \\[-0.3em]
$d_{\rm BAL}$[2] & $\sigma_{d_{\rm BAL}}$[2] &
EW[1] & $\sigma_{\rm EW}$[1]   & 
EW[2] & $\sigma_{\rm EW}$[2] & 
$\Delta \rm{EW}$ & $\sigma_{\Delta EW}$ & 
$\frac {\Delta\rm{EW}} {\langle \rm{EW} \rangle}$ & 
$\sigma_{\frac {\Delta\rm{EW}}  {\langle \rm{EW} \rangle}}$  \\
& & (\AA) & (\AA)& (\AA)& (\AA)& (\AA)& (\AA)
\\[1em] 
\tableline\\

0.181 & 0.009 & 5.10 & 0.40 & 4.75 & 0.31 & $-$0.35 & 0.51 & $-$0.07 & 0.14\\
0.136 & 0.008 & 2.17 & 0.49 & 3.00 & 0.31 & 0.83 & 0.58 & 0.32 & 0.32\\
0.164 & 0.011 & 2.31 & 0.35 & 2.78 & 0.26 & 0.47 & 0.43 & 0.18 & 0.24\\
0.267 & 0.007 & 10.12 & 1.11 & 13.65 & 0.96 & 3.53 & 1.47 & 0.30 & 0.18\\
0.167 & 0.007 & 1.95 & 0.11 & 2.38 & 0.07 & 0.43 & 0.14 & 0.20 & 0.09\\

 \\[-0.6em]
\hline \\[0.6em]
\end{tabular}
\\ \textbf{Notes. }Throughout this table [1] indicates the first-epoch spectra and [2] indicates 
the second-epoch spectra.
\\ (This table is available in its entirety in a machine-readable form in the online journal. 
A portion is shown here for guidance regarding its form and content.)}

\end{center}
\end{table*}

{
\begin{table*}
\caption{C\,{\sc iv}$_{\rm s0}$ Troughs} \label{mcs0}
\begin{center}
{\scriptsize 
\begin{tabular}{ccrrrrrrrrrrrrrrrrrrrrrrrrrrrrrrrrrrrrrrr}
\tableline\tableline \\[-0.3em]

Quasar & C\,{\sc iv}$_{s0}$ &
\multicolumn{1}{c}{$v_{\rm max}$} &
\multicolumn{1}{c}{$v_{\rm min}$} & 
\multicolumn{1}{c}{$\Delta v$} & 
\multicolumn{1}{c}{$v_{\rm cent}$[1]} & 
\multicolumn{1}{c}{$v_{\rm cent}$[2] }&
$d_{\rm BAL}$[1] & $\sigma_{d_{\rm BAL}}$[1] & 
 \nodata\\
ID & Trough ID &  (km\,s$^{-1}$) & ( km\,s$^{-1}$) &  (km\,s$^{-1}$)&  (km\,s$^{-1}$) &  (km\,s$^{-1}$)& \\
\\[-0.3em]
\tableline\\ 
Q6 & Cs0-1 & $-19685.5$ & $-17001.8$ & $2683.6$ & $-18364.8$ & $-18301.9$ & $0.271$ & 0.022 \\
Q7 & Cs0-2 & $-8706.6$ & $-4632.3$ & $4074.4$ & $-7007.8$ & $-6966.5$ & $0.644$ & 0.036 \\
Q9 & Cs0-3 & $-11106.9$ & $-8037.0$ & $3070.0$ & $-9649.7$ & $-9640.8$ & $0.386$ & 0.033 \\
Q11 & Cs0-4 & $-20000.0$ & $-12762.0$ & $7238.0$ & $-16286.1$ & $-16213.2$ & $0.231$ & 0.008 \\
Q14 & Cs0-5 & $-7718.3$ & $-4045.6$ & $3672.7$ & $-5927.5$ & $-6022.1$ & $0.419$ & 0.023 \\
Q15 & Cs0-6 & $-20000.0$ & $-16994.0$ & $3006.0$ & $-18419.0$ & $-18327.8$ & $0.277$ & 0.018 \\
Q16 & Cs0-7 & $-18275.7$ & $-15476.7$ & $2799.0$ & $-16788.8$ & $-16843.7$ & $0.357$ & 0.035 \\
Q17 & Cs0-8 & $-7513.5$ & $-4592.3$ & $2921.1$ & $-5896.5$ & $-5936.5$ & $0.640$ & 0.040 \\
Q18 & Cs0-9 & $-20000.0$ & $-16264.3$ & $3735.7$ & $-18157.3$ & $-17886.8$ & $0.368$ & 0.024 \\
Q24 & Cs0-10 & $-20000.0$ & $-13729.8$ & $6270.2$ & $-16870.4$ & $-16856.6$ & $0.275$ & 0.012 \\
\\[-0.6em]
\hline \\[0.6em]
\end{tabular}

\begin{tabular}{rrrrrrrrrrrrrrrrrrrrrrrrrrrrrrrrrrrrrrrrr}
\tableline\tableline \\[-0.3em]
$d_{\rm BAL}$[2] & $\sigma_{d_{\rm BAL}}$[2] &
EW[1] & $\sigma_{\rm EW}$[1]   & 
EW[2] & $\sigma_{\rm EW}$[2] & 
$\Delta \rm{EW}$ & $\sigma_{\Delta EW}$ & 
$\frac {\Delta\rm{EW}} {\langle \rm{EW} \rangle}$ & 
$\sigma_{\frac {\Delta\rm{EW}}  {\langle \rm{EW} \rangle}}$  \\
& & (\AA) & (\AA)& (\AA)& (\AA)& (\AA)& (\AA)\\
\\[-0.3em]
\tableline\\
0.210 & 0.014 & 3.65 & 0.31 & 2.83 & 0.18 & $-0.82$ & 0.36 & $-0.25$ & 0.14 \\
0.548 & 0.034 & 13.79 & 0.34 & 11.11 & 0.14 & $-2.68$ & 0.36 & $-0.22$ & 0.04 \\
0.335 & 0.032 & 5.92 & 0.20 & 5.24 & 0.11 & $-0.67$ & 0.23 & $-0.12$ & 0.06 \\
0.208 & 0.010 & 8.29 & 0.17 & 7.45 & 0.23 & $-0.84$ & 0.28 & $-0.11$ & 0.05 \\
0.388 & 0.024 & 7.93 & 0.19 & 7.06 & 0.09 & $-0.86$ & 0.21 & $-0.12$ & 0.04 \\
0.281 & 0.021 & 4.18 & 0.30 & 4.29 & 0.12 & $0.12$ & 0.32 & $0.03$ & 0.10 \\
0.200 & 0.027 & 4.82 & 0.32 & 2.50 & 0.15 & $-2.32$ & 0.35 & $-0.63$ & 0.11 \\
0.673 & 0.036 & 9.45 & 0.25 & 10.13 & 0.09 & $0.68$ & 0.27 & $0.07$ & 0.04 \\
0.369 & 0.027 & 6.37 & 0.37 & 6.11 & 0.19 & $-0.26$ & 0.42 & $-0.04$ & 0.09 \\
0.183 & 0.010 & 8.53 & 0.42 & 5.73 & 0.32 & $-2.80$ & 0.53 & $-0.39$ & 0.10 \\
 \\[-0.6em]
\hline \\[0.6em]
\end{tabular}
\\ \textbf{Notes. }Throughout this table [1] indicates the first-epoch spectra and [2] indicates the second-epoch spectra.
\\ (This table is available in its entirety in a machine-readable form in the online journal. A portion is shown here 
for guidance regarding its form and content.)}
\end{center}
\end{table*}

{
\begin{table*}
\caption{C\,{\sc iv}$_{\rm sa}$ Troughs} \label{mcsa}
\begin{center}
{\scriptsize 
\begin{tabular}{ccrrrrrrrrrrrrrrrrrrrrrrrrrrrrrrrrrrrrrrr}
\tableline\tableline \\[-0.3em]

Quasar & C\,{\sc iv}$_{sa}$ &
\multicolumn{1}{c}{$v_{\rm max}$} &
\multicolumn{1}{c}{$v_{\rm min}$} & 
\multicolumn{1}{c}{$\Delta v$} & 
\multicolumn{1}{c}{$v_{\rm cent}$[1]} & 
\multicolumn{1}{c}{$v_{\rm cent}$[2] }&
$d_{\rm BAL}$[1] & $\sigma_{d_{\rm BAL}}$[1] & 
 \nodata\\
ID & Trough ID &  (km\,s$^{-1}$) & ( km\,s$^{-1}$) &  (km\,s$^{-1}$)&  (km\,s$^{-1}$) &  (km\,s$^{-1}$)& \\
\\[-0.3em]
\tableline\\ 

Q68 & Csa-1 & $-16520.3$ & $-7967.2$ & $8553.2$ & $-12538.7$ & $-12413.4$ & $0.470$ & 0.019 \\
Q74 & Csa-2 & $-12104.4$ & $-5549.0$ & $6555.4$ & $-8424.5$ & $-8381.3$ & $0.542$ & 0.025 \\
Q92 & Csa-3 & $-15812.8$ & $-7274.5$ & $8538.3$ & $-11413.4$ & $-11523.5$ & $0.307$ & 0.018 \\
Q95 & Csa-4 & $-20000.0$ & $-7477.7$ & $12522.3$ & $-14343.8$ & $-13767.6$ & $0.408$ & 0.015 \\
Q102 & Csa-5 & $-15573.0$ & $-9850.9$ & $5722.1$ & $-12641.0$ & $-12533.8$ & $0.327$ & 0.017 \\
Q121 & Csa-6 & $-16532.5$ & $-7118.1$ & $9414.4$ & $-11798.3$ & $-11969.2$ & $0.335$ & 0.016 \\
Q122 & Csa-7 & $-12716.2$ & $-4706.0$ & $8010.1$ & $-8953.3$ & $-9023.5$ & $0.310$ & 0.017 \\
Q192 & Csa-8 & $-16109.3$ & $-11388.8$ & $4720.5$ & $-13675.9$ & $-13745.2$ & $0.098$ & 0.009 \\
Q217 & Csa-9 & $-6388.4$ & $-4238.7$ & $2149.7$ & $-5038.8$ & $-5092.9$ & $0.711$ & 0.053 \\
Q240 & Csa-10 & $-11158.8$ & $-7751.6$ & $3407.3$ & $-9243.6$ & $-9357.1$ & $0.415$ & 0.036 \\

\\[-0.6em]
\hline \\[0.6em]
\end{tabular}

\begin{tabular}{rrrrrrrrrrrrrrrrrrrrrrrrrrrrrrrrrrrrrrrrr}
\tableline\tableline \\[-0.3em]
$d_{\rm BAL}$[2] & $\sigma_{d_{\rm BAL}}$[2] &
EW[1] & $\sigma_{\rm EW}$[1]   & 
EW[2] & $\sigma_{\rm EW}$[2] & 
$\Delta \rm{EW}$ & $\sigma_{\Delta EW}$ & 
$\frac {\Delta\rm{EW}} {\langle \rm{EW} \rangle}$ & 
$\sigma_{\frac {\Delta\rm{EW}}  {\langle \rm{EW} \rangle}}$  \\
& & (\AA) & (\AA)& (\AA)& (\AA)& (\AA)& (\AA)\\
\\[-0.3em]
\tableline\\
0.225 & 0.015 & 20.15 & 0.82 & 9.04 & 0.41 & $-11.10$ & 0.91 & $-0.76$ & 0.07 \\
0.317 & 0.024 & 18.23 & 0.53 & 8.10 & 0.46 & $-10.12$ & 0.71 & $-0.77$ & 0.07 \\
0.348 & 0.016 & 12.82 & 0.96 & 15.08 & 0.57 & $2.26$ & 1.12 & $0.16$ & 0.11 \\
0.152 & 0.006 & 25.10 & 1.20 & 9.16 & 0.59 & $-15.94$ & 1.34 & $-0.93$ & 0.09 \\
0.386 & 0.019 & 8.51 & 0.66 & 10.35 & 0.43 & $1.84$ & 0.79 & $0.20$ & 0.12 \\
0.295 & 0.013 & 13.48 & 0.76 & 13.89 & 0.44 & $0.41$ & 0.87 & $0.03$ & 0.09 \\
0.339 & 0.016 & 12.08 & 0.95 & 13.79 & 0.27 & $1.71$ & 0.99 & $0.13$ & 0.10 \\
0.200 & 0.008 & 1.52 & 0.43 & 4.71 & 0.51 & $3.19$ & 0.66 & $1.02$ & 0.29 \\
0.626 & 0.056 & 8.15 & 0.25 & 6.81 & 0.11 & $-1.35$ & 0.27 & $-0.18$ & 0.05 \\
0.225 & 0.022 & 7.18 & 0.22 & 3.42 & 0.13 & $-3.76$ & 0.25 & $-0.71$ & 0.06 \\
 \\[-0.6em]
\hline \\[0.6em]
\end{tabular}
\\ \textbf{Notes. }Throughout this table [1] indicates the first-epoch spectra and [2] indicates the second-epoch spectra.
\\ (This table is available in its entirety in a machine-readable form in the online journal. A portion is shown here for guidance regarding its form and content.)}
\end{center}
\end{table*} 

{
\begin{table*}
\caption{C\,{\sc iv}$_{\rm Sa}$ Troughs} \label{mcssa}
\begin{center}
{\scriptsize 
\begin{tabular}{ccrrrrrrrrrrrrrrrrrrrrrrrrrrrrrrrrrrrrrrr}
\tableline\tableline \\[-0.3em]

Quasar & C\,{\sc iv}$_{Sa}$ &
\multicolumn{1}{c}{$v_{\rm max}$} &
\multicolumn{1}{c}{$v_{\rm min}$} & 
\multicolumn{1}{c}{$\Delta v$} & 
\multicolumn{1}{c}{$v_{\rm cent}$[1]} & 
\multicolumn{1}{c}{$v_{\rm cent}$[2] }&
$d_{\rm BAL}$[1] & $\sigma_{d_{\rm BAL}}$[1] & 
 \nodata\\
ID & Trough ID &  (km\,s$^{-1}$) & ( km\,s$^{-1}$) &  (km\,s$^{-1}$)&  (km\,s$^{-1}$) &  (km\,s$^{-1}$)& \\
\\[-0.3em]
\tableline\\ 

Q2 & CSa-1 & $-20000.0$ & $-7997.7$ & $12002.3$ & $-14588.0$ & $-14972.5$ & $0.409$ & 0.018 \\
Q19 & CSa-2 & $-9532.0$ & $-3000.0$ & $6532.0$ & $-7219.7$ & $-7519.0$ & $0.537$ & 0.026 \\
Q21 & CSa-3 & $-20000.0$ & $-7452.5$ & $12547.5$ & $-14198.9$ & $-14177.9$ & $0.698$ & 0.015 \\
Q33 & CSa-4 & $-20000.0$ & $-6346.1$ & $13653.9$ & $-14370.2$ & $-14044.3$ & $0.448$ & 0.020 \\
Q39 & CSa-5 & $-10772.3$ & $-3000.0$ & $7772.3$ & $-8305.6$ & $-8191.4$ & $0.592$ & 0.026 \\
Q43 & CSa-6 & $-20000.0$ & $-7389.2$ & $12610.8$ & $-15426.7$ & $-15715.8$ & $0.548$ & 0.021 \\
Q45 & CSa-7 & $-5099.9$ & $-3000.0$ & $2099.9$ & $-4180.0$ & $-4273.3$ & $0.473$ & 0.032 \\
Q46 & CSa-8 & $-10647.4$ & $-4855.0$ & $5792.4$ & $-8520.5$ & $-8436.7$ & $0.791$ & 0.028 \\
Q49 & CSa-9 & $-12874.5$ & $-7052.6$ & $5821.9$ & $-9976.5$ & $-10007.5$ & $0.404$ & 0.022 \\
Q50 & CSa-10 & $-6340.5$ & $-3018.0$ & $3322.5$ & $-4813.8$ & $-5072.8$ & $0.705$ & 0.039 \\
\\[-0.6em]
\hline \\[0.6em]
\end{tabular}

\begin{tabular}{rrrrrrrrrrrrrrrrrrrrrrrrrrrrrrrrrrrrrrrrr}
\tableline\tableline \\[-0.3em]
$d_{\rm BAL}$[2] & $\sigma_{d_{\rm BAL}}$[2] &
EW[1] & $\sigma_{\rm EW}$[1]   & 
EW[2] & $\sigma_{\rm EW}$[2] & 
$\Delta \rm{EW}$ & $\sigma_{\Delta EW}$ & 
$\frac {\Delta\rm{EW}} {\langle \rm{EW} \rangle}$ & 
$\sigma_{\frac {\Delta\rm{EW}}  {\langle \rm{EW} \rangle}}$  \\
& & (\AA) & (\AA)& (\AA)& (\AA)& (\AA)& (\AA)\\
\\[-0.3em]
\tableline\\
0.406 & 0.018 & 21.82 & 1.52 & 24.36 & 0.51 & $2.54$ & 1.60 & $0.11$ & 0.09 \\
0.591 & 0.029 & 19.20 & 0.45 & 20.23 & 0.31 & $1.03$ & 0.54 & $0.05$ & 0.04 \\
0.703 & 0.015 & 43.90 & 0.15 & 44.02 & 0.15 & $0.11$ & 0.22 & $0.00$ & 0.01 \\
0.498 & 0.016 & 29.36 & 1.19 & 33.83 & 0.63 & $4.47$ & 1.35 & $0.14$ & 0.06 \\
0.641 & 0.022 & 24.24 & 0.60 & 25.89 & 0.77 & $1.64$ & 0.97 & $0.07$ & 0.05 \\
0.574 & 0.023 & 34.78 & 1.03 & 36.19 & 0.33 & $1.41$ & 1.08 & $0.04$ & 0.04 \\
0.653 & 0.034 & 5.24 & 0.16 & 7.42 & 0.08 & $2.18$ & 0.18 & $0.34$ & 0.04 \\
0.788 & 0.028 & 23.43 & 0.11 & 23.31 & 0.12 & $-0.12$ & 0.17 & $0.00$ & 0.01 \\
0.413 & 0.018 & 11.73 & 0.65 & 12.15 & 0.28 & $0.42$ & 0.71 & $0.03$ & 0.08 \\
0.703 & 0.045 & 12.35 & 0.37 & 11.87 & 0.07 & $-0.48$ & 0.38 & $-0.04$ & 0.04 \\
 \\[-0.6em]
\hline \\[0.6em]
\end{tabular}
\\ \textbf{Notes. }Throughout this table [1] indicates the first-epoch spectra and [2] indicates the second-epoch spectra.
\\ (This table is available in its entirety in a machine-readable form in the online journal. A portion is shown here for guidance regarding its form and content.)}
\end{center}
\end{table*}


In this study, we generally focus on the  C\,{\sc iv}, Si\,{\sc iv}, and Al\,{\sc iii} BAL regions. 
We do not analyze in detail the BAL regions of the other transitions listed in Section~\ref{hmlintro} 
for  several reasons: 
(1) Due to the limited wavelength coverage of the observed spectra (\hbox{3800--9200~\AA} for SDSS,  
and \hbox{3600--10000~\AA} for BOSS), only a small number of quasars  have suitable wavelength 
coverage to investigate all of these transitions. 
(2)  The strong Ly$\alpha$ and N\,{\sc v}  lines  blend with each other and  their blueshifted absorption 
lines blend with Ly$\alpha$ forest, where numerous absorption features from intervening gas  are found. 
(3) A proper investigation  of Mg\,{\sc ii} BALs would require modeling and subtraction of broad Fe\,{\sc ii} emission 
in addition to underlying continuum estimation. Moreover, only a small fraction of our targeted quasars have 
the wavelength coverage to investigate simultaneously  C\,{\sc iv}, Si\,{\sc iv}, and Mg\,{\sc ii} BAL troughs.

\subsection{Measurements of BAL Troughs}\label{ident2}

We measure the rest-frame timescales, $\Delta t$,  for our main sample; $\Delta t$ is between  
0.85 and 4.13~yr with a mean of 2.55 yr. Given the established connection between BAL-variation 
strength and timescale \citep[e.g.,][]{gibson10,ak13}, we compared the  $\Delta t$ distributions for 
C\,{\sc iv}$_{\rm 00}$, C\,{\sc iv}$_{\rm S0}$, and C\,{\sc iv}$_{\rm SA}$ troughs.  In Figure~\ref{fig4}, 
we show the $\Delta t$ distributions for all C\,{\sc iv} troughs in our main sample and for 
C\,{\sc iv}$_{\rm 00}$, C\,{\sc iv}$_{\rm S0}$, and C\,{\sc iv}$_{\rm SA}$ troughs. 
Given their definition, the sum of the total number of C\,{\sc iv}$_{\rm 00}$, C\,{\sc iv}$_{\rm S0}$, and 
C\,{\sc iv}$_{\rm SA}$ troughs is not equal to the total number of C\,{\sc iv} troughs in our sample. 
We compared these $\Delta t$ distributions using two-sample AD tests. 
The test results show no significant differences between the distributions.

\begin{figure}[!h!]
\epsscale{1.18}
\plotone{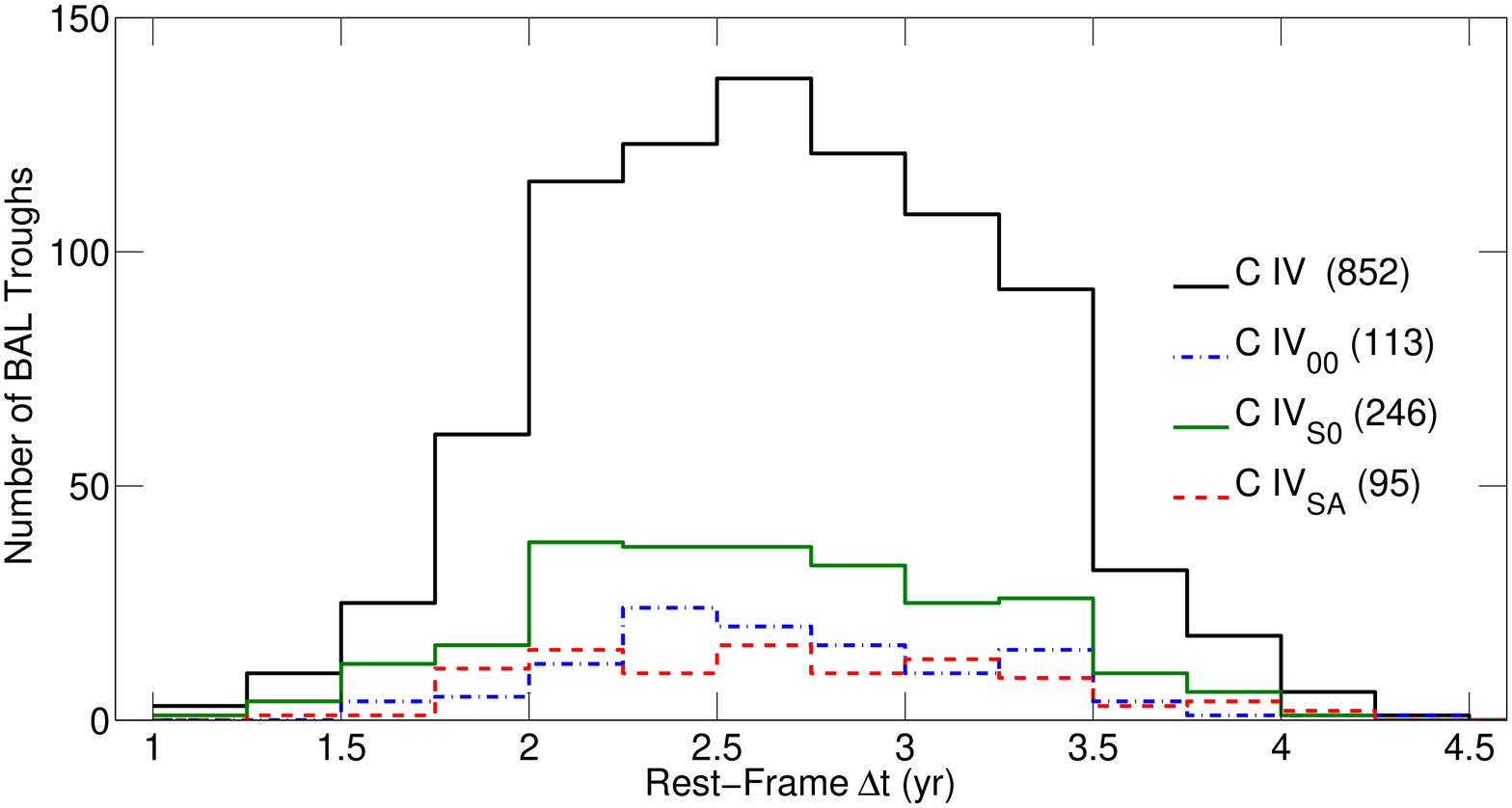}
\caption{The rest-frame $\Delta t$ distributions for all C\,{\sc iv} troughs in our main sample (solid black) and 
C\,{\sc iv}$_{\rm 00}$  (dot-dashed blue), C\,{\sc iv}$_{\rm S0}$ (dashed green), and C\,{\sc iv}$_{\rm SA}$ 
(solid red) BAL troughs. The total number of BAL troughs in each sample is given in parentheses. 
After renormalization for the number of each type of trough, there are no statistically significant differences between 
the distributions. Given their definitions, the sum of the total numbers of C\,{\sc iv}$_{\rm 00}$, C\,{\sc iv}$_{\rm S0}$, 
and C\,{\sc iv}$_{\rm SA}$ troughs are not necessarily equal to total number of  C\,{\sc iv} troughs in our main sample.
The Anderson-Darling test results  indicate no significant $\Delta{t}$ differences between the samples.}
\label{fig4}
\end{figure}

Considering that BAL troughs are sometimes isolated and occasionally appear in complexes in which single troughs 
may split or adjacent troughs may merge over time, we define minimum and maximum velocities for each BAL trough 
following the BAL-trough identification algorithm for multi-epoch observations described in Section 3.2 of \citet{ak13}.
We set $v_{\rm min}$  to be the minimum red-edge velocity  and $v_{\rm max}$ to be the 
maximum blue-edge velocity of the associated absorption complex in all available epochs.  
For BAL troughs that reach beyond our adopted velocity limits (see Equation~\ref{ehml1}), 
we simply truncate the BAL trough $v_{\rm max}$  to be  $-20,000~\rm{km\,s^{-1}}$
and $v_{\rm min}$  to be  $-3000~\rm{km\,s^{-1}}$.

We measure the rest-frame EW of each BAL trough in each epoch. To calculate uncertainties on EWs, we 
propagate observational errors for each contributing pixel  and continuum-estimation errors \citep[see][]{ak12,ak13}
using Equations 1 and 2 of \citet{kaspi02}.  We calculate EW variations, $\Delta$EW, fractional EW variations, 
$\Delta$EW$/\langle{\rm EW}\rangle$, and uncertainties on these quantities, $\sigma_{\Delta {\rm EW}}$ and 
$\sigma_{\Delta{\rm EW}/\langle{\rm EW} \rangle}$, following Equations 3 and 4 of \citet{ak13}, respectively. 
In this study, positive values of $\Delta$EW and $\Delta$EW$/\langle{\rm EW}\rangle$ indicate strengthening 
troughs and negative values indicate weakening troughs.  In addition, we measure the average depth, $d_{\rm BAL}$, for 
each trough by calculating the mean distance of each contributing data point  from the normalized continuum level. We 
calculate a BAL-trough velocity width, $\Delta v$, and a weighted centroid velocity, $v_{\rm cent}$, which is the mean 
velocity where each data point is weighted with its distance from the normalized continuum level.  
We adapt redshift values from \citet{hw10} for all velocity calculations. 

\subsection{Comparisons with {\rm Mg\,{\sc ii}}, {\rm Fe\,{\sc ii}}, and {\rm P\,{\sc v}} } \label{mgp}

We have compared our adopted classification of C\,{\sc iv} BAL troughs with the standard subtypes of HiBALs, 
LoBALs,  and FeLoBALs (see Section~\ref{hmlintro}). 
All 113 {C\,{\sc iv}$_{\rm 00}$} and 246 {C\,{\sc iv}$_{\rm S0}$} troughs fit into the definition of HiBALs where 
only BAL troughs of high-ionization transitions are present in the spectra. Standard classes of LoBALs  and 
FeLoBALs are identified with the presence of Mg\,{\sc ii} and Fe\,{\sc ii} absorption, respectively. 
We visually investigated the Mg\,{\sc ii} $\lambda\lambda$2797, 2804~\AA\, and 
Fe\,{\sc ii} $\lambda\lambda$2400, 2600~\AA\, absorption-line regions for 
95 {C\,{\sc iv}$_{\rm SA}$} troughs to examine the correspondence to the standard  LoBAL and FeLoBAL definitions. 
The spectra of  34 quasars with {C\,{\sc iv}$_{\rm SA}$} troughs do not have coverage of the Mg\,{\sc ii} region, 
and these are not utilized in our correspondence checking.  We find that 
$\approx 74\%$ (45 out of 61) of quasars with {C\,{\sc iv}$_{\rm SA}$} troughs exhibit Mg\,{\sc ii} BAL/mini-BAL 
troughs at corresponding velocities; four of them also exhibit Fe\,{\sc ii} BAL/mini-BAL troughs. 
There are 16 quasars with {C\,{\sc iv}$_{\rm SA}$} troughs that have no Mg\,{\sc ii} 
absorption. Due to the differences between their ionization potentials (28.4~eV for Al\,{\sc iii} and 15.0~eV for Mg\,{\sc ii}),
Al\,{\sc iii} absorption is expected to be slightly more common than Mg\,{\sc ii} absorption given that higher 
ionization troughs are found more frequently than lower ionization troughs in BAL quasars \citep[e.g.,][]{hall02}.

Previous investigations of P\,{\sc v} $\lambda\lambda1118, 1128$~\AA\, absorption lines  corresponding in velocity with 
C\,{\sc iv} and  Si\,{\sc iv} BAL troughs have shown that P\,{\sc v} absorption lines are an important indicator of  line 
saturation \citep[e.g.,][]{hamann98,arav01}; note that the ionization potentials of P\,{\sc v} (65.0~eV) and C\,{\sc iv} 
(64.5~eV) are very similar. We thus visually investigate the P\,{\sc v} BAL regions that align with our 
{C\,{\sc iv}$_{\rm 00}$}, {C\,{\sc iv}$_{\rm S0}$}, and {C\,{\sc iv}$_{\rm SA}$} troughs. 
The spectral coverage is sufficient to investigate the P\,{\sc v} BAL region for  40  
{C\,{\sc iv}$_{\rm 00}$} troughs, 113 {C\,{\sc iv}$_{\rm S0}$} troughs, and 47  {C\,{\sc iv}$_{\rm SA}$} troughs. 
Our visual inspection reveals that spectra of a large fraction ($\approx88\%$) of {C\,{\sc iv}$_{\rm SA}$} troughs exhibit 
visually detectable P\,{\sc v} absorption features; a majority ($\approx70\%$) of {C\,{\sc iv}$_{\rm SA}$}  troughs are 
accompanied by moderate-to-strong P\,{\sc v} absorption. P\,{\sc v} absorption that aligns with {C\,{\sc iv}$_{\rm S0}$} 
troughs is generally weaker than that  aligning  with {C\,{\sc iv}$_{\rm SA}$} troughs. Approximately half of 
the {C\,{\sc iv}$_{\rm S0}$} troughs are accompanied by  detectable P\,{\sc v} absorption; however, 
only $\approx 10\%$ of those {C\,{\sc iv}$_{\rm S0}$} troughs align with  moderate-to-strong P\,{\sc v} absorption. 
Only a small fraction ($\approx12\%$) of 
{C\,{\sc iv}$_{\rm 00}$} troughs are accompanied by detectable P\,{\sc v} absorption, and the  strength of this 
P\,{\sc v} absorption is generally   weak. These results show that quasars 
possessing detectable  P\,{\sc v} absorption in their spectra are also more likely to present lower-ionization transitions; 
therefore their {C\,{\sc iv}} troughs are likely to be  {C\,{\sc iv}$_{\rm SA}$} or to a  smaller extent {C\,{\sc iv}$_{\rm S0}$} 
troughs.

\section{Results}\label{hmlresults}

In this section, we present the observational results of our investigation. Utilizing the two-epoch 
observations for 113  C\,{\sc iv}$_{\rm 00}$, 246  C\,{\sc iv}$_{\rm S0}$, and 95  C\,{\sc iv}$_{\rm SA}$
BAL troughs, we  investigate the C\,{\sc iv}, Si\,{\sc iv}, and Al\,{\sc iii}  BAL profiles (Section~\ref{profile}), 
the C\,{\sc iv} trough strengths (Section~\ref{properties}), 
the C\,{\sc iv} trough velocities (Section~\ref{properties2}),
the C\,{\sc iv} trough variation profiles (Section~\ref{vprofile}), 
the C\,{\sc iv} trough EW variation characteristics (Section~\ref{EWvar}), and 
the C\,{\sc iv}, Si\,{\sc iv}, and Al\,{\sc iii} trough EW variation correlations  (Section~\ref{correlation}).

\subsection{BAL-Trough Profiles}\label{profile}

It is well known that BAL troughs from different transitions show different profiles; for instance, Al\,{\sc iii} 
BAL troughs tend to be narrower and  align with lower velocity portions of corresponding C\,{\sc iv} BAL 
troughs \citep[e.g.,][]{wey91,voit93,trump06}.  Moreover, previous studies \citep[e.g.,][]
{wey91,reichard03,allen11} have demonstrated  
that C\,{\sc iv} BAL troughs tend to be stronger in quasars exhibiting  Al\,{\sc iii} and/or Mg\,{\sc ii} BAL troughs 
(i.e., LoBAL quasars). To compare the typical properties of C\,{\sc iv}$_{\rm 00}$, C\,{\sc iv}$_{\rm S0}$, and 
C\,{\sc iv}$_{\rm SA}$ BAL troughs, we calculate composite mean profile shapes for each C\,{\sc iv} group. 
Figure~\ref{fig5} shows the mean profile shapes for 113 C\,{\sc iv}$_{\rm 00}$, 246 
C\,{\sc iv}$_{\rm S0}$, and 95 C\,{\sc iv}$_{\rm SA}$ BAL troughs as a function of 
outflow velocity relative to $v_{\rm min}$. 
{We calculate the mean profiles by defining an outflow velocity, \hbox{$v_{\rm t} = v - v_{\rm min}$}, 
which is set to 0 at its  $v_{\rm min}$ and runs from 0 to $\Delta v$ (where $\Delta v$ is the trough velocity width).}
{We fix the normalized flux density, $F_{\lambda}$, to 1 at velocities higher than $v_{\rm max}$ 
for each C\,{\sc iv} BAL trough. Given the BAL-trough definition, $F_{\lambda}$ is less than 0.9 for 
$0 < v_{\rm t} < \Delta v$.}
We also present mean profile shapes of Si\,{\sc iv}  and Al\,{\sc iii} BAL troughs  in overlapping 
velocity ranges for comparison. 

A comparison of the resulting composite profiles in Figure ~\ref{fig5}
indicates  that  C\,{\sc iv}$_{\rm 00}$ troughs 
tend to be shallower and narrower than both C\,{\sc iv}$_{\rm S0}$ and C\,{\sc iv}$_{\rm SA}$ troughs. 
The C\,{\sc iv}$_{\rm S0}$ and C\,{\sc iv}$_{\rm SA}$ trough depths change in a characteristic manner 
as a function of velocity within the trough:  troughs are generally deeper  at  lower velocities. 
Consistent with previous studies, the composite profile shapes show that  C\,{\sc iv} BALs are usually  deepest  
when there is an Al\,{\sc iii} BAL at  corresponding velocities.  Similarly, a Si\,{\sc iv} BAL-trough profile change 
is apparent  between  the C\,{\sc iv}$_{\rm S0}$ and C\,{\sc iv}$_{\rm SA}$ samples.
We also find that Al\,{\sc iii} BAL troughs tend to be narrower and  be found at lower 
velocities than corresponding C\,{\sc iv}$_{\rm SA}$ BAL troughs.

We find qualitatively consistent results when computing median composites instead of mean composites.
This indicates that outliers are not strongly affecting our composites. 

\begin{figure}[h]
\epsscale{1.2}
\plotone{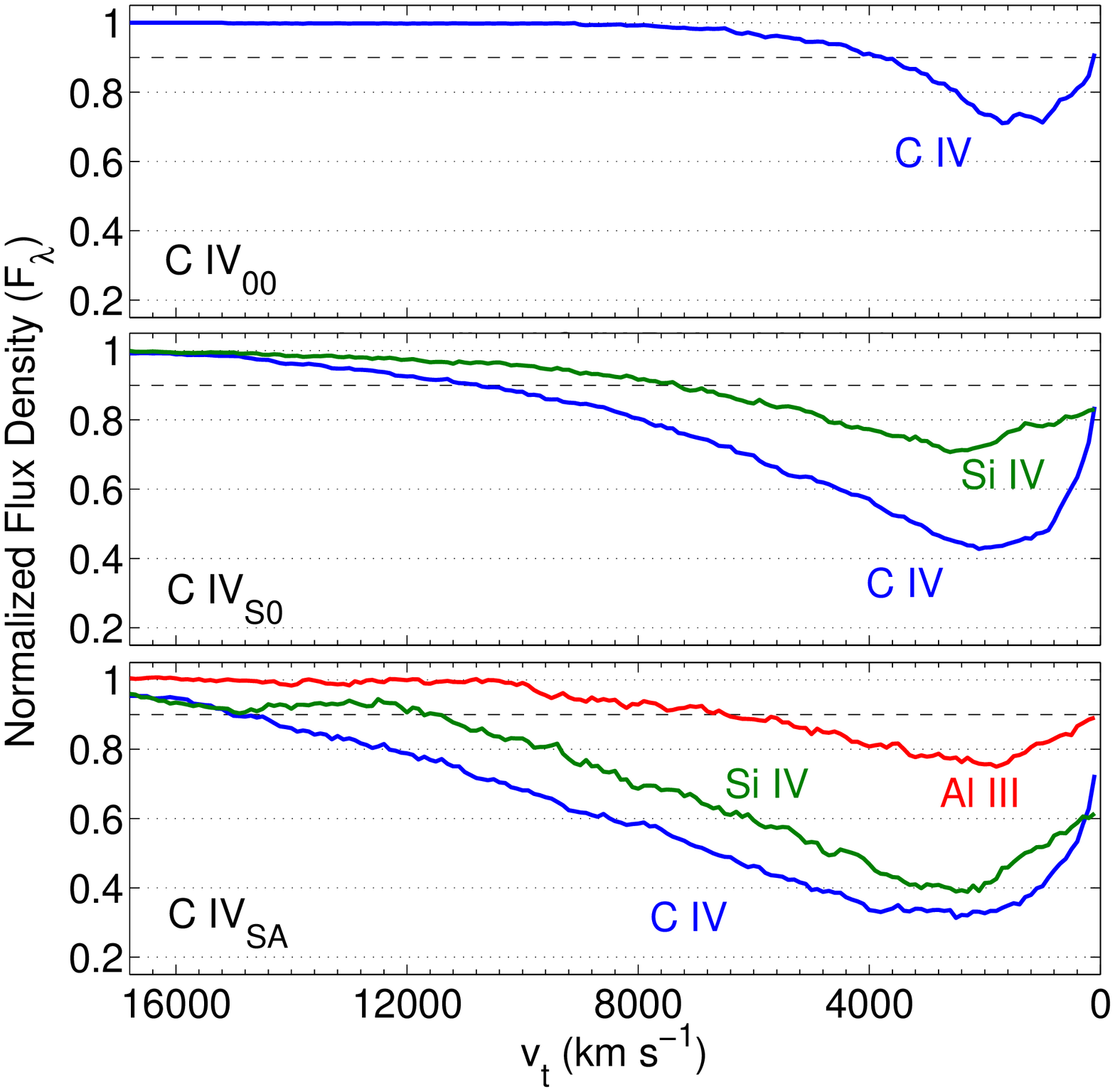} 
\caption{Composite BAL-trough profiles for C\,{\sc iv}$_{\rm 00}$ (top), C\,{\sc iv}$_{\rm S0}$ (middle), and 
C\,{\sc iv}$_{\rm SA}$ (bottom) BAL troughs. Solid curves show mean normalized flux density for  
C\,{\sc iv} (blue), Si\,{\sc iv} (green), and Al\,{\sc iii} (red) BAL troughs at overlapping velocities {as a function of 
outflow velocity relative to $v_{\rm min}$, $v_{\rm t}$ = $v - v_{\rm min}$. }
The horizontal dashed black lines in each panel show the level for 
10\% under the continuum, a level important for BAL-trough identification (See Equation~\ref{ehml1}). The composite 
profile shapes of C\,{\sc iv}$_{\rm 00}$, C\,{\sc iv}$_{\rm S0}$, and C\,{\sc iv}$_{\rm SA}$ troughs differ significantly; 
stronger and wider  C\,{\sc iv} troughs are found when BAL troughs from lower ionization transitions are present.}
\label{fig5}
\end{figure}


\subsection{C\,{\sc iv} BAL-Trough Strengths}\label{properties}

For a more quantitative comparison between  the C\,{\sc iv}$_{\rm 00}$, C\,{\sc iv}$_{\rm S0}$, 
and C\,{\sc iv}$_{\rm SA}$ groups, we assess differences between measured C\,{\sc iv} BAL-trough 
properties in this subsection and the next.  Figure~\ref{fig7} shows the distributions of average 
{EW from two-epoch observations, }
$\langle$EW$\rangle$, for C\,{\sc iv}$_{\rm 00}$, C\,{\sc iv}$_{\rm S0}$, and C\,{\sc iv}$_{\rm SA}$ 
BAL troughs. Considering that the total number of BAL troughs  increases with decreasing 
$\langle$EW$\rangle$,  the fraction of BAL troughs with given 
$\langle$EW$\rangle$ is also displayed in Figure~\ref{fig7}. We calculate the fractions as the ratio  of 
the number of C\,{\sc iv}$_{\rm 00}$, C\,{\sc iv}$_{\rm S0}$, and C\,{\sc iv}$_{\rm SA}$ BAL troughs 
to all 852 main-sample C\,{\sc iv} troughs; they therefore need not sum to unity in a given bin.

The mean $\langle$EW$\rangle$ is $15.46 \pm 0.42~{\rm \AA}$ for all 852 C\,{\sc iv} BAL troughs 
in our main sample, whereas it is $4.76 \pm 0.25~{\rm \AA}$ for C\,{\sc iv}$_{\rm 00}$, 
$19.29 \pm 0.62~{\rm \AA}$ for C\,{\sc iv}$_{\rm S0}$, and $32.38 \pm 1.35~{\rm \AA}$ for 
C\,{\sc iv}$_{\rm SA}$  troughs as given in Table~\ref{tab1}.  Uncertainties on the mean are 
calculated using the standard $\sigma/\sqrt{N}$ formula. 
The median $\langle$EW$\rangle$ is $11.51~{\rm \AA}$ for all  C\,{\sc iv}, 
$4.44~{\rm \AA}$ for C\,{\sc iv}$_{\rm 00}$, $17.58~{\rm \AA}$ for 
C\,{\sc iv}$_{\rm S0}$, and $31.14~{\rm \AA}$ for C\,{\sc iv}$_{\rm SA}$  BAL troughs.
These results confirm and quantify the increase of C\,{\sc iv} BAL-trough strength with the 
existence of absorption lines from lower ionization-level transitions.

\begin{figure*}[t]
\epsscale{0.75}
\plotone{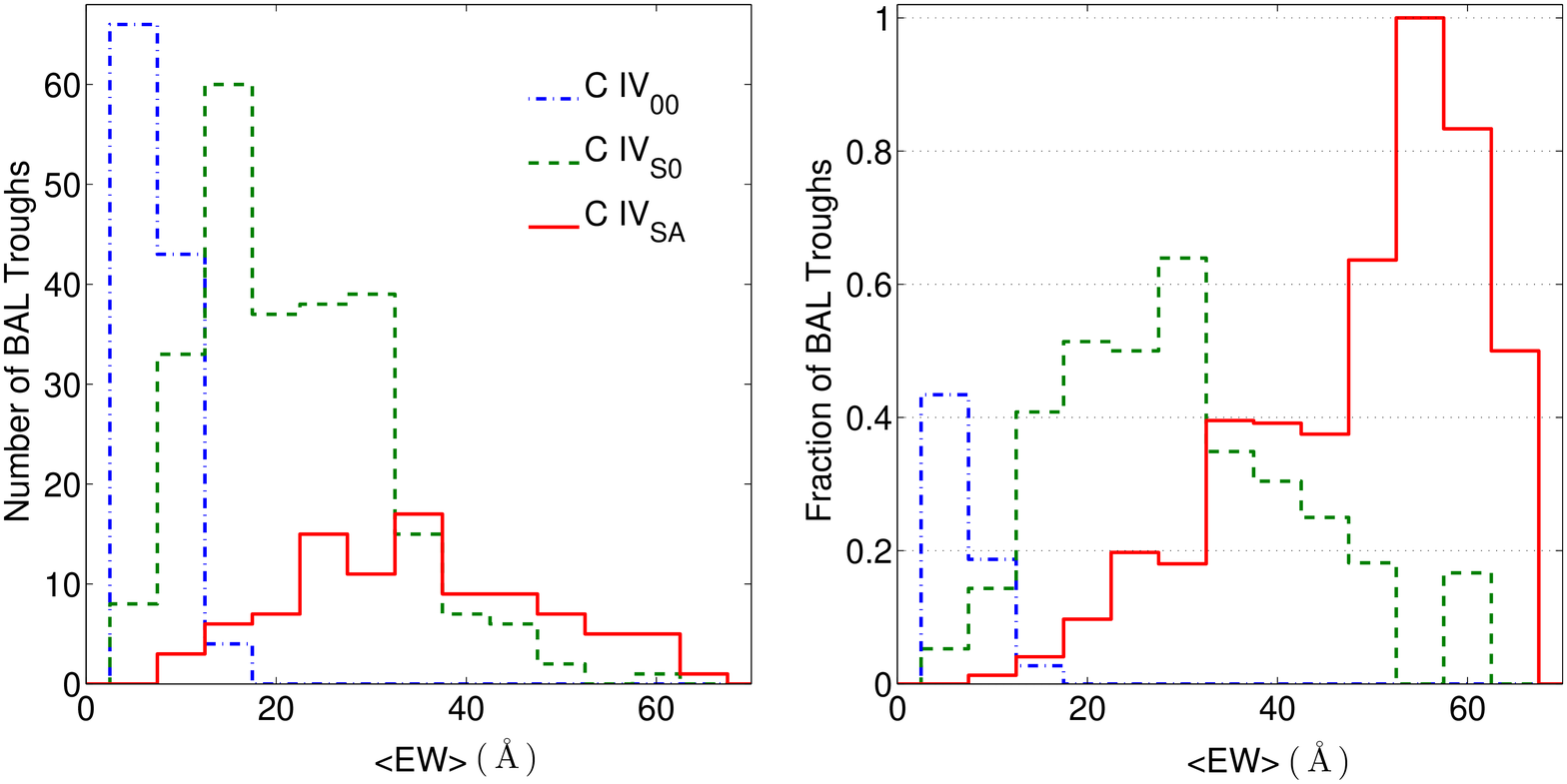} 
\caption{Average EW, $\langle$EW$\rangle$,  distributions for C\,{\sc iv}$_{\rm 00}$ 
(dot-dashed blue), C\,{\sc iv}$_{\rm S0}$ (dashed green), and C\,{\sc iv}$_{\rm SA}$ 
(solid red) BAL troughs. The right panel shows the fraction of C\,{\sc iv}$_{\rm 00}$, 
C\,{\sc iv}$_{\rm S0}$, and C\,{\sc iv}$_{\rm SA}$ BAL troughs to all 852 C\,{\sc iv} 
BAL troughs in our sample. Since troughs with mini-BALs are excluded, the numbers 
in the right panel need not  sum to unity for a given bin. The $\langle$EW$\rangle$ 
distributions of the three C\,{\sc iv} groups are significantly (99.9\%) different from each other.}
\label{fig7}
\end{figure*}

In order to determine the contributions of the depth and width components of BAL EWs, we assess  average 
depth from two-epoch observations, $\langle d_{\rm BAL}\rangle$, and velocity width, $\Delta v$, distributions 
for  C\,{\sc iv}$_{\rm 00}$, C\,{\sc iv}$_{\rm S0}$, and C\,{\sc iv}$_{\rm SA}$ troughs (see Figure~\ref{fig8}). 
Table~\ref{tab1} presents  the mean $\langle d_{\rm BAL}\rangle$ and $\Delta v$ values for all three C\,{\sc iv} 
BAL-trough groups. We compare these distributions using an AD test and find  that both the $\langle d_{\rm BAL}\rangle$ 
and $\Delta v$ distributions  for  C\,{\sc iv}$_{\rm 00}$, C\,{\sc iv}$_{\rm S0}$, and C\,{\sc iv}$_{\rm SA}$ BAL 
troughs are significantly different (at a confidence level of $>99.9\%$) from each other. Our findings indicate 
that  C\,{\sc iv}$_{\rm SA}$ BAL troughs tend to be the deepest and widest BAL troughs, while C\,{\sc iv}$_{\rm 00}$ 
troughs tend to be the shallowest and narrowest. 
The ranges of the $\langle d_{\rm BAL}\rangle$ and $\Delta v$ values for  C\,{\sc iv}$_{\rm 00}$, 
C\,{\sc iv}$_{\rm S0}$, and C\,{\sc iv}$_{\rm SA}$ BAL troughs demonstrate that the contributions 
of the depth and the width  to the differences between BAL-trough EWs are  comparable; the 
$\langle d_{\rm BAL}\rangle$ values change by a factor of $\approx 2.3$ and the $\Delta v$ 
values change by a factor of $\approx 3$ between the  C\,{\sc iv}$_{\rm 00}$ and C\,{\sc iv}$_{\rm SA}$ samples. 

{Note that for BAL troughs extending beyond our adopted BAL-trough definition velocity limits, the 
$v_{\rm min}$ and $v_{\rm max}$ values are  truncated at $-$3000 and $-$20,000~km~s$^{-1}$, respectively. 
Due to this truncation process, the measured $\Delta v$ values are only lower limits for such BAL troughs. 
However, the results discussed in Section~\ref{properties2} indicate that such truncation does not have a 
strong effect on our main conclusions.}

 \begin{table*}[h!]
\caption{Average values of BAL-trough properties for C\,{\sc iv}$_{\rm 00}$, 
C\,{\sc iv}$_{\rm S0}$, C\,{\sc iv}$_{\rm SA}$, and all C\,{\sc iv} BAL troughs}
\begin{center}
\begin{tabular}{c c c c c}
\hline \hline \\[-.3em]
& C\,{\sc iv}$_{\rm 00}$& C\,{\sc iv}$_{\rm S0}$&C\,{\sc iv}$_{\rm SA}$  & All C\,{\sc iv} \\[.6em] \hline \\[-0.3em]
$\langle$EW$\rangle$    (\AA)       & 4.76 $\pm$ 0.25 & 19.29 $\pm$ 0.62 & 32.38 $\pm$ 1.35 & $15.46\pm 0.42$\\
$\langle d_{\rm BAL}\rangle$ & 0.25 $\pm$ 0.009 &  0.47 $\pm$ 0.009 & 0.56 $\pm$ 0.012 &$0.41 \pm 0.006$\\
$\Delta v$         (km s$^{-1}$)              & 3940 $\pm$ 182 & 8600 $\pm$ 255 & 11668 $\pm$ 428  &$7195 \pm 146$\\[0.5em]
\hline \\[-0.5em]
Number of Data Points & 113 & 246 & 95 & 852\\[0.5em]
\hline 
\end{tabular}\\
Uncertainties on the mean are calculated using the standard $\sigma/\sqrt{N}$ formula.
\end{center}
\label{tab1}
\end{table*}%

\begin{figure*}[t!]
\epsscale{0.75}
\plotone{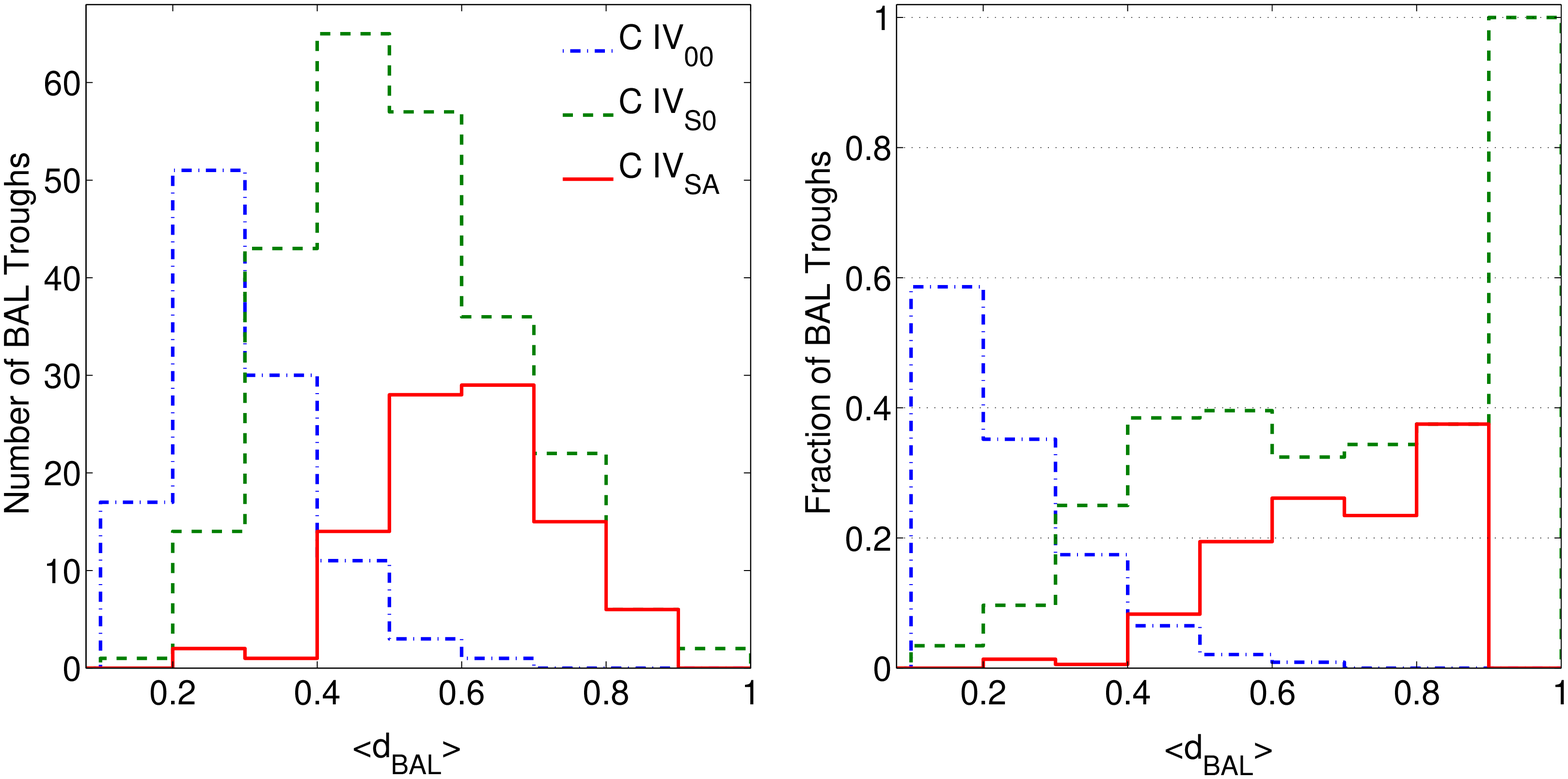} 
\plotone{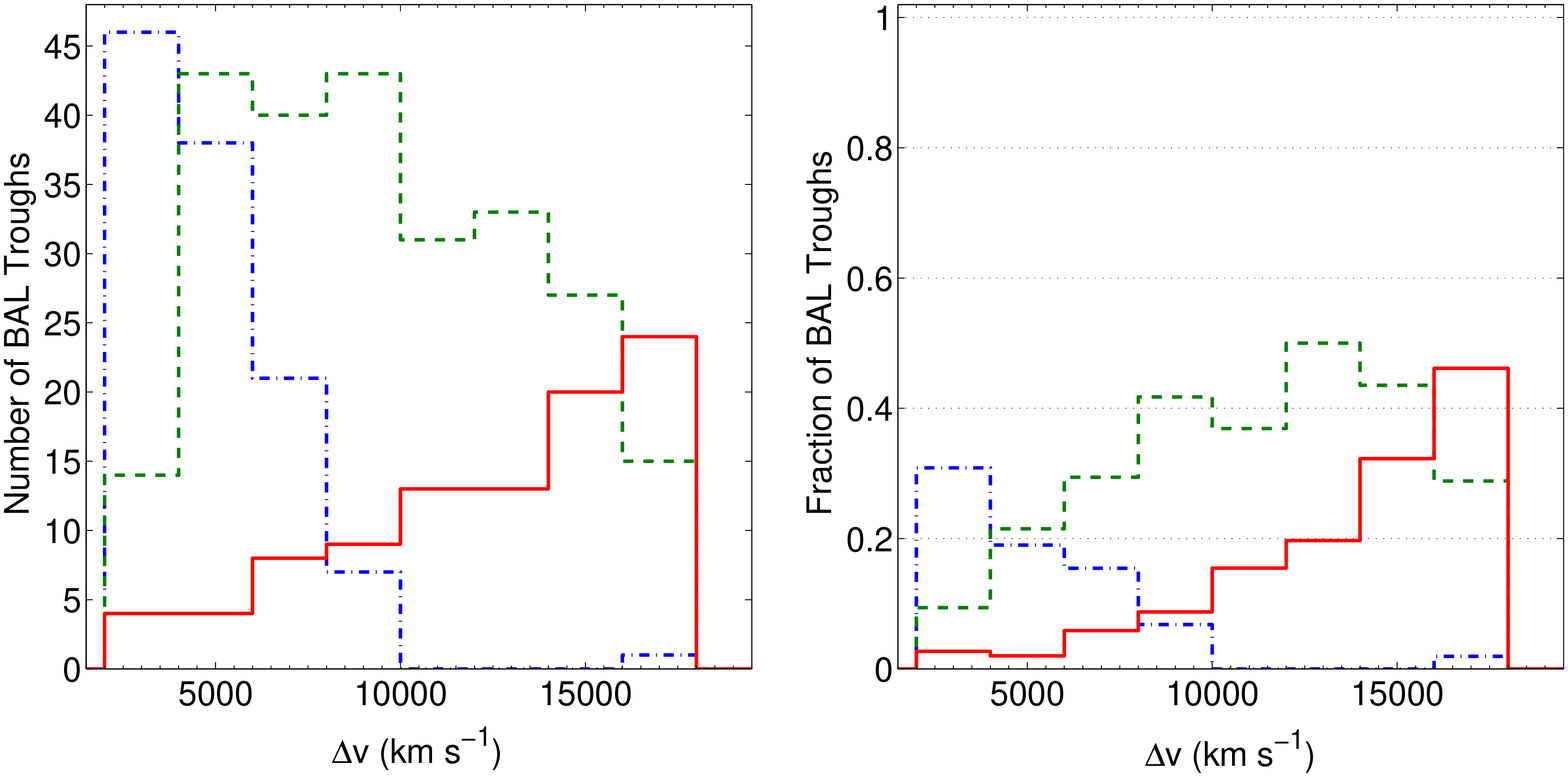} 
\caption{Average depth, $\langle d_{\rm BAL}\rangle$ (upper panels), and 
velocity width, $\Delta v$ (lower panels), distributions for C\,{\sc iv}$_{\rm 00}$ (dot-dashed blue), 
 C\,{\sc iv}$_{\rm S0}$ (dashed green), and C\,{\sc iv}$_{\rm SA}$ 
(solid red) BAL troughs. The right panels show the fraction of C\,{\sc iv}$_{\rm 00}$, C\,{\sc iv}$_{\rm S0}$, 
and C\,{\sc iv}$_{\rm SA}$  BAL troughs relative to all  C\,{\sc iv} BAL troughs in our main sample. 
The $\langle d_{\rm BAL}\rangle$ and   $\Delta v$ distributions for C\,{\sc iv}$_{\rm 00}$, 
C\,{\sc iv}$_{\rm S0}$, and C\,{\sc iv}$_{\rm SA}$ BAL troughs are significantly (99.9\%) different from each other.  }
\label{fig8}
\end{figure*}

As can be seen in Figures~\ref{fig5}, the strength of Si\,{\sc iv} BAL troughs  is also larger when 
 Al\,{\sc iii} BAL troughs are present.
The mean $\langle$EW$\rangle$ is $7.35 \pm 0.49~{\rm \AA}$ for Si\,{\sc iv} BAL troughs in 
the C\,{\sc iv}$_{\rm S0}$ sample and $20.84 \pm 1.02~{\rm \AA}$ for Si\,{\sc iv} BAL troughs in 
the C\,{\sc iv}$_{\rm SA}$ sample. The mean $\langle$EW$\rangle$ values change by a factor of $\approx 2.8$, 
which is even stronger than the corresponding change for C\,{\sc iv} of $\approx1.7$.

\subsection{C\,{\sc iv} BAL-Trough Velocities}\label{properties2}		
		
To assess  differences in  BAL-trough velocities, we investigate minimum velocity, $v_{\rm min}$,  
maximum velocity, $v_{\rm max}$,  and average centroid velocity from two-epoch observations, 
$\langle v_{\rm cent} \rangle = 1/2 (v_{\rm cent_1} + v_{\rm cent_2})$, for C\,{\sc iv}$_{\rm 00}$, C\,{\sc iv}$_{\rm S0}$, and 
C\,{\sc iv}$_{\rm SA}$ BAL troughs. {We present these distributions in  Figure~\ref{fig9}.}

\begin{figure*}[!p!]
\epsscale{0.75}
\plotone{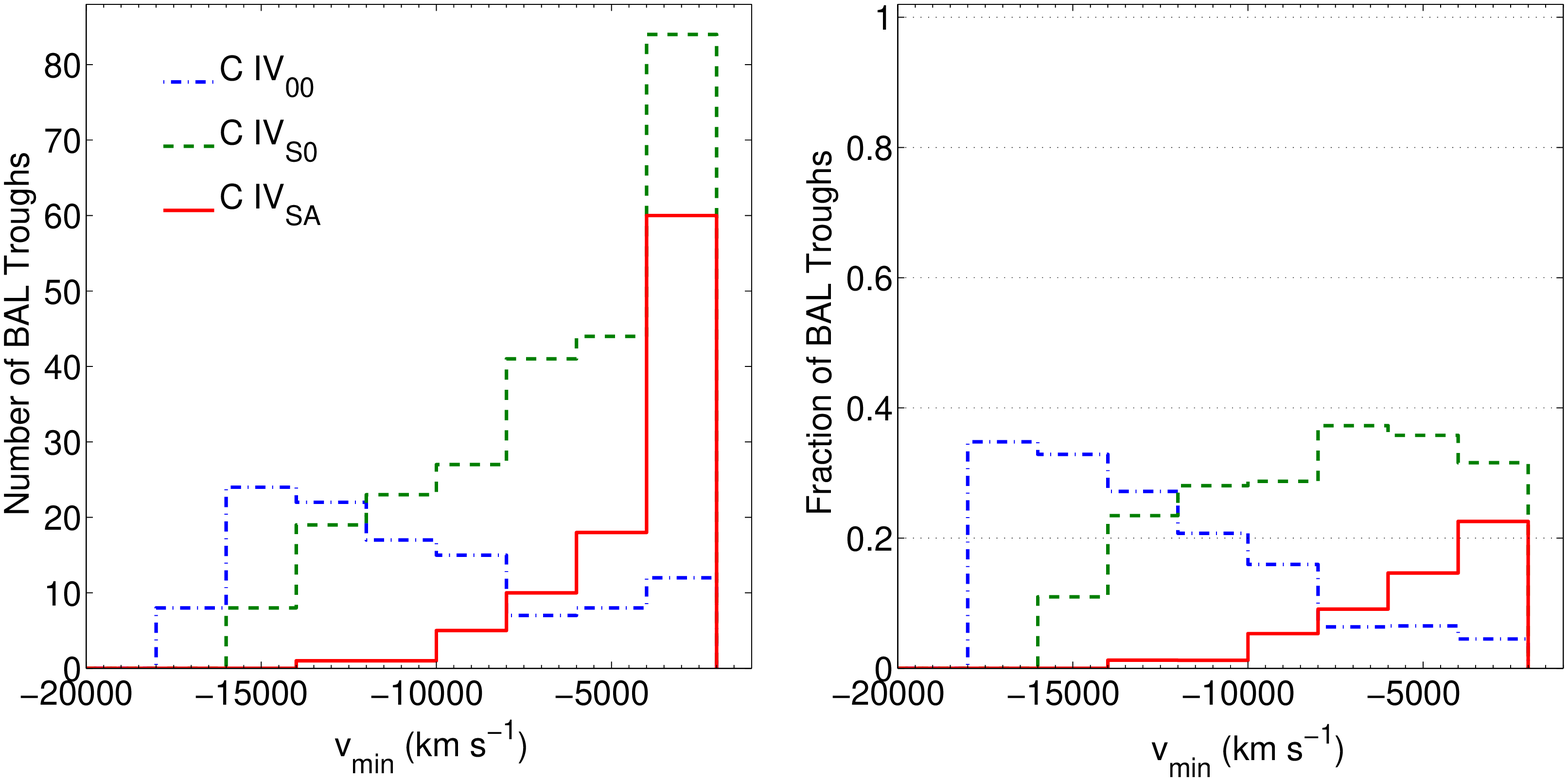} 
\plotone{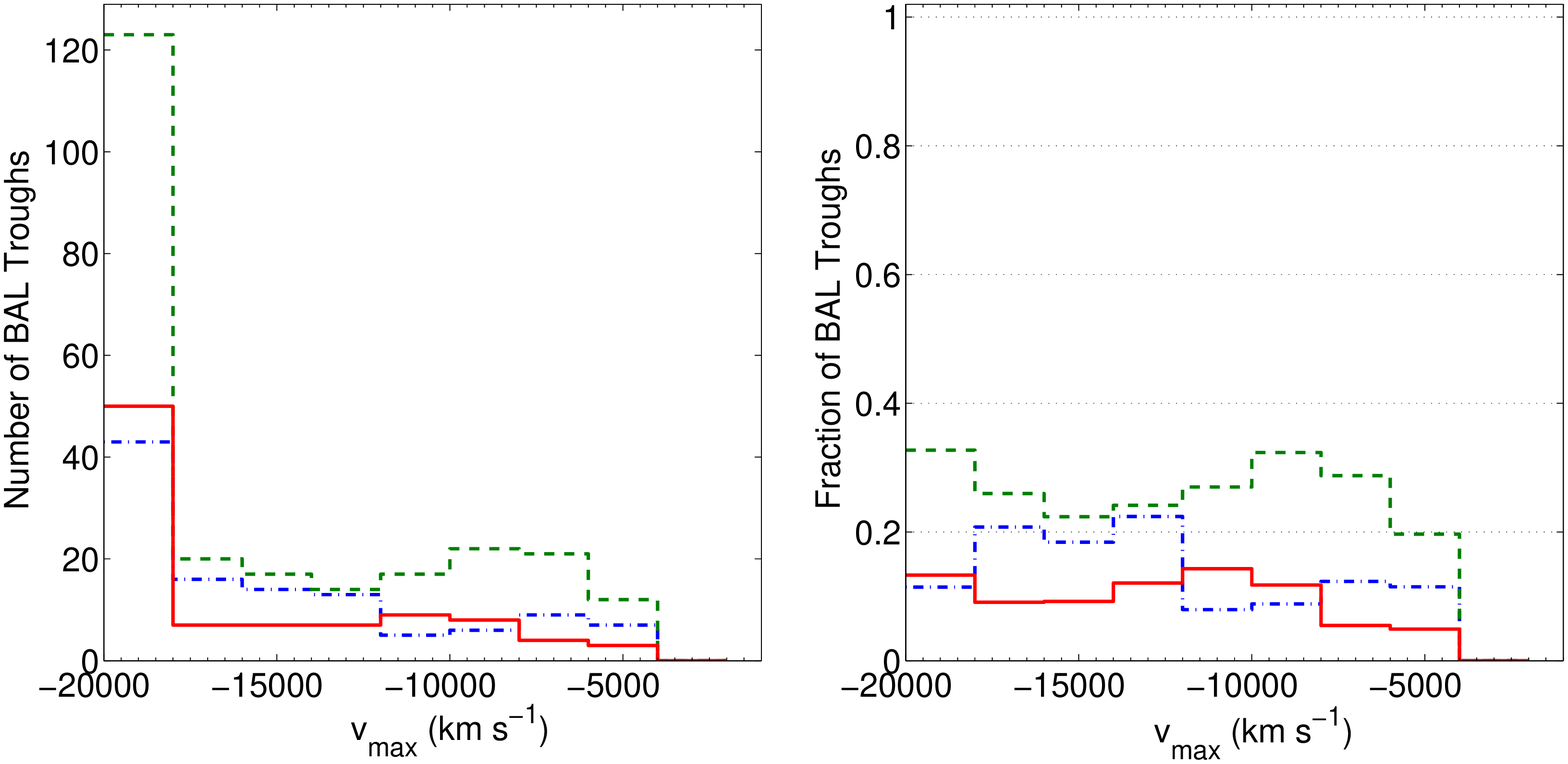} 
\plotone{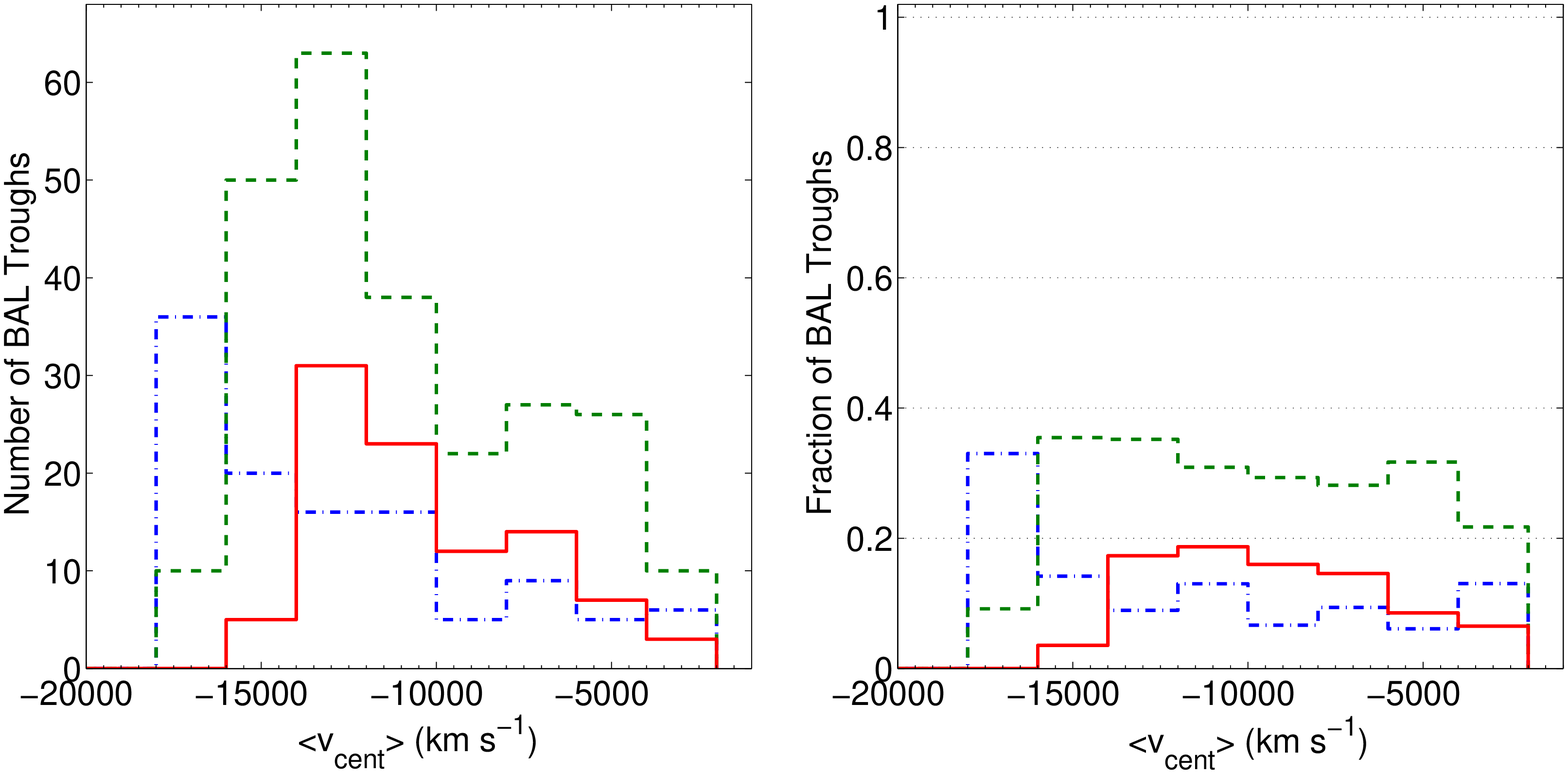} 
\caption{Minimum velocity, $v_{\rm min}$ (upper panels), maximum velocity, $v_{\rm max}$ (middle panels), 
and average centroid velocity, $\langle v_{\rm cent} \rangle$ (lower panels), 
distributions for C\,{\sc iv}$_{\rm 00}$ (dot-dashed blue),  C\,{\sc iv}$_{\rm S0}$ (dashed green), and C\,{\sc iv}$_{\rm SA}$ 
(solid red) BAL troughs. The right panels display the fraction of C\,{\sc iv}$_{\rm 00}$, C\,{\sc iv}$_{\rm S0}$, 
and C\,{\sc iv}$_{\rm SA}$  BAL troughs relative to all  C\,{\sc iv} BAL troughs in our main sample. 
The $v_{\rm min}$ and $\langle v_{\rm cent} \rangle$ distributions for C\,{\sc iv}$_{\rm 00}$, 
C\,{\sc iv}$_{\rm S0}$, and C\,{\sc iv}$_{\rm SA}$ BAL troughs are significantly (99.9\%) different from each other, 
whereas the $v_{\rm max}$ distributions do not show highly significant differences.}
\label{fig9}
\end{figure*}
		
 \begin{table*}[h!]
\caption{Average values of BAL-trough velocities for C\,{\sc iv}$_{\rm 00}$, 
C\,{\sc iv}$_{\rm S0}$, C\,{\sc iv}$_{\rm SA}$, and all C\,{\sc iv} BAL troughs}
\begin{center}
\begin{tabular}{c c c c c}
\hline \hline \\[-.3em]
& C\,{\sc iv}$_{\rm 00}$& C\,{\sc iv}$_{\rm S0}$&C\,{\sc iv}$_{\rm SA}$  & All C\,{\sc iv} \\[.6em] \hline \\[-0.3em]
$v_{\rm min}$  (km s$^{-1}$) & $-11898 \pm 400$ &  $-7434 \pm 238$ & $-4838 \pm 241$ &  $-8424 \pm 154$\\
${v_{\rm max}}$ (km s$^{-1}$)  &$-15836 \pm 426$ &  $-16035 \pm 308$ & $-16506 \pm 453$ &  $-15619 \pm 168$\\
$\langle v_{\rm cent}\rangle$ (km s$^{-1}$)  
&  $-$13830 $\pm$ 402 & $-$11998 $\pm$ 235 & $-$11405 $\pm$ 291  & $ -12229 \pm 140$\\[.5em]
\hline \\[-0.5em]
Number of Data Points & 113 & 246 & 95 & 852\\[.5em]
\hline
\end{tabular}\\
Uncertainties on the mean are calculated using the standard $\sigma/\sqrt{N}$ formula. 
Note that values \\ in this table are affected by censoring  (see the text).\\
\end{center}
\label{tab2}
\end{table*}%

 {Given our adopted BAL-trough definition, $v_{\rm min}$ and $v_{\rm max}$ values are 
 truncated at $-$3000 and $-$20,000~km~s$^{-1}$, respectively, for BAL troughs exceeding 
 these limits. It is apparent from Figure~\ref{fig9} that  this effect is present for the
$v_{\rm max}$ and less severely for the $v_{\rm min}$ distributions for C\,{\sc iv}$_{\rm 00}$, 
C\,{\sc iv}$_{\rm S0}$, and C\,{\sc iv}$_{\rm SA}$ BAL troughs. We thus compare $v_{\rm min}$ 
and $v_{\rm max}$ distributions using a two-sample Peto-Prentice test \citep[PP; e.g.,][]{latta81},
implemented in the Astronomy Survival Analysis  \citep[ASURV; e.g.,][]{lavalley92} package,
which considers such censoring effects in the samples. 
The PP test results show that all three $v_{\rm min}$ distributions are significantly  different 
(at a level of $>99.9\%$) from each other 
and all three $v_{\rm max}$ distributions do not show highly significant differences 
($P$ is 30.0\% for the C\,{\sc iv}$_{\rm 00}$ vs. C\,{\sc iv}$_{\rm S0}$ samples, 
35.2\% for the C\,{\sc iv}$_{\rm S0}$ vs. C\,{\sc iv}$_{\rm SA}$ samples, 
and 75.8\% for the C\,{\sc iv}$_{\rm 00}$ vs. C\,{\sc iv}$_{\rm SA}$ samples). 
We also compare  $v_{\rm min}$ and $v_{\rm max}$ distributions for the three C\,{\sc iv} 
groups and find similar results; the consistency of both the PP and AD tests suggests that 
the truncation of $v_{\rm min}$ and $v_{\rm max}$ does not have a significant effect on our 
main results. Still,  we focus our discussion on $v_{\rm min}$  as per its importance for our 
comparative assessments (see Section~\ref{hmlsummary}) rather than the more affected 
$v_{\rm max}$.}

{The mean $v_{\rm min}$,  $v_{\rm max}$,  and $\langle v_{\rm cent} \rangle$ values are 
given in Table~\ref{tab2}.} C\,{\sc iv}$_{\rm 00}$ BAL troughs tend to have 
higher onset velocities than  C\,{\sc iv}$_{\rm S0}$ and C\,{\sc iv}$_{\rm SA}$ troughs. 
The mean $v_{\rm min}$ values change by a factor of $\approx 2.5$ from C\,{\sc iv}$_{\rm 00}$ 
to C\,{\sc iv}$_{\rm SA}$ troughs.
Such differences between the $v_{\rm min}$ distributions for C\,{\sc iv}$_{\rm 00}$, 
C\,{\sc iv}$_{\rm S0}$, and C\,{\sc iv}$_{\rm SA}$ BAL troughs are perhaps expected given 
that  the average outflow velocity is higher for weak BAL troughs \citep[e.g.,][]{wey91}. 
{Unlike the $v_{\rm min}$ distributions, the $v_{\rm max}$ distributions do not show 
strong apparent  differences between the C\,{\sc iv} groups. }

AD test results show that the $\langle v_{\rm cent} \rangle$  distributions are significantly  different 
(at a level of $>99.9\%$) from each other.  C\,{\sc iv}$_{\rm SA}$ BAL troughs tend to be found 
at the lowest such  velocities, and C\,{\sc iv}$_{\rm 00}$ troughs tend to be found at the highest such velocities, 
although the mean $\langle v_{\rm cent} \rangle$ values differ by only $\approx25\%$.

\subsection{C\,{\sc iv} BAL-Variation Profiles} \label{vprofile}

We assess the overall differences between C\,{\sc iv} BAL-variation characteristics 
by comparing composite variation profiles for C\,{\sc iv}$_{\rm 00}$, C\,{\sc iv}$_{\rm S0}$, 
and C\,{\sc iv}$_{\rm SA}$ troughs. 
{
Since BAL-trough widths cover a considerable  range for each C\,{\sc iv} group 
(see Figure~\ref{fig8}), we  divide each group into three bins considering $\Delta v$ values. 
The bin limits are 
$\Delta v <3000$, $3000 < \Delta v <5000$, and $\Delta v>5000$~km~s$^{-1}$ for C\,{\sc iv}$_{\rm 00}$ troughs;
$\Delta v<8000$, $8000 <\Delta v <12,000$, and $\Delta v>12,000$~km~s$^{-1}$ for C\,{\sc iv}$_{\rm S0}$ troughs; and
$\Delta v<10,000$, $10,000 <\Delta v< 14,000$, and $\Delta v>14,000$~km~s$^{-1}$ for C\,{\sc iv}$_{\rm SA}$ troughs.}

{First, we calculate the absolute value of the depth variation between the SDSS and 
BOSS spectra for each data point of each trough as a function of 
outflow velocity, $v_{\rm t}$.  Second, we calculate the mean absolute depth variation 
for BAL troughs in each bin by interpolating the data  points for given $v_{\rm t}$ values. 
Figure~\ref{fig10} displays  the mean composites of depth-variation profiles for 
C\,{\sc iv}$_{\rm 00}$ troughs with $3000 < \Delta v < 5000$~km~s$^{-1}$, 
C\,{\sc iv}$_{\rm S0}$ troughs with $8000 < \Delta v < 12,000$~km~s$^{-1}$, 
and C\,{\sc iv}$_{\rm SA}$ troughs with $10000 < \Delta v < 14,000$~km~s$^{-1}$. }
We find qualitatively consistent results when computing median composites instead of mean 
composites.   Figure~\ref{fig10} also presents the mean composite for absolute fractional 
variations which is the variation of depth between SDSS and BOSS  divided by the average 
depth {as a function of $v_{\rm t}$.}

\begin{figure*}[t!]
\epsscale{1.1}
\plotone{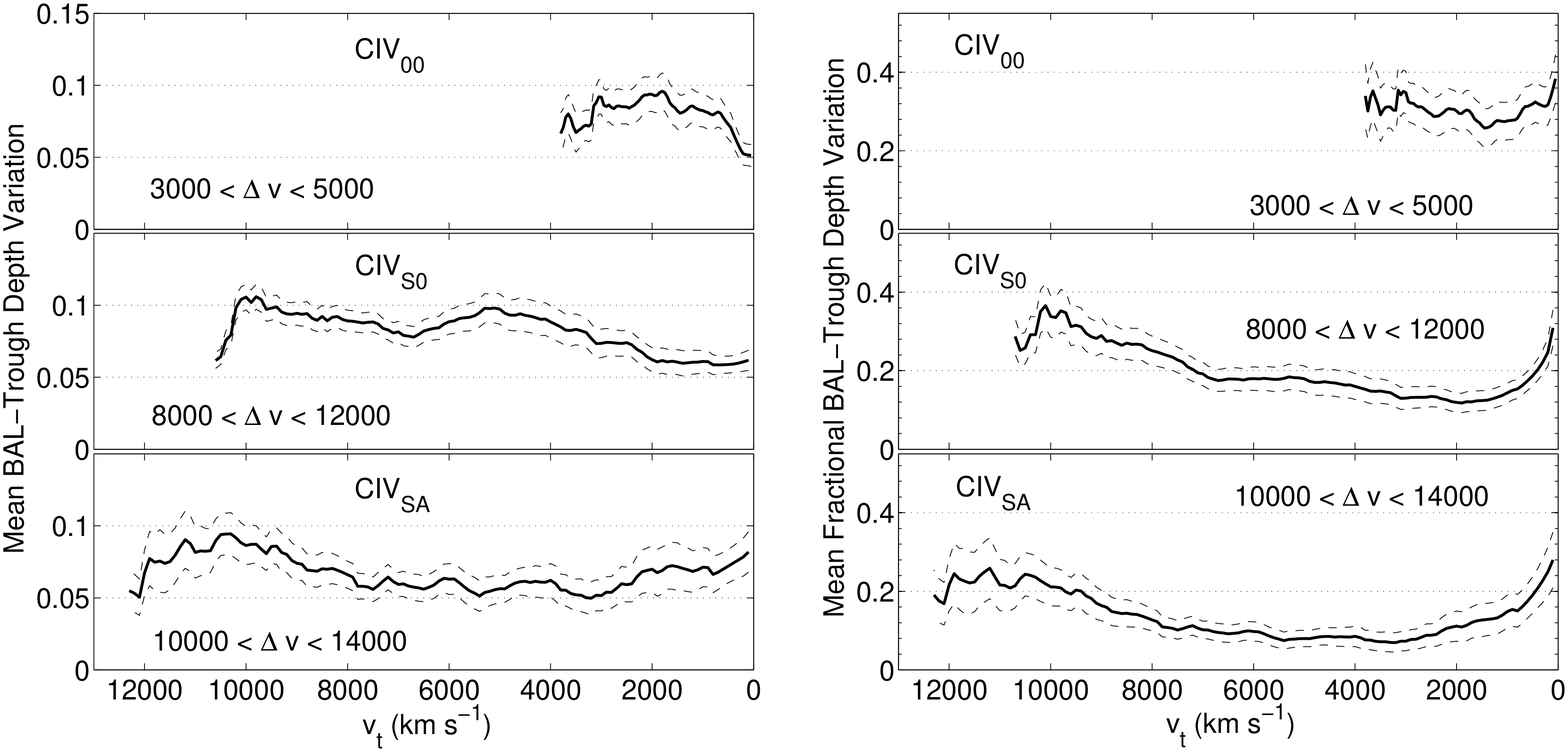} 
\caption{{BAL-trough variation profiles for C\,{\sc iv}$_{\rm 00}$ troughs with 
$3000 < \Delta v < 5000$ (top panels), C\,{\sc iv}$_{\rm S0}$ troughs with 
$8000 < \Delta v < 12000$ (middle panels), and C\,{\sc iv}$_{\rm SA}$ troughs with 
$10000 < \Delta v < 14000$ (bottom panels) in km~s$^{-1}$.} 
Left panels: Solid-black curves show the mean composite profiles for 
absolute depth variations of C\,{\sc iv} BAL troughs {as a function of outflow velocity, 
$v_{\rm t}$.} The dashed curves show the error on the mean. 
Right panels: Mean composite profiles for absolute fractional depth variations of 
C\,{\sc iv} BAL troughs,  where the fractional variation is the  depth variation 
divided by the average depth. While the C\,{\sc iv}$_{\rm 00}$ troughs show 
similar variability across the entire trough, the C\,{\sc iv}$_{\rm S0}$ and 
C\,{\sc iv}$_{\rm SA}$ troughs show less variability at lower velocities where 
the Si\,{\sc iv} and Al\,{\sc iii} absorption tend to be the strongest.}
\label{fig10}
\end{figure*}

{The composite variation profiles demonstrate that  the level of variation 
generally tends to be smaller for the lower velocity portions of C\,{\sc iv} troughs. 
Note that  in each panel of Figure~\ref{fig10}, the trough number statistics are limited for 
the highest velocity end. 
%
%
Figure~\ref{fig10} (left panels) shows that the lower velocity portions of 
C\,{\sc iv}$_{\rm S0}$ troughs (up to $\approx 30$\% of the trough width) 
tend to be less variable than the higher velocity portions, and these less 
variable portions usually correspond to regions where the Si\,{\sc iv} BAL 
troughs are strongest (see Figure~\ref{fig5}).
The variation profile for C\,{\sc iv}$_{\rm SA}$ troughs is strongly different from the 
variation profiles for C\,{\sc iv}$_{\rm 00}$ and C\,{\sc iv}$_{\rm S0}$ troughs. 
For C\,{\sc iv}$_{\rm SA}$ troughs,  the level of variability is notably low for up 
to $\approx 80$\% of the trough width; this result is highly statistically significant, 
considering the number of independent spectral data points and their error bars. 
Figure~\ref{fig5} demonstrates that these lower velocity 
portions of C\,{\sc iv}$_{\rm SA}$ troughs correspond to regions where 
Al\,{\sc iii} BAL troughs are generally found. }

{Composite fractional-variation profiles further highlight the differences
between the variation characteristics of C\,{\sc iv}$_{\rm 00}$, 
C\,{\sc iv}$_{\rm S0}$, and C\,{\sc iv}$_{\rm SA}$ BAL troughs. These 
arguably allow the best possible comparison of overall variability between 
these C\,{\sc iv} groups, since they account for the significantly different 
strengths of C\,{\sc iv}$_{\rm 00}$, C\,{\sc iv}$_{\rm S0}$,
and C\,{\sc iv}$_{\rm SA}$ troughs (see Section~\ref{properties}).
Overall, we find that C\,{\sc iv}$_{\rm 00}$ troughs tend to be significantly 
more fractionally variable than both C\,{\sc iv}$_{\rm S0}$ and 
C\,{\sc iv}$_{\rm SA}$ troughs, and likewise C\,{\sc iv}$_{\rm S0}$ troughs 
tend to be somewhat more fractionally variable than C\,{\sc iv}$_{\rm SA}$
troughs. Variability is again smallest in the lower velocity portions of
C\,{\sc iv}$_{\rm S0}$ and C\,{\sc iv}$_{\rm SA}$ troughs. 
However, this behavior is not apparent for C\,{\sc iv}$_{\rm 00}$ troughs.}

\subsection{C\,{\sc iv} BAL EW Variability} \label{EWvar}
		\subsubsection{ EW Variability as a Function of {\rm $\langle$EW$\rangle$} } \label{EWvar1}

In this section, we investigate the characteristics of BAL  EW variability assessing the differences between  
C\,{\sc iv}$_{\rm 00}$, C\,{\sc iv}$_{\rm S0}$, and C\,{\sc iv}$_{\rm SA}$ troughs. Figure~\ref{fig11} presents 
the  EW variation, $\Delta$EW, between the two-epoch spectra  as a function of average EW, 
$\langle$EW$\rangle$. For comparison, the  standard-deviation curves for these three 
C\,{\sc iv} groups are also given; the standard-deviation curves are calculated using  a sliding window where 
each window contains  10\% of the total number of BAL troughs found in a given group 
(i.e., 11 for C\,{\sc iv}$_{\rm 00}$, 25 for C\,{\sc iv}$_{\rm S0}$, and 10 for C\,{\sc iv}$_{\rm SA}$). 
We statistically remove the mean EW error in each window from the standard deviation, so the dispersion shown in 
Figure~\ref{fig11} refers to the intrinsic dispersion. 

Consistent with Section~4.5 of \citet{ak13}, we find that the standard deviation of $\Delta$EW generally increases 
with increasing $\langle$EW$\rangle$ for all C\,{\sc iv} groups. Given that  such a trend exists for all 
C\,{\sc iv} groups, a proper intergroup comparison requires consideration of  EW variation  for troughs 
with similar $\langle$EW$\rangle$ values.
The C\,{\sc iv}$_{\rm S0}$ $\Delta$EW spread appears less than that for C\,{\sc iv}$_{\rm 00}$ BALs with similar 
$\langle$EW$\rangle$ values (i.e., \hbox{$7<\langle$EW$\rangle <10~{\rm \AA}$}).
Similarly, the C\,{\sc iv}$_{\rm SA}$ $\Delta$EW spread appears less than that for C\,{\sc iv}$_{\rm S0}$ BALs 
for $12<\langle$EW$\rangle <35~{\rm \AA}$. These results suggest that C\,{\sc iv}  EW variation has  a 
dependence upon the existence of absorption lines from Si\,{\sc iv} and Al\,{\sc iii} transitions.

\begin{figure*}[!t!]
\epsscale{0.8}
\plotone{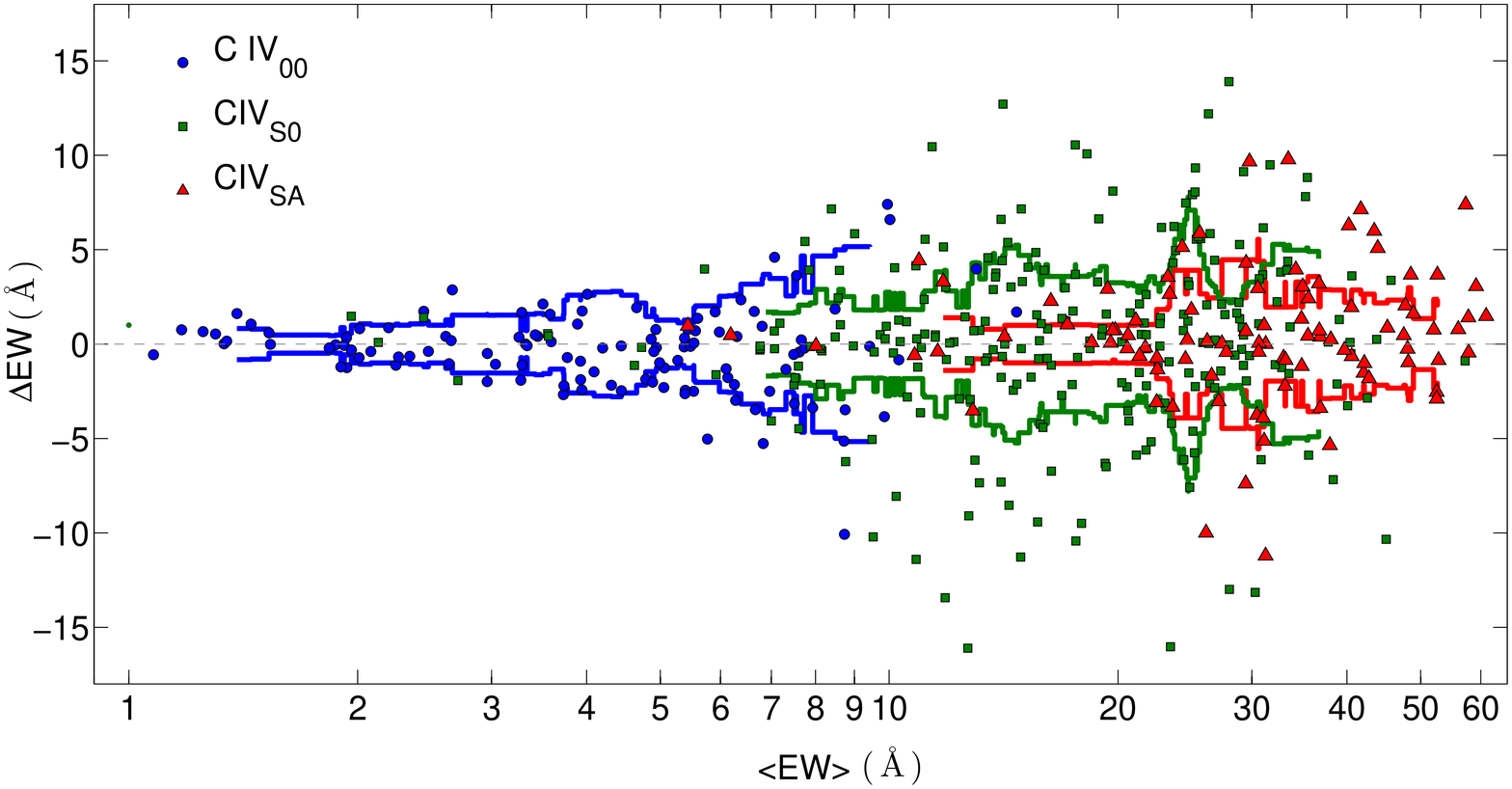} 
\caption{$\Delta$EW vs. $\langle$EW$\rangle$ for  C\,{\sc iv}$_{\rm 00}$ (blue circles), 
C\,{\sc iv}$_{\rm S0}$ (green squares), and C\,{\sc iv}$_{\rm SA}$ (red triangles) BAL troughs. 
Solid curves show rms values calculated with a sliding window containing 10\% of the total data points. 
The spread of $\Delta$EW  generally increases with increasing $\langle$EW$\rangle$ for  
C\,{\sc iv}$_{\rm 00}$, C\,{\sc iv}$_{\rm S0}$, and C\,{\sc iv}$_{\rm SA}$ troughs. In overlapping 
$\langle{\rm EW} \rangle$ ranges, C\,{\sc iv}$_{\rm 00}$ troughs tend to be more variable than  
C\,{\sc iv}$_{\rm S0}$ troughs, and similarly C\,{\sc iv}$_{\rm S0}$ troughs tend to be  more variable 
than  C\,{\sc iv}$_{\rm SA}$ troughs. }
\label{fig11}
\end{figure*}

In Figure~\ref{fig12}, we show the fractional EW variation, $\Delta$EW$/\langle{\rm EW} \rangle$, as a function 
of $\langle$EW$\rangle$ for C\,{\sc iv}$_{\rm 00}$, C\,{\sc iv}$_{\rm S0}$, and C\,{\sc iv}$_{\rm SA}$ troughs. 
Similarly to Figure~\ref{fig11}, solid curves represent the sliding-window standard deviations. 
{The overall trend in $\Delta$EW$/\langle{\rm EW} \rangle$  vs. $\langle$EW$\rangle$ 
appears mainly due to the increase of $\langle$EW$\rangle$.}
The curves indicate that the spread of $\Delta$EW$/\langle{\rm EW} \rangle$ for  C\,{\sc iv}$_{\rm 00}$ 
is larger than that for C\,{\sc iv}$_{\rm S0}$, and the $\Delta$EW$/\langle{\rm EW} \rangle$ spread for 
C\,{\sc iv}$_{\rm S0}$ is larger than that for C\,{\sc iv}$_{\rm SA}$ in matched $\langle{\rm EW} \rangle$ ranges.

Previous BAL-variability studies \citep{gibson08,gibson10,cap11,ak13} have demonstrated  that BAL EW variability 
increases with increasing timescale. However, the trends in Figures~\ref{fig11} and \ref{fig12} are not caused or 
amplified by this effect; the testing in Section~\ref{ident} showed that the timescale distributions  for 
C\,{\sc iv}$_{\rm 00}$, C\,{\sc iv}$_{\rm S0}$, and C\,{\sc iv}$_{\rm SA}$ troughs do not have any significant 
differences from each other (see Figure~\ref{fig4}).  

Considering  that the $\langle$EW$\rangle$ distributions are significantly different between the three 
C\,{\sc iv} groups (see Figure~\ref{fig7}) and that  there are strong correlations between $\Delta$EW or \linebreak
$|\Delta$EW$/\langle{\rm EW} \rangle|$  and $\langle$EW$\rangle$ for all  C\,{\sc iv} troughs 
\citep[also see Section~4.5 of] []{ak13}, we investigate the EW variation distributions 
of these three groups using   matched samples of troughs with similar EWs.
		
\begin{figure*}[!t!]
\epsscale{0.8}
\plotone{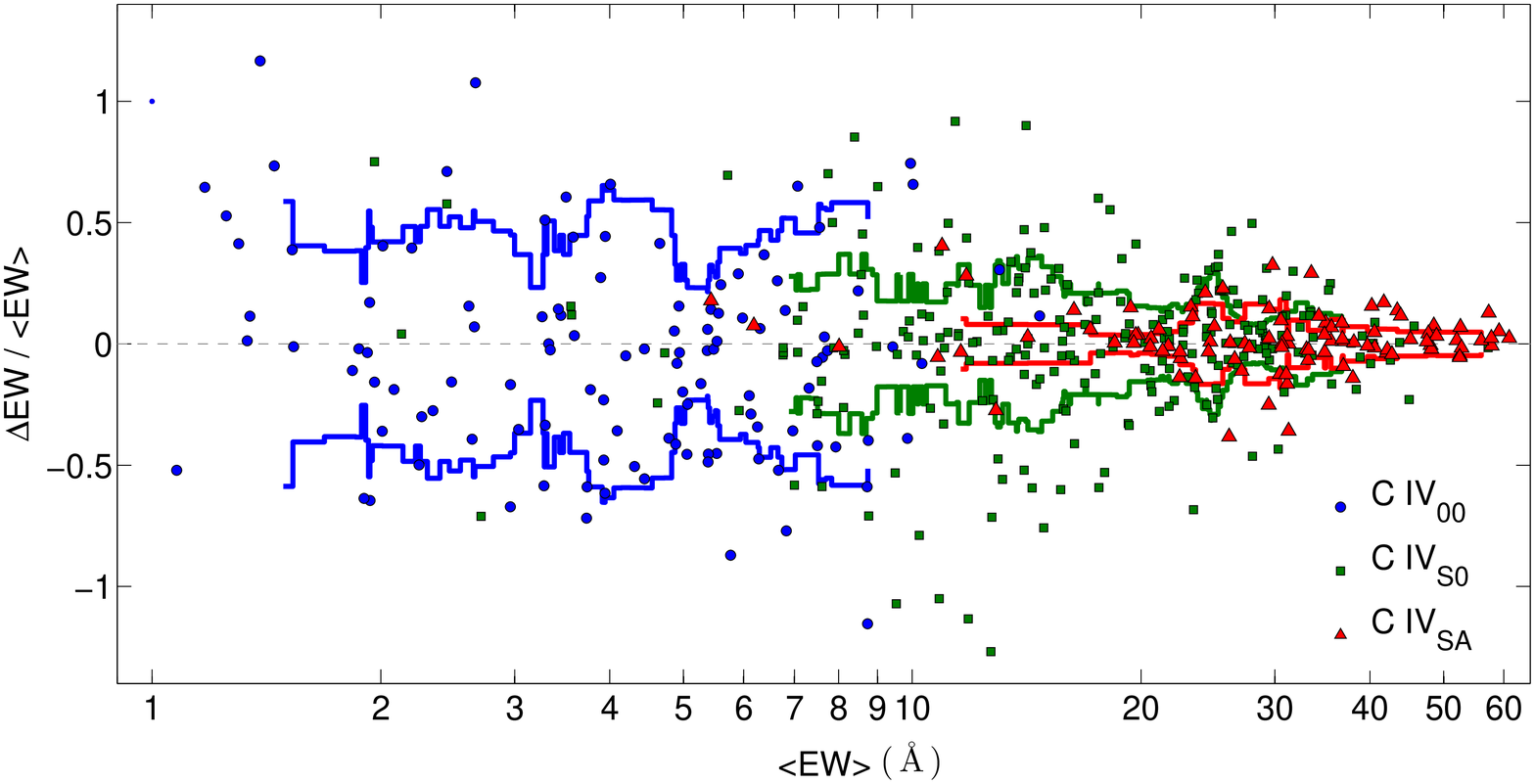} 
\caption{Same as Figure~\ref{fig11} but for $\Delta$EW$/\langle{\rm EW} \rangle$. The spread of 
$\Delta$EW$/\langle{\rm EW} \rangle$ decreases with increasing $\langle{\rm EW} \rangle$. In overlapping 
$\langle{\rm EW} \rangle$ ranges, C\,{\sc iv}$_{\rm 00}$ troughs tend to be fractionally more variable than  
C\,{\sc iv}$_{\rm S0}$ troughs,  and similarly C\,{\sc iv}$_{\rm S0}$ troughs tend to be fractionally more variable 
than  C\,{\sc iv}$_{\rm SA}$ troughs. }
\label{fig12}
\end{figure*}

\subsubsection{BAL-Trough Samples with Matching EWs}  \label{EWvar2}

We investigated the EW-variation characteristics of C\,{\sc iv}$_{\rm 00}$ vs.\,C\,{\sc iv}$_{\rm S0}$  and 
C\,{\sc iv}$_{\rm S0}$ vs.\,C\,{\sc iv}$_{\rm SA}$ troughs using matched samples. 
These matched samples contain the same number of BAL troughs from both groups matched by 
first-epoch EW, EW$_1$. Requiring  similar EW$_1$ for each trough pair allows comparison of  
variation behaviors for C\,{\sc iv} groups having similar  initial conditions. 
We first  randomly select a C\,{\sc iv}$_{\rm 00}$ trough for each C\,{\sc iv}$_{\rm S0}$ 
trough such that their EWs are in  agreement to within $1\sigma$. Due to significant differences in 
the EW distributions, we cannot match all data points of a group. In the matching procedure,  we sample 
any given BAL trough only once. Second,  we applied a two-sample AD test to compare the $\Delta$EW 
distributions. We repeated the matching  and AD testing 1000 times  and found the median 
results given in Table~\ref{tab3}. 

Comparing the $\Delta$EW distributions of C\,{\sc iv}$_{\rm 00}$ and C\,{\sc iv}$_{\rm S0}$ BAL troughs 
with matched EWs (where each distribution has 40 troughs), we found that the two distributions  significantly 
($P > 99.9\%$) differ from each other. Similarly, we repeat the matching  for C\,{\sc iv}$_{\rm S0}$ and 
C\,{\sc iv}$_{\rm SA}$ troughs. The test results for the comparison between the $\Delta$EW distributions of
C\,{\sc iv}$_{\rm S0}$ and C\,{\sc iv}$_{\rm SA}$  troughs with similar EWs (where each  has 65 troughs) 
show that the two distributions  differ  with a significance of $97.8 \%$. Figure~\ref{fig13} presents the 
$\Delta$EW distributions for C\,{\sc iv}$_{\rm 00}$ and C\,{\sc iv}$_{\rm S0}$  BAL troughs with matched 
EWs  in the left panel, and  C\,{\sc iv}$_{\rm S0}$ and C\,{\sc iv}$_{\rm SA}$ BAL troughs with matched  
EWs in the right panel. 

As given in Table~\ref{tab3},  the mean of $|\Delta$EW$|$  for C\,{\sc iv}$_{\rm 00}$ troughs is 
$\approx 1.3$ times larger than that for matched C\,{\sc iv}$_{\rm S0}$ troughs.
Likewise, the mean of $|\Delta$EW$|$  for C\,{\sc iv}$_{\rm S0}$ troughs is $\approx 1.8$ times
larger than that for matched C\,{\sc iv}$_{\rm SA}$ troughs.
The standard deviation of  the $\Delta$EW  distribution for C\,{\sc iv}$_{\rm 00}$ troughs is 
$\approx 1.1$ times larger than that for matched C\,{\sc iv}$_{\rm S0}$ troughs,  and 
for C\,{\sc iv}$_{\rm S0}$ troughs is $\approx 1.7$ times larger than that for matched 
C\,{\sc iv}$_{\rm SA}$ troughs.

\begin{table*}[t]
\caption{Two-Sample Anderson-Darling Test Results for BAL-Trough Samples with Matching First-Epoch 
EWs\tablenotemark{a}}
\begin{center}
\begin{tabular}{crrcrr}
\hline \hline
& C\,{\sc iv}$_{\rm 00}$  & C\,{\sc iv}$_{\rm S0}$&  & C\,{\sc iv}$_{\rm S0}$ & C\,{\sc iv}$_{\rm SA}$ \\ \hline
Mean $\Delta {\rm EW}$ & $-1.13 \pm 0.47$ & $1.33 \pm 0.49$ & & $-0.70 \pm 0.64$ & $0.75 \pm 0.38$ \\
Mean $|\Delta {\rm EW}|$ & $2.34 \pm 0.34$ & $1.84 \pm 0.45$ & & $3.78 \pm 0.44$ & $2.11 \pm 0.29$ \\
$\sigma _{\Delta {\rm EW}}$ & $ 3.42 \pm 0.34$ & $3.14\pm 0.35$ && $5.16 \pm 0.45$ & $3.09 \pm 0.27$ \\
$p_{\rm AD}$ & \multicolumn{2}{c}{$>99.9\%$ } & & \multicolumn{2}{c}{$97.8\%$ } \\
Sample Size &  \multicolumn{2}{c}{40 } & &  \multicolumn{2}{c}{65 }\\
\hline
\end{tabular}
\\[1em]
{$^{\rm a}$Entries are the median values calculated for 1000 random draws. 
Uncertainties on the mean and standard deviation are calculated using the standard 
$\sigma/\sqrt{N}$ formula and Equation~14.1.3 of \citet{press}, respectively. }
\end{center}
\label{tab3}
\end{table*}%

\begin{figure}[h!]
\epsscale{1.2}
\plotone{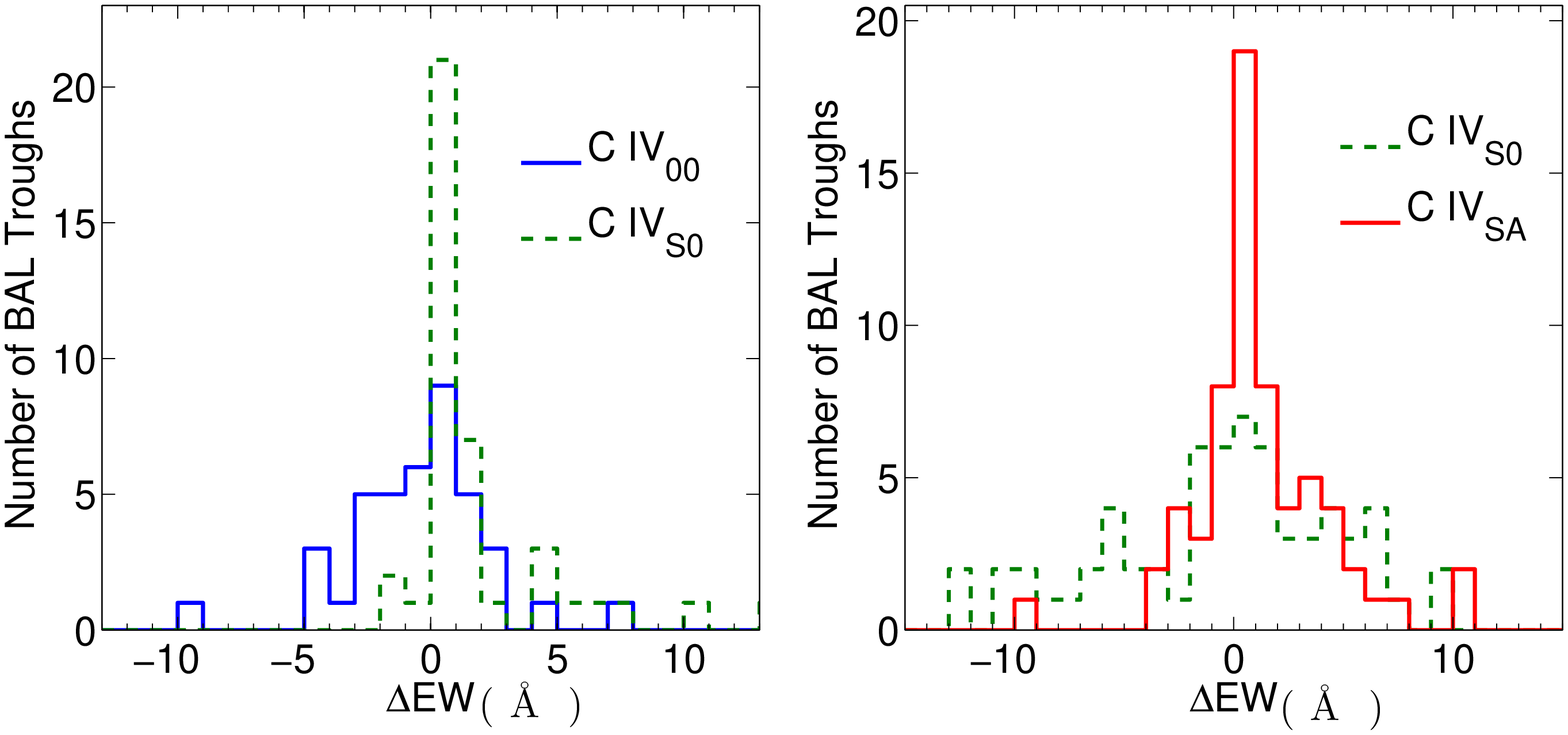} 
\caption{$\Delta$EW distributions for C\,{\sc iv}$_{\rm 00}$  vs. C\,{\sc iv}$_{\rm S0}$ troughs (left panel) 
and for C\,{\sc iv}$_{\rm S0}$ vs. C\,{\sc iv}$_{\rm SA}$ troughs (right panel), where the samples show  
BAL troughs with similar EWs. C\,{\sc iv}$_{\rm 00}$ troughs are more variable than C\,{\sc iv}$_{\rm S0}$ troughs, 
while C\,{\sc iv}$_{\rm S0}$ troughs are  in turn  more variable than C\,{\sc iv}$_{\rm SA}$ troughs.}
\label{fig13}
\end{figure}

Although the matched C\,{\sc iv}$_{\rm S0}$ and C\,{\sc iv}$_{\rm SA}$ BAL troughs have both a larger 
sample size (65) and difference between the $\sigma _{\Delta {\rm EW}}$ values (\hbox{$5.16-3.09 = 2.07$}) 
than the matched C\,{\sc iv}$_{\rm 00}$ 
and C\,{\sc iv}$_{\rm S0}$ troughs, the AD-test results indicate a  more 
significant overall difference for the $\Delta$EW distributions of C\,{\sc iv}$_{\rm 00}$ and C\,{\sc iv}$_{\rm S0}$ 
troughs. This result is mainly due to a more significant difference between the mean $\Delta$EW values of 
matching  C\,{\sc iv}$_{\rm 00}$ and C\,{\sc iv}$_{\rm S0}$ troughs ($3.6\sigma$) than that for matching  
C\,{\sc iv}$_{\rm S0}$ and C\,{\sc iv}$_{\rm SA}$ troughs ($1.9\sigma$). 
Another notable result of these comparisons is that the means of the $\Delta$EW distributions for both 
the C\,{\sc iv}$_{\rm 00}$ and C\,{\sc iv}$_{\rm S0}$ troughs with similar EWs differ from zero at more 
than $2\sigma$: \hbox{$-1.13 \pm 0.47~{\rm \AA}$}   and \hbox{$1.33 \pm 0.49~{\rm \AA}$}, respectively.
Moreover, the means of the $\Delta$EW distributions for C\,{\sc iv}$_{\rm S0}$ and 
C\,{\sc iv}$_{\rm SA}$ troughs with similar EWs also differ from zero at more than $1\sigma$. 
The means of the $\Delta$EW distributions for C\,{\sc iv}$_{\rm S0}$ in the  two different matching samples 
change from a positive value to a negative value. 
These results are initially surprising given the fact that the mean of the $\Delta$EW distribution 
for all 852 C\,{\sc iv} BAL troughs in our main sample is consistent with zero (\hbox{$-0.0055 \pm 0.0373~{\rm \AA}$}; 
also see Section~4.4 of \citeauthor{ak13} 2013). 
Indeed, the weakening and strengthening of BAL troughs are expected to be balanced  to maintain an equilibrium 
population of  BAL troughs in quasar spectra.

The  differences between the mean values of the $\Delta$EW distributions can occur, for 
instance, if not only $|\Delta$EW$|$ but also $\Delta$EW has a  dependence on BAL-trough strength. 
To assess such dependency, we show  $\Delta$EW as a function of  first-epoch EW for  
C\,{\sc iv}$_{\rm 00}$,  C\,{\sc iv}$_{\rm S0}$, and C\,{\sc iv}$_{\rm SA}$ troughs in Figure~\ref{fig14}. 
The best fits  of basic linear-regression models are plotted in each panel of Figure~\ref{fig14} to demonstrate
the apparent trends. 
Using Spearman rank-correlation tests \citep[see][]{press}, we find connections between  $\Delta$EW and 
${\rm EW}_1$  for C\,{\sc iv}$_{\rm 00}$ ($P>99.9\%$) and likely also for C\,{\sc iv}$_{\rm S0}$ ($P=98.4\%$) troughs. 

\begin{figure}[h!]
\epsscale{1.18}
\plotone{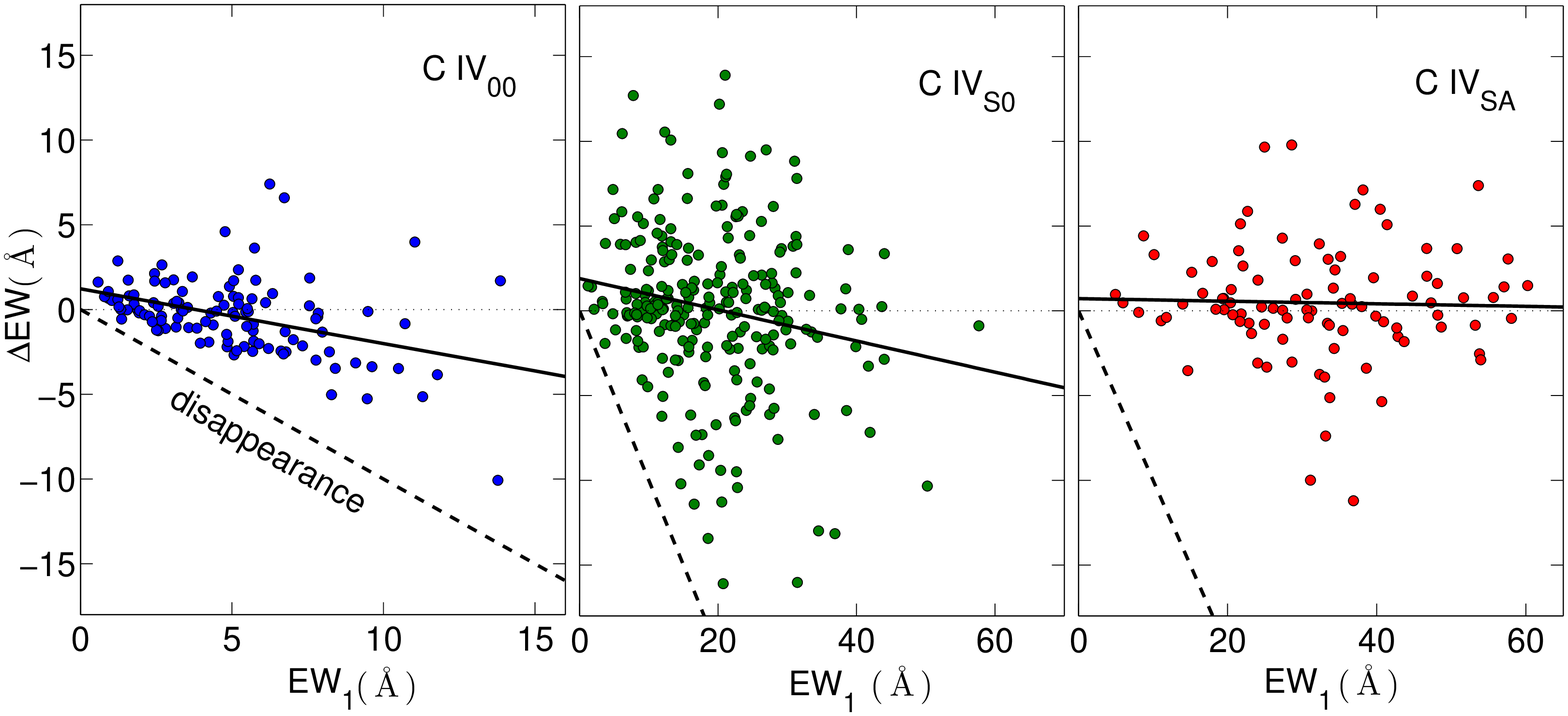} 
\caption{$\Delta$EW as a function of  ${\rm EW}_1$ for C\,{\sc iv}$_{\rm 00}$ (left),  C\,{\sc iv}$_{\rm S0}$ (middle), 
and   C\,{\sc iv}$_{\rm SA}$ (right) BAL troughs.
The solid lines show the best fits  of basic linear-regression models in each panel. 
The dashed lines denote where   $\Delta$EW $ = - {\rm EW}_1$ (corresponding to BAL disappearance).
The apparent connection between $\Delta$EW and ${\rm EW}_1$ arises primarily because a BAL trough cannot 
weaken by more than its first-epoch EW. }
\label{fig14}
\end{figure}

The apparent $\Delta$EW vs.  ${\rm EW}_1$ relation primarily arises  because the BAL-trough EW 
variation range is limited. Given that a BAL trough cannot weaken by more than its first-epoch EW 
and cannot strengthen by more than its second-epoch EW, ${\rm EW}_2$, the variation strength is 
restricted to  be ${\rm EW_2} \geq \Delta$EW $ \geq - {\rm EW}_1$. 
In addition, we exclude quasars that do not satisfy our necessary quasar-selection criterion of  
\hbox{BI$_{3}^{20} > 0$} in either epoch (see Section~\ref{colobs}). 
Thus, this criterion leads to exclusion of  disappearing (i.e., \hbox{$\Delta$EW $ \approx - {\rm EW}_1$}; 
e.g., \citeauthor{ak12} 2012) and 
emerging (i.e., \hbox{$\Delta$EW $ \approx {\rm EW}_2$}) BAL troughs from our main sample.  

In our matched samples, we must compare the $\Delta$EW distribution of the  stronger  C\,{\sc iv}$_{\rm 00}$ 
troughs with that of the weaker C\,{\sc iv}$_{\rm S0}$ troughs, and that of the  stronger  C\,{\sc iv}$_{\rm S0}$ troughs with 
that of the weaker C\,{\sc iv}$_{\rm SA}$ troughs. Therefore the differences in the means of these $\Delta$EW distributions 
are   expected given the relations shown in Figure~\ref{fig14}.

\subsection{EW Variation Correlations}  \label{correlation}

Previous studies \citep[e.g.,][]{gibson10,cap12,ak13} have demonstrated  that  Si\,{\sc iv} BAL troughs  tend to vary in 
concert with C\,{\sc iv}  troughs at corresponding velocities and that the strength of fractional EW variations 
tends to be larger for Si\,{\sc iv} BAL troughs. In order to investigate similar  relations for C\,{\sc iv}$_{\rm S0}$ 
(Section~\ref{correlation1}) and C\,{\sc iv}$_{\rm SA}$  (Section~\ref{correlation2}) troughs, we assess  variation 
correlations between C\,{\sc iv}  troughs and corresponding Si\,{\sc iv} and Al\,{\sc iii} troughs. 

\subsubsection{ EW Variation Correlations for the {\rm C\,{\sc iv}$_{\rm S0}$} Sample}\label{correlation1}

Figure~\ref{fig16} presents the $\Delta$EW and $\Delta$EW$/\langle$EW$\rangle$ correlations  for 
C\,{\sc iv}$_{\rm S0}$ and the corresponding Si\,{\sc iv} troughs. Consistent with the findings of \citet{ak13} for all 
C\,{\sc iv} troughs, the Spearman-test results show that both the $\Delta$EW and $\Delta$EW$/\langle$EW$\rangle$ 
correlations are highly significant ($P > 99.9\%$).  Given the apparent  
scatter in both panels of Figure~\ref{fig16}, we determine the best fits using a Bayesian linear-regression model that 
considers the intrinsic scatter of the sample \citep{kelly07}. We found the following relations, calculating the mean and the 
standard deviation of the linear-regression model parameters for 10{,}000 random draws from the sample:
\begin{eqnarray}
\label{ehml2}
\Delta\rm{EW}_{\rm{C\,IV_{S0}}} = (1.20 \pm 0.082) \times \Delta\rm{EW}_{\rm{Si\,IV}}   \\  \nonumber 
+ (0.051 \pm 0.216),
~~~ \sigma_{\rm IS} = 3.4~{\rm \AA}
\end{eqnarray}
\begin{eqnarray}
\label{ehml3}
\frac{\Delta\rm{EW}}{\langle\rm{EW}\rangle}_{\rm{C\,IV_{S0}}} = (0.515 \pm 0.023) \times 
\frac{\Delta\rm{EW}}{\langle\rm{EW}\rangle}_{\rm{Si\,IV}}   \\  \nonumber  + (0.006 \pm 0.011),
~~~ \sigma_{\rm IS} = 0.15.
\end{eqnarray}
Here $\sigma_{\rm IS}$ is the standard deviation of the intrinsic scatter. 
The slopes of the $\Delta$EW and $\Delta$EW$/\langle$EW$\rangle$ correlations are consistent with the findings 
in Section~4.6 of \citet{ak13} within $1\sigma$, indicating that the EW variations of C\,{\sc iv}$_{\rm S0}$ troughs are  
comparable to those  of Si\,{\sc iv} troughs, whereas the fractional EW variations of C\,{\sc iv}$_{\rm S0}$ 
troughs are approximately  half of those of Si\,{\sc iv} troughs. 
{The EW variations for C\,{\sc iv}$_{\rm S0}$ vs. corresponding Si\,{\sc iv}  troughs 
are  not highly inconsistent with being equal; the deviation between the best fit and equality 
is $\approx 2.4\sigma$ and the intrinsic scatter is large.}

\begin{figure*}[t!]
\epsscale{0.75}
\plotone{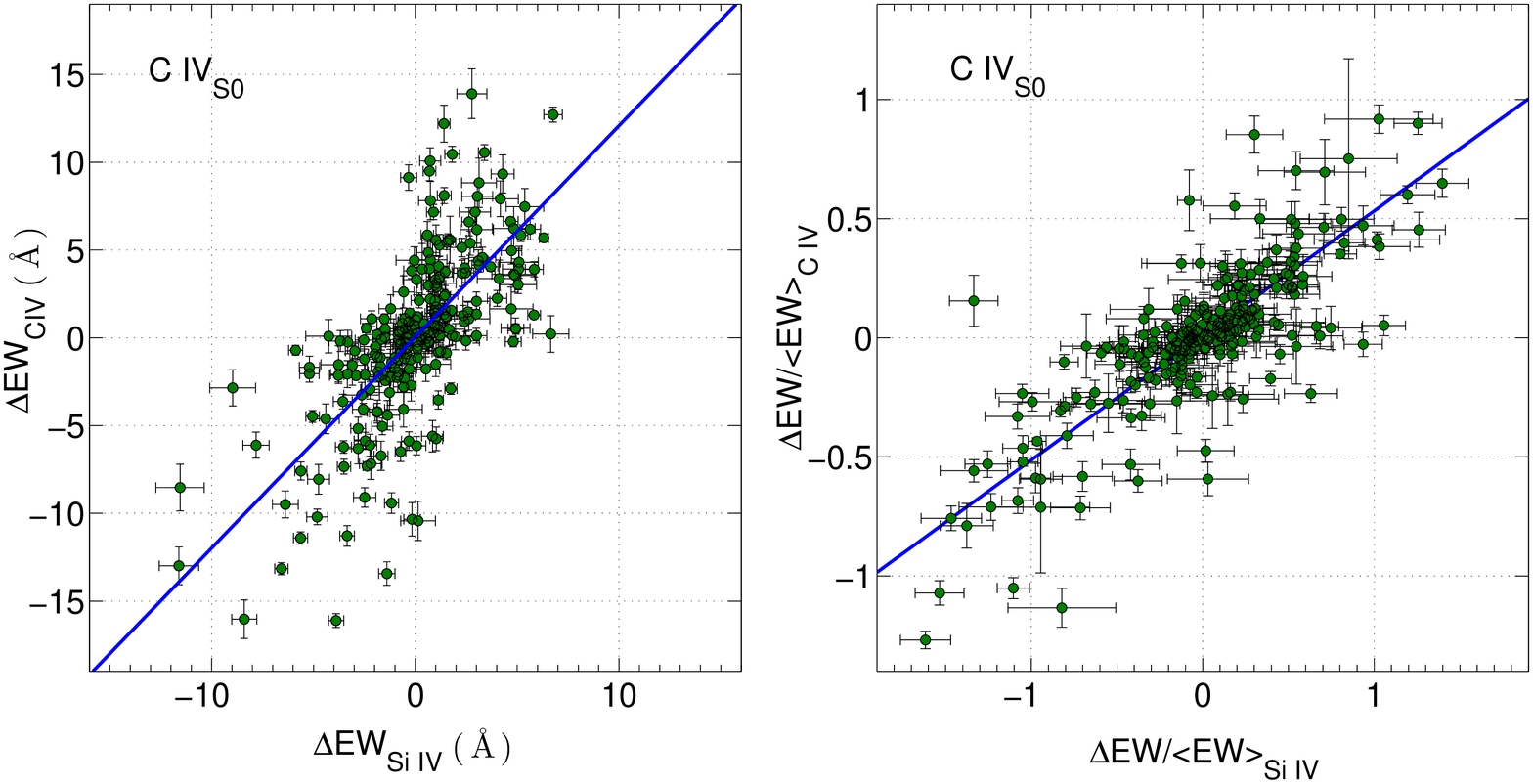} 
\caption{The $\Delta$EW (left panel)  and  $\Delta$EW$/\langle$EW$\rangle$ (right panel) correlations between  
C\,{\sc iv}$_{\rm S0}$ BAL troughs and the corresponding  Si\,{\sc iv}  BAL troughs. Spearman-test results show highly
significant correlations ($P > 99.9\%$) for both panels. The solid-blue lines show the best fit found using a Bayesian 
linear-regression model.}
\label{fig16}
\end{figure*}

\subsubsection{ EW Variation Correlations for the {\rm C\,{\sc iv}$_{\rm SA}$} Sample}\label{correlation2}

Figure~\ref{fig18} displays the $\Delta$EW and $\Delta$EW$/\langle$EW$\rangle$ relations for 
C\,{\sc iv}$_{\rm SA}$ BAL troughs and the corresponding Si\,{\sc iv} and Al\,{\sc iii} BAL troughs. 
The Spearman-test results indicate  highly significant ($P > 99.9\%$) 
correlations for both  $\Delta$EW and $\Delta$EW$/\langle$EW$\rangle$  between C\,{\sc iv}$_{\rm SA}$ and the 
corresponding Si\,{\sc iv} troughs. The $\Delta$EW correlation between C\,{\sc iv}$_{\rm SA}$ and the 
corresponding Al\,{\sc iii} troughs is marginally significant  ($P = 99.8\%$),  while the $\Delta$EW$/\langle$EW$\rangle$
correlation is highly significant  ($P > 99.9\%$). 
The $\Delta$EW correlation between C\,{\sc iv}$_{\rm SA}$  and Al\,{\sc iii} troughs is not  highly significant, possibly 
because of the small range of  $\Delta$EW values for both ions. In addition to this effect, the relatively small number 
of data points in the C\,{\sc iv}$_{\rm SA}$ sample, and the apparently  large intrinsic scatter of the  $\Delta$EW values, 
may hide any possible correlation.
The following relations are found  using the Bayesian linear-regression fit:
\begin{eqnarray}\label{ehml6}
\Delta\rm{EW}_{\rm{C\,IV_{SA}}} = (0.646 \pm 0.081) \times \Delta\rm{EW}_{\rm{Si\,IV}}   \\  \nonumber  + (0.080 \pm 0.280),
~~~ \sigma_{\rm IS} = 2.5~{\rm \AA}
\end{eqnarray}
\begin{eqnarray}\label{ehml7}
\frac{\Delta\rm{EW}}{\langle\rm{EW}\rangle}_{\rm{C\,IV_{SA}}} = (0.355 \pm 0.044) \times 
\frac{\Delta\rm{EW}}{\langle\rm{EW}\rangle}_{\rm{Si\,IV}}   \\  \nonumber  - (0.002 \pm 0.008), 
~~~ \sigma_{\rm IS} = 0.08
\end{eqnarray}
\begin{eqnarray}\label{ehml8}
\Delta\rm{EW}_{\rm{C\,IV_{SA}}} = (0.172 \pm 0.171) \times \Delta\rm{EW}_{\rm{Al\,III}}  \\  \nonumber + (0.493 \pm 0.348),
~~~ \sigma_{\rm IS} = 3.3~{\rm \AA}
\end{eqnarray}
\begin{eqnarray}\label{ehml9}
\frac{\Delta\rm{EW}}{\langle\rm{EW}\rangle}_{\rm{C\,IV_{SA}}} = (0.111 \pm 0.035) \times 
\frac{\Delta\rm{EW}}{\langle\rm{EW}\rangle}_{\rm{Al\,III}}   \\  \nonumber  + (0.005 \pm 0.011), ~~~ \sigma_{\rm IS} = 0.10.
\end{eqnarray}

\begin{figure*}[t!]
\epsscale{0.75}
\plotone{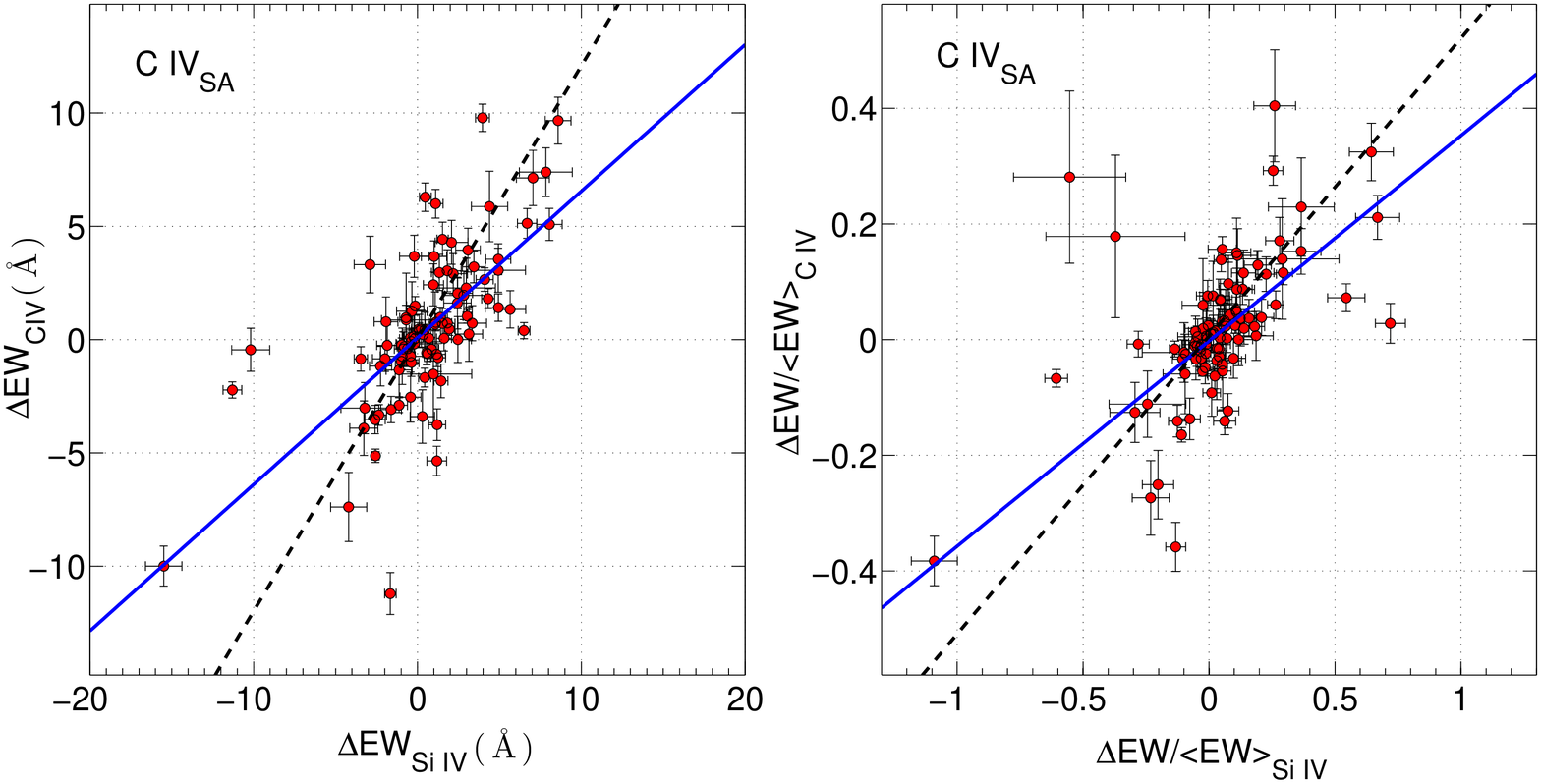} 
\plotone{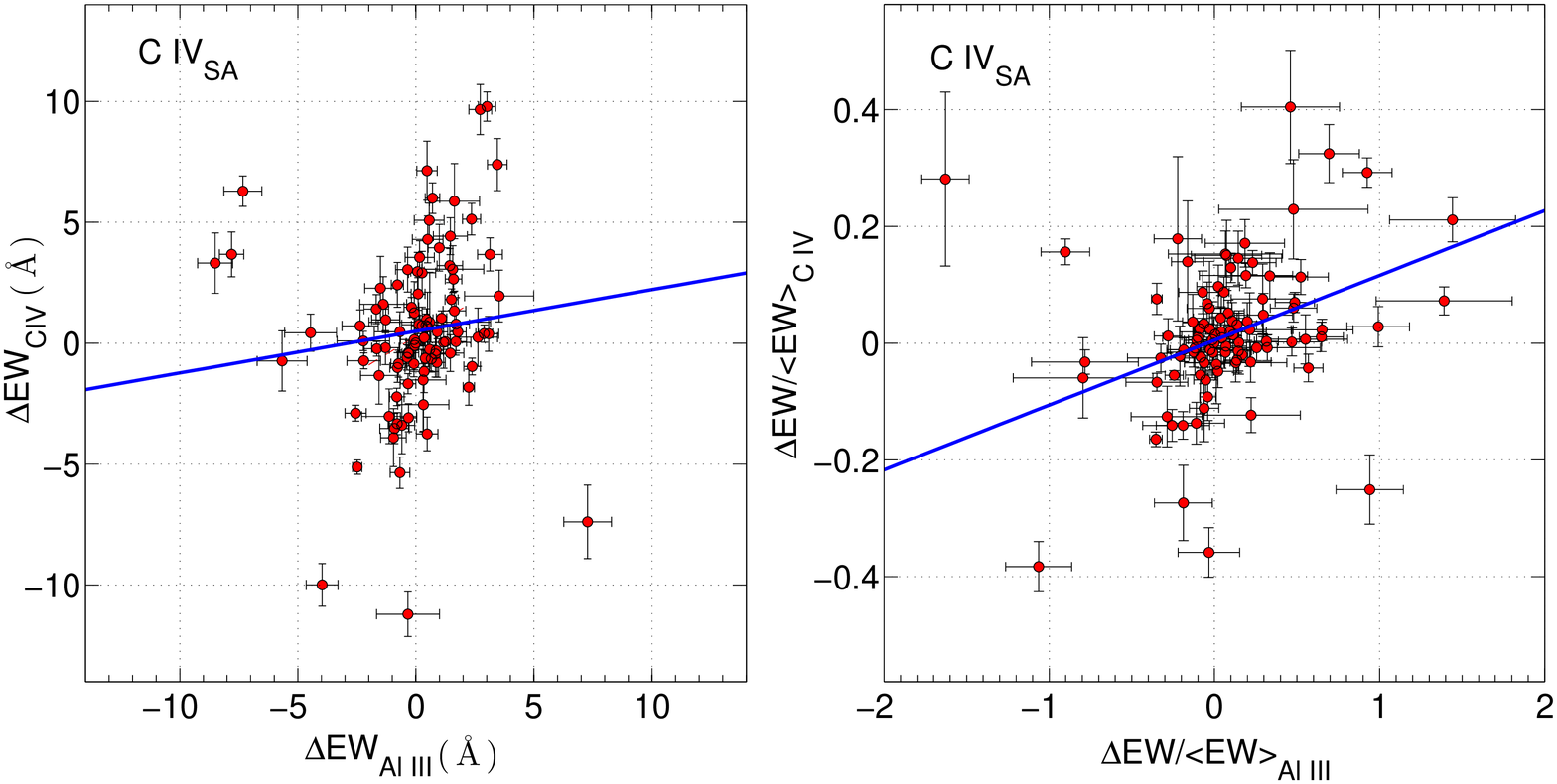} 
\caption{The $\Delta$EW (left panels)  and  $\Delta$EW$/\langle$EW$\rangle$ (right panels) correlations for 
C\,{\sc iv}$_{\rm SA}$ vs. Si\,{\sc iv} troughs (top panels) and C\,{\sc iv}$_{\rm SA}$  vs. Al\,{\sc iii}  troughs 
(bottom panels). Spearman-test results show highly significant  (both with $P > 99.9\%$) $\Delta$EW and 
$\Delta$EW$/\langle$EW$\rangle$ correlations for  C\,{\sc iv}$_{\rm SA}$ vs. Si\,{\sc iv} troughs, and a significant 
$\Delta$EW$/\langle$EW$\rangle$ correlation for C\,{\sc iv}$_{\rm SA}$ vs. Al\,{\sc iii} troughs, whereas the
$\Delta$EW correlation for C\,{\sc iv}$_{\rm SA}$ vs. Al\,{\sc iii} troughs is marginally significant with $P = 99.8\%$. 
The solid-blue lines show the best-fit relations in each panel. The dashed-black lines in the  top two panels 
show the best-fit models for C\,{\sc iv}$_{\rm S0}$ BAL troughs (Equations~\ref{ehml2} 
and~\ref{ehml3}) for comparison.}
\label{fig18}
\end{figure*}

Figure~\ref{fig18} presents the best fits of the C\,{\sc iv} and Si\,{\sc iv} variation relations both for the C\,{\sc iv}$_{\rm S0}$ 
and the C\,{\sc iv}$_{\rm SA}$ samples for comparison purposes. Consistent with  the results of our investigations in 
Section~\ref{EWvar}, the ranges of  $\Delta\rm{EW}_{\rm{C\,IV}}$ and 
${\Delta\rm{EW}}{/\langle\rm{EW}\rangle}_{\rm{C\,IV}}$ in Figures~\ref{fig16} and \ref{fig18} 
indicate that C\,{\sc iv}$_{\rm SA}$ troughs tend to be less variable than C\,{\sc iv}$_{\rm S0}$ troughs. 
Moreover, the ranges of $\Delta\rm{EW}_{\rm{Si\,IV}}$ and ${\Delta\rm{EW}}{/\langle\rm{EW}\rangle}_{\rm{Si\,IV}}$ 
 in Figures~\ref{fig16} and \ref{fig18} suggest  that the Si\,{\sc iv} BAL troughs of  the C\,{\sc iv}$_{\rm SA}$ 
sample show less variation than those of the  C\,{\sc iv}$_{\rm S0}$ sample.
The flat slopes of Equations~\ref{ehml6} and \ref{ehml8} indicate that the C\,{\sc iv}$_{\rm SA}$ troughs tend to show very 
small variations. 
The slopes of Equations~\ref{ehml7} and \ref{ehml9} indicate that Al\,{\sc iii} troughs tend to be more 
fractionally variable than both the corresponding C\,{\sc iv}$_{\rm SA}$ and Si\,{\sc iv} troughs, and Si\,{\sc iv} troughs 
tend to be more fractionally variable than the corresponding C\,{\sc iv}$_{\rm SA}$ troughs. 

\section{Summary of Results, Discussion,  and Future Work}\label{hmlsummary}

We have investigated the profiles, standard characteristic properties, and variation behaviors of C\,{\sc iv} BAL troughs,  
considering how these change when BAL troughs from Si\,{\sc iv} and Al\,{\sc iii} are present at 
corresponding velocities. We have utilized a sample of 852 C\,{\sc iv} BAL troughs; 
113 of these have no detection of any corresponding Si\,{\sc iv} or Al\,{\sc iii} BALs or mini-BALs in both epochs 
(C\,{\sc iv}$_{\rm 00}$ troughs), 
246 of these are accompanied by a Si\,{\sc iv} BAL trough but have no corresponding 
Al\,{\sc iii} BALs or mini-BALs (C\,{\sc iv}$_{\rm S0}$ troughs), and 
95 of these are accompanied by both Si\,{\sc iv} and Al\,{\sc iii} BALs (C\,{\sc iv}$_{\rm SA}$ troughs). 
The main observational findings of our study are the following:

\begin{enumerate}

\item \label{obf1}
The composite profiles of C\,{\sc iv}$_{\rm 00}$, C\,{\sc iv}$_{\rm S0}$, and C\,{\sc iv}$_{\rm SA}$ 
troughs differ significantly; stronger C\,{\sc iv} troughs are found when accompanying BAL troughs from 
lower ionization transitions are present. Furthermore, the composite profiles for C\,{\sc iv}$_{\rm S0}$ 
and C\,{\sc iv}$_{\rm SA}$ troughs are deeper at the lower velocities where Al\,{\sc iii} and, to a lesser
extent, Si\,{\sc iv} troughs are preferentially found. See Section~\ref{profile}.

\item \label{obf2}
The two-epoch average EW, $\langle$EW$\rangle$, distributions for C\,{\sc iv}$_{\rm 00}$, 
C\,{\sc iv}$_{\rm S0}$, and C\,{\sc iv}$_{\rm SA}$ BAL troughs are significantly ($>99.9\%$) different. 
Generally, C\,{\sc iv}$_{\rm 00}$ troughs have small EWs, C\,{\sc iv}$_{\rm S0}$ troughs have moderate 
EWs, and C\,{\sc iv}$_{\rm SA}$ troughs have large EWs. We find that increases in both depth and 
velocity width contribute comparably to the increase in EW from C\,{\sc iv}$_{\rm 00}$ to 
C\,{\sc iv}$_{\rm S0}$ to C\,{\sc iv}$_{\rm SA}$ troughs. See Section~\ref{properties}.

\item \label{obf3}
The minimum and average centroid velocities decrease from C\,{\sc iv}$_{\rm 00}$ to  
C\,{\sc iv}$_{\rm S0}$ to C\,{\sc iv}$_{\rm SA}$ troughs; this decrease is most notable 
for the minimum velocity, which changes on average by a factor of $\approx 2.5$, while 
the decrease is mild ($\approx 25\%$) for the average centroid velocity. 
See Section~\ref{properties2}.

\item \label{obf4}
Composite depth-variation and fractional-depth-variation profiles have been used to investigate the
relative variability of C\,{\sc iv}$_{\rm 00}$, C\,{\sc iv}$_{\rm S0}$, and C\,{\sc iv}$_{\rm SA}$ 
BAL troughs. BAL variability generally decreases from C\,{\sc iv}$_{\rm 00}$ to C\,{\sc iv}$_{\rm S0}$ 
to C\,{\sc iv}$_{\rm SA}$ BALs, particularly in a fractional sense. For C\,{\sc iv}$_{\rm S0}$ 
and C\,{\sc iv}$_{\rm SA}$ troughs, the lower velocity portions of the troughs tend to be the 
least variable, and these are the regions where Al\,{\sc iii} and, to a lesser extent, Si\,{\sc iv} 
troughs are preferentially found. See Section~\ref{vprofile}.

\item \label{obf5}
The spread of $\Delta$EW generally increases with increasing $\langle {\rm EW} \rangle$, and the spread of 
$\Delta$EW$/ \langle {\rm EW} \rangle$ generally decreases with increasing $\langle {\rm EW} \rangle$,
for C\,{\sc iv}$_{\rm 00}$, C\,{\sc iv}$_{\rm S0}$, and C\,{\sc iv}$_{\rm SA}$ BAL troughs; 
this result is consistent with the general behavior of all C\,{\sc iv} BAL troughs \citep[e.g.,][]{ak13}. 
{In overlapping ranges of $\langle {\rm EW} \rangle$, C\,{\sc iv}$_{\rm 00}$ troughs appear to 
vary more strongly than C\,{\sc iv}$_{\rm S0}$ troughs ($P > 99.9\%$), and similarly C\,{\sc iv}$_{\rm S0}$ troughs
appear to vary more strongly than C\,{\sc iv}$_{\rm SA}$ troughs ($P=97.8\%$)}. See Section~\ref{EWvar1}.

\item \label{obf6}
For a proper comparison of the variation characteristics of the three C\,{\sc iv} groups, we compare 
$\Delta$EW distributions of samples of BAL troughs with matched first-epoch EWs.  
C\,{\sc iv}$_{\rm S0}$ troughs 
are somewhat less variable than C\,{\sc iv}$_{\rm 00}$ troughs with matched EWs, and 
C\,{\sc iv}$_{\rm SA}$ troughs are substantially less variable than 
C\,{\sc iv}$_{\rm S0}$ troughs with matched EWs. See Section~\ref{EWvar2}.

\item \label{obf7}
The Si\,{\sc iv} BAL troughs associated with the C\,{\sc iv}$_{\rm S0}$  sample show EW 
and fractional EW variations that are generally in concert with those of the corresponding 
C\,{\sc iv}$_{\rm S0}$ troughs.  We quantify the relevant correlations and find that Si\,{\sc iv} troughs
show similar EW and larger fractional EW variations than corresponding C\,{\sc iv}$_{\rm S0}$ troughs. 
See Section~\ref{correlation1}.

\item \label{obf8}
The EW and fractional EW variations of Al\,{\sc iii} troughs are less clearly linked with those of 
C\,{\sc iv}$_{\rm SA}$ troughs, although correlation testing does indicate  some correspondence. 
Al\,{\sc iii} troughs show larger EW and fractional EW variations than  corresponding C\,{\sc iv}$_{\rm SA}$ and 
Si\,{\sc iv} troughs.  See Section~\ref{correlation2}.

\end{enumerate}

We now examine the implications of our observational findings considering the 
best-developed model for quasar BAL outflows, that of an equatorial radiation-driven 
disk wind \citep[e.g.,][see also Section~\ref{hmlintro}]{murray95,proga00,higginbottom}. 
While other scenarios for BAL outflows also exist, such as those proposing
that BALs primarily are formed at large distances \hbox{(0.1--10 kpc)}
from the SMBH (e.g., \citealt{arav13}; also see \citealt{faucher12}; but see Section 
5.3 of \citealt{lucy14} for a critique),
these have not been developed via numerical simulations to the point
where robust comparisons with our observational findings are possible.
Of course, any model for BAL winds, current or future, can be usefully
constrained by our observational results presented above.

Figure~\ref{fig20} shows density and poloidal velocity maps of the disk-wind model. In this figure, 
two lines-of-sight are marked, corresponding to different viewing inclinations, along which an 
observer would see C\,{\sc iv}$_{\rm 00}$ and  C\,{\sc iv}$_{\rm SA}$ 
troughs; in parallel with our subscript notation for C\,{\sc iv} troughs, we will refer to these 
lines-of-sight (LOS) as LOS$_{\rm 00}$ and LOS$_{\rm SA}$, respectively, throughout. 
We expect generally correlated changes of ionization level, kinematics, and column density as our 
line-of-sight is varied from LOS$_{\rm 00}$ to LOS$_{\rm SA}$ (see Section~\ref{hmlintro}), 
and this is relevant to our discussion below.\footnote{We appreciate that such correlated 
changes will have scatter owing to, e.g., time-dependent phenomena  leading to local 
inhomogeneities (see Figure~3 of \citealt{proga00}). Overcoming such scatter is a prime 
driver for our utilization of large samples in this study.}
We recognize that there is strong observational evidence supporting intrinsic object-to-object 
differences as well as inclination effects among BAL quasars 
\citep[e.g.,][]{boroson92,turnshek94,zhang10,dipompeo12}.
For example, the probability of observing a C\,{\sc iv}$_{\rm SA}$ trough will be higher for 
objects having a larger global covering factor of low-ionization gas (cf., the weak [O\,{\sc iii}] 
objects discussed in \citealt{boroson92}).\footnote{The evidence for high global covering factors 
of low-ionization gas has been most notably presented for BAL quasars with detected Mg\,{\sc ii} 
absorption \citep[e.g.,][]{boroson92,turnshek94,zhang10}. Although 
we primarily use Al\,{\sc iii} absorption to identify lines-of-sight with low-ionization gas,
$\approx 75\%$ of the objects in our C\,{\sc iv}$_{\rm SA}$ sample show Mg\,{\sc ii} 
absorption at corresponding velocities (see Section~\ref{colobs}).}
However, such global-covering-factor effects do not affect our main reasoning below which 
is focused upon the typical measured properties of C\,{\sc iv}$_{\rm 00}$ vs. 
C\,{\sc iv}$_{\rm S0}$ vs. C\,{\sc iv}$_{\rm SA}$ troughs rather than how often each of these 
trough types is observed.

\begin{figure*}[p!] 
   \centering
   \includegraphics[trim=3.1cm 0.2cm 0.2cm 0.5cm, clip=true,scale=0.5,angle=90]{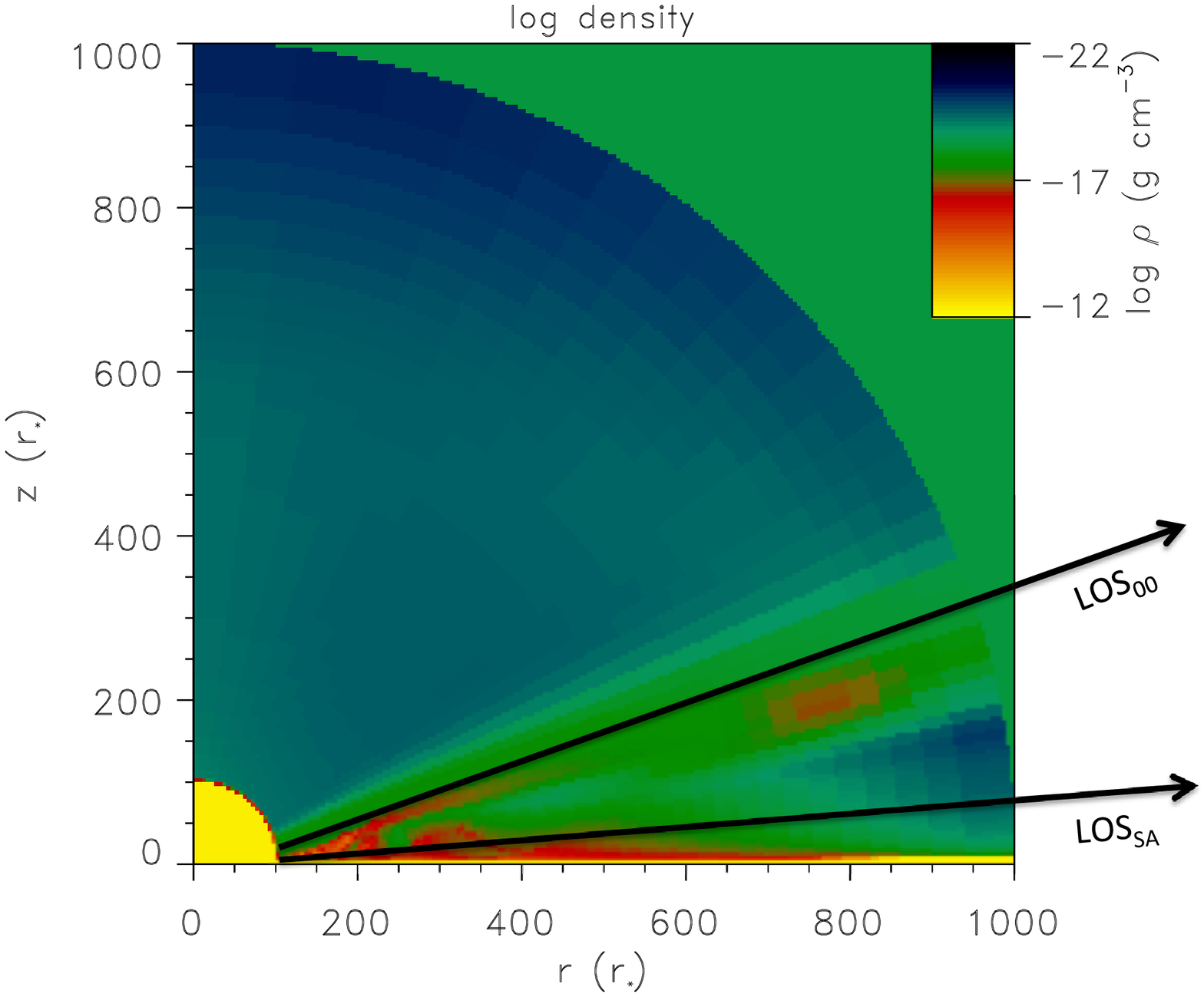} 
      \includegraphics[trim=3.1cm 0.2cm 0.2cm 0.5cm, clip=true,scale=0.5, angle=90 ]{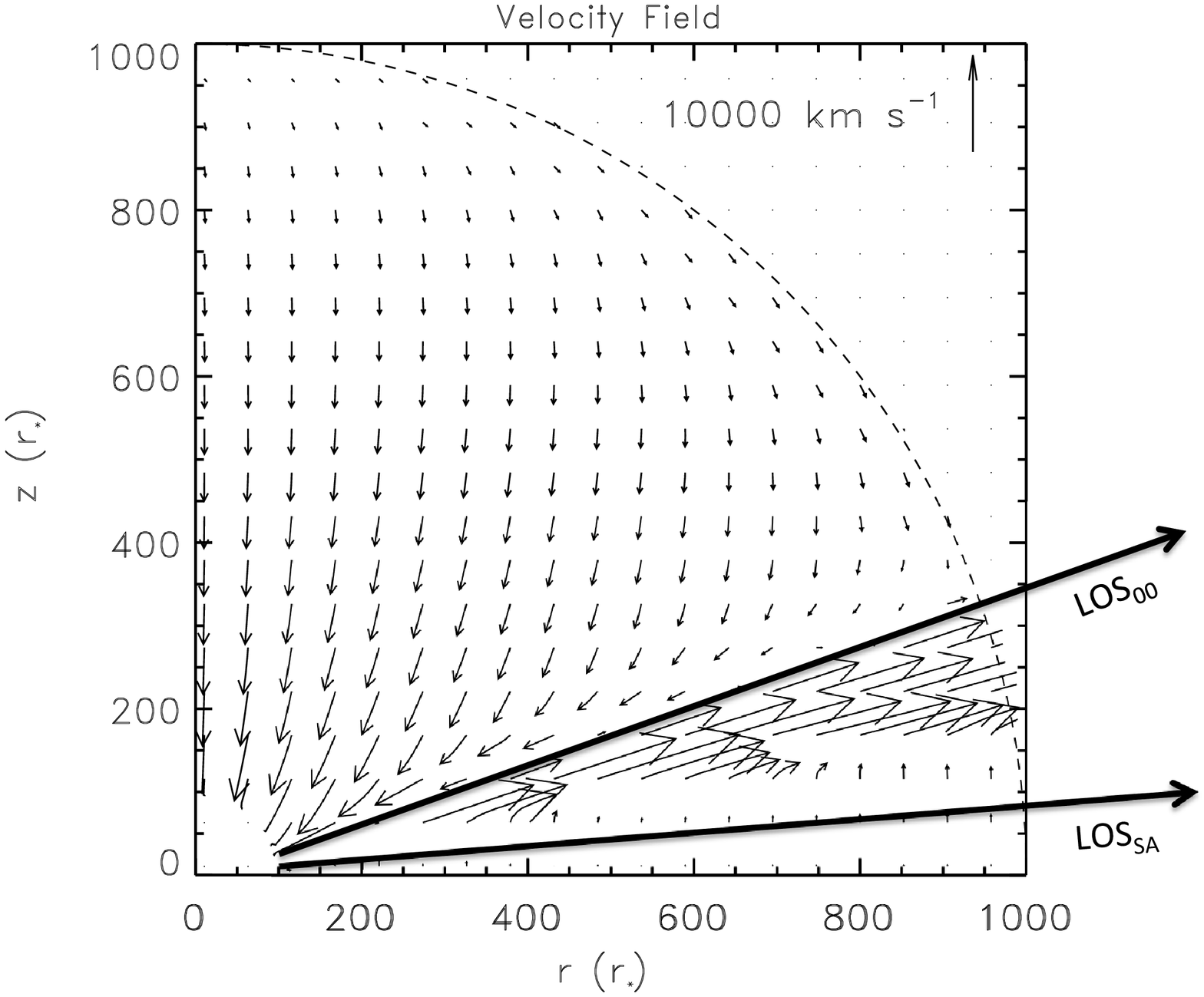} 
      \caption{Density  (top) and poloidal velocity (bottom) maps of the disk-wind model
      \citep[adapted from Figure~2 of][]{proga00}. The SMBH is located at (0, 0). 
      The $x$ and $y$-axes show the distance  from the SMBH in units of the radius at the  inner 
      edge of the disk, where $r_{*} = 3 \times r_{S}$ and 
     $r_{S}$ is the Schwarzschild radius of a black hole. 
     The bold black arrows indicate 
   lines-of-sight with different viewing inclinations; LOS$_{\rm 00}$ and LOS$_{\rm SA}$ are 
   the two lines-of-sight along which an observer would see C\,{\sc iv}$_{\rm 00}$ 
   and  C\,{\sc iv}$_{\rm SA}$ troughs, respectively. }
   \label{fig20}
\end{figure*}

Considering the LOS$_{\rm 00}$ 
and LOS$_{\rm SA}$ lines-of-sight in Figure~\ref{fig20}, we 
present below a comparative assessment of the expected properties of C\,{\sc iv}$_{\rm 00}$ 
and C\,{\sc iv}$_{\rm SA}$ troughs, relating these to the observational findings above (we then 
discuss C\,{\sc iv}$_{\rm S0}$ troughs as an intermediate case). 
Specifically, we consider BAL-trough profile properties (e.g., depth, width, EW, velocity, and
profile shape) and BAL-variability characteristics (e.g., EW and fractional EW variation 
strengths, depth variation profiles). Our comparative assessment points are the following: 

\begin{enumerate}

\item 
The column density of outflowing gas is considerably larger along LOS$_{\rm SA}$ than along
LOS$_{\rm 00}$, while the line-of-sight covering factors need not differ substantially. 
If the column density plays a role in setting trough depth, it is expected that 
C\,{\sc iv}$_{\rm SA}$ troughs will  generally be deeper than C\,{\sc iv}$_{\rm 00}$ troughs
(observational findings \ref{obf1} and \ref{obf2} above). 
While some BAL quasars are known to have highly 
saturated C\,{\sc iv} troughs with depths largely set by the line-of-sight covering factor 
(rather than column density; see Section~\ref{hmlintro}), it is not clear 
that all C\,{\sc iv} troughs are highly saturated. Indeed, variability studies suggest 
that some C\,{\sc iv} troughs are not highly saturated (see Section~\ref{hmlintro}). 
The C\,{\sc iv} troughs with detailed previous 
studies showing strong saturation are C\,{\sc iv}$_{\rm S0}$ or C\,{\sc iv}$_{\rm SA}$ 
troughs, while, to our knowledge, no C\,{\sc iv}$_{\rm 00}$ troughs have been demonstrated 
to be highly saturated. 
Broadly consistent with this, we note that P\,{\sc v}  absorption corresponding to 
C\,{\sc iv}$_{\rm 00}$ troughs is rare and weak (see Section~\ref{mgp}).

\item 
Given the poloidal velocity field of the model, we expect
C\,{\sc iv}$_{\rm 00}$ troughs to have generally higher minimum
outflow velocities than C\,{\sc iv}$_{\rm SA}$ troughs, as
observed (see observational finding~\ref{obf3}). This is because
LOS$_{\rm 00}$ intersects gas with a high outflow velocity
without intersecting much gas with a low outflow velocity.
LOS$_{\rm SA}$, on the other hand, primarily intersects gas
with a low outflow velocity. LOS$_{\rm SA}$ might also intersect
gas with a high outflow velocity if such gas extends close to the 
accretion-disk surface (i.e., prompt acceleration) at small
radii, as appears required by our results showing high $v_{\rm max}$ 
values for C\,{\sc iv}$_{\rm SA}$ troughs. These same considerations
can also explain the larger velocity widths of C\,{\sc iv}$_{\rm SA}$
troughs compared to C\,{\sc iv}$_{\rm 00}$ troughs (see observational
finding~\ref{obf2}). {Figure~\ref{fig20} (lower panel) shows that 
the velocities for LOS$_{\rm 00}$ appear to be 5000--10000 km~s$^{-1}$ 
along essentially the entire line-of-sight, and this is comparable to our 
measured $v_{\rm min}$ values for C\,{\sc iv}$_{\rm 00}$ troughs.}

\item 
Given that the EW of a trough is set by a combination of its depth and width, 
we expect from the two comparative assessment points above that C\,{\sc iv}$_{\rm SA}$ 
troughs will have larger EWs than C\,{\sc iv}$_{\rm 00}$ troughs, as observed
(see observational findings~\ref{obf1} and \ref{obf2}).

\item 
From the model we expect that Al\,{\sc iii} BALs will be mainly formed in the 
region close to the disk with high density and small poloidal velocity, while 
C\,{\sc iv} BALs will be formed within both high-velocity and low-velocity gas. 
Thus, we expect that Al\,{\sc iii} troughs will reside within the lower
velocity portions of C\,{\sc iv} troughs, as observed  \citep[observational finding~\ref{obf1}
and][]{voit93}. 

\item 
The model shows that the optical depth is velocity-dependent, and that it is 
generally higher for low poloidal velocities. Therefore, the low-velocity portions of 
C\,{\sc iv}$_{\rm SA}$ troughs that align with corresponding Al\,{\sc iii} troughs 
are likely to be more saturated than the high-velocity portions, perhaps
partly leading to their larger depths (observational finding~\ref{obf1}). 
Therefore these portions should be  less variable than the high-velocity portions. 
This behavior is observed (observational finding~\ref{obf4}).

\item 
Previous studies have presented evidence that ionization-level changes likely 
have a significant role in driving some BAL variability (see Section~\ref{hmlintro}). 
Strongly saturated lines will be 
less susceptible to variability driven by ionization-level changes. Given that
C\,{\sc iv}$_{\rm SA}$ troughs are likely more saturated than C\,{\sc iv}$_{\rm 00}$ 
troughs, they are expected to be less variable. This is observed in an 
absolute sense and even more strongly in a fractional sense
(observational findings~\ref{obf4}--\ref{obf6}).

 \item 
The model  indicates that the C\,{\sc iv} optical depth along LOS$_{\rm SA}$ is substantially 
larger than the Al\,{\sc iii} optical depth. Therefore, Al\,{\sc iii} troughs are expected to be 
more variable than C\,{\sc iv}$_{\rm SA}$ troughs, {as they should be less saturated. }
This behavior is observed in an absolute sense and even more strongly in a fractional 
sense (observational finding~\ref{obf8}).

\end{enumerate}

The basic expectations of the disk-wind model for the characteristics of the C\,{\sc iv}$_{\rm 00}$ and 
C\,{\sc iv}$_{\rm SA}$ samples show    qualitative agreement with our observational findings. 
In this model, a line-of-sight along which an observer would see a C\,{\sc iv}$_{\rm S0}$ trough 
is expected to intercept at least some less ionized gas than LOS$_{\rm 00}$. 
Consistent with this expectation, our 
observational findings show that the C\,{\sc iv}$_{\rm S0}$ sample is an intermediate 
case between the C\,{\sc iv}$_{\rm 00}$ and C\,{\sc iv}$_{\rm SA}$ samples (observational findings 1--7).
\footnote{{We have
also performed basic testing with the C\,{\sc iv}$_{\rm s0}$,  C\,{\sc iv}$_{\rm sa}$, and 
C\,{\sc iv}$_{\rm Sa}$  samples described in Section~\ref{identity2}, and we generally find them to show suitably
intermediate properties as well.}}

There are a number of ways the results above might be advanced, and here
we highlight four that appear particularly promising. First, for the
reasons discussed in Section~\ref{hmlintro}, our current work has made use of the
strong C\,{\sc iv}, Si\,{\sc iv}, and Al\,{\sc iii} BAL transitions as a basic measure of
average line-of-sight ionization level. These transitions, spanning a
factor of $\approx 2.5$ in ionization potential, have served effectively
for our work. However, this ionization-potential range could be
expanded with the use of additional lower ionization
(e.g., Mg\,{\sc ii}, Fe\,{\sc ii}, and Fe\,{\sc iii}) and higher
ionization (e.g., Ne\,{\sc vii}, N\,{\sc v}, O\,{\sc vi}) transitions, thereby presumably 
probing wind
zones even closer to and further from, respectively, the accretion
disk. Second, the profiles and variability of the emission lines for
large samples of BAL quasars with C\,{\sc iv}$_{\rm 00}$, 
C\,{\sc iv}$_{\rm S0}$, and C\,{\sc iv}$_{\rm SA}$
troughs should be measured systematically and compared with
predictions (e.g., Murray \& Chiang 1997; Flohic et~al. 2012).
For a flattened Broad Line Region geometry, to first order one might
expect the emission lines for BAL quasars with C\,{\sc iv}$_{\rm SA}$ troughs to be
generally the broadest (though different emission lines,
tracing different phases of the Broad Line Region, may behave
differently). Third, the ongoing BOSS ancillary project and
upcoming SDSS-IV Time Domain Spectroscopic Survey (TDSS)\footnote{The current
planning for SDSS-IV is briefly described at http://www.sdss3.org/future/}
observations will both enlarge the sample size and improve the
temporal sampling pattern for BAL quasars. This will allow variability
to be used even more effectively as a tool for assessing correlated changes
of ionization level, kinematics, and column density. Finally, while we
have found generally good qualitative agreement with expectations for
the disk-wind model, our ability to perform quantitative comparisons
has been limited by the available simulation results. Future
simulations capable of predicting trough profiles and variability 
(e.g., Higginbottom et~al. 2013; D. Proga 2013, priv. comm.) can be
quantitatively tested and constrained using large-sample measurements
such as those provided here. Alternatives to the disk-wind model should also be developed 
to the point where quantitative testing is possible. 

\vspace{9mm}

We gratefully acknowledge financial support from National Science Foundation grant 
AST-1108604 (N.F.A., W.N.B., D.P.S.) and from NSERC (P.B.H.). We thank  N. Arav and 
M. Eracleous for helpful discussions. We thank D. Proga for allowing us to adapt Figure~2 
of \citet{proga00} as our Figure~\ref{fig20}. 
{We also thank the anonymous referee for constructive feedback.}

Funding for SDSS-III has been provided by the Alfred P. Sloan Foundation, the Participating 
Institutions, the National Science Foundation, and the U.S. Department of Energy Office of Science. 
The SDSS-III web site is http://www.sdss3.org/.

SDSS-III is managed by the Astrophysical Research Consortium for the Participating Institutions 
of the SDSS-III Collaboration including the University of Arizona, the Brazilian Participation Group, 
Brookhaven National Laboratory, Carnegie Mellon University, University of Florida, the French 
Participation Group, the German Participation Group, Harvard University, the Instituto de Astrofisica 
de Canarias, the Michigan State/Notre Dame/JINA Participation Group, Johns Hopkins University, 
Lawrence Berkeley National Laboratory, Max Planck Institute for Astrophysics, Max Planck Institute 
for Extraterrestrial Physics, New Mexico State University, New York University, Ohio State University, 
Pennsylvania State University, University of Portsmouth, Princeton University, the Spanish Participation 
Group, University of Tokyo, University of Utah, Vanderbilt University, University of Virginia, University of 
Washington, and Yale University.

\bibliographystyle{apj}

		\end{document}